\documentclass[final,3p,times]{elsarticle}

\usepackage{amssymb}
\usepackage{amsmath}
\usepackage{amsthm}

\usepackage{amsgen,amstext}
\usepackage{amsfonts,bm}
\usepackage{dsfont}
\usepackage{mathrsfs}
\usepackage{latexsym}
\usepackage[utf8]{inputenc} 
\usepackage[T1]{fontenc}    
\usepackage{hyperref}       
\usepackage{booktabs}       
\usepackage{nicefrac}       
\usepackage{microtype}      
\usepackage{caption,subcaption}
\usepackage{xcolor}
\usepackage[section]{algorithm}
\usepackage{braket}
\usepackage{algpseudocode}

\newcommand{\cE}{{\mathcal{E}}}

\newcommand{\cL}{{\mathcal{L}}}

\newcommand{\cP}{{\mathcal{P}}}

\newcommand{\cT}{{\mathcal{T}}}
\newcommand{\cW}{{\mathcal{W}}}

\newcommand{\A}{\mathbf{A}}
\newcommand{\B}{\mathbf{B}}
\newcommand{\C}{\mathbf{C}}
\newcommand{\D}{\mathbf{D}}
\newcommand{\bE}{\mathbf{E}}
\newcommand{\F}{\mathbf{F}}
\newcommand{\G}{\mathbf{G}}
\newcommand{\bH}{\mathbf{H}}
\newcommand{\I}{\mathbf{I}}
\newcommand{\J}{\mathbf{J}}
\newcommand{\K}{\mathbf{K}}
\newcommand{\bL}{\mathbf{L}}
\newcommand{\bO}{\mathbf{O}}
\newcommand{\bP}{\mathbf{P}}
\newcommand{\Q}{\mathbf{Q}}
\newcommand{\R}{\mathbf{R}}
\newcommand{\bS}{\mathbf{S}}
\newcommand{\T}{\mathbf{T}}
\newcommand{\U}{\mathbf{U}}
\newcommand{\V}{\mathbf{V}}
\newcommand{\W}{\mathbf{W}}
\newcommand{\X}{\mathbf{X}}
\newcommand{\Y}{\mathbf{Y}}
\newcommand{\Z}{\mathbf{Z}}
\newcommand{\ba}{\mathbf{a}}
\newcommand{\bb}{\mathbf{b}}

\newcommand{\f}{\mathbf{f}}
\newcommand{\bg}{\mathbf{g}}

\newcommand{\bk}{\mathbf{k}}
\newcommand{\be}{{\mathbf{e}}}
\newcommand{\ones}{\mathbf{1}}
\newcommand{\bu}{\mathbf{u}}
\newcommand{\bv}{\mathbf{v}}
\newcommand{\x}{\mathbf{x}}
\newcommand{\bw}{\mathbf{w}}
\newcommand{\bz}{\mathbf{z}}
\newcommand{\y}{\mathbf{y}}
\newcommand{\bzero}{\mathbf{0}}
\newcommand{\M}{\mathbf{M}}
\newcommand{\uu}{\mathbf{u}}
\newcommand{\ff}{\mathbf{f}}
\newcommand{\yy}{\mathbf{y}}

\newcommand{\Ints}{{\mathbb{Z}}}
\newcommand{\Nats}{{\mathbb{N}}}
\newcommand{\Reals}{\mathbb{R}}
\newcommand{\Complex}{{\mathbb{C}}}

\newcommand{\bbC}{\mathbb{C}}
\newcommand{\braopket}[3]{\langle #1 | #2 | #3\rangle} 

\newcommand{\bpi}{{\boldsymbol\pi}}
\newcommand{\balpha}{{\boldsymbol{\alpha}}}

\newtheorem{theorem}{Theorem}[section]
\newtheorem{lemma}[theorem]{Lemma}
\newtheorem{remark}[theorem]{Remark}

\newcommand{\E}{\mathbb{E}}

\newcommand{\pr}{\mathbb{P}}

\newcommand{\poly}{\mathrm{poly}}

\newcommand{\dotleq}{\dot{\leq}}

\newcommand{\mss}[1]{{#1}}
\newcommand{\su}[1]{{#1}}

\begin{document}

\begin{frontmatter}



\title{On Efficient Solutions of \mss{General} Structured Markov Processes \\ in Quantum Computational Environments}


\author{Vasileios Kalantzis, Mark S.\ Squillante, Shashanka Ubaru} 

\affiliation{
organization={Mathematics of Computation, IBM Research},
            addressline={Thomas J.\ Watson Research Center}, 
            city={Yorktown Heights},
            postcode={10598}, 
            state={NY},
            country={USA}}

\begin{abstract}
We study 
\mss{from a theoretical viewpoint}
the fundamental problem of efficiently computing the stationary distribution of general classes of structured Markov processes.
In
strong contrast with previous work, we consider this \mss{fundamental} problem within the context of quantum computational environments
{from a mathematical perspective}
and devise the first quantum algorithms for computing the stationary distribution of \mss{general} structured Markov processes.
We derive a mathematical analysis of the computational
properties
of our quantum algorithms together with related theoretical results, establishing that our quantum algorithms provide
{the potential for significant computational improvements}
over {that of} the best-known 
and most-efficient
classical algorithms in
{various}
settings of both theoretical and practical importance.
Although motivated by \mss{general} structured Markov processes, our quantum algorithms
\mss{can be exploited}
to address a much larger class of numerical computation problems\mss{, as well as to potentially play the role of a subroutine as part of solving larger computational problems involving the stationary distribution on a quantum computer}.
\end{abstract}




\end{frontmatter}



\section{Introduction}
Multidimensional Markov processes arise 
\mss{naturally and often}
in many 
aspects of the mathematical performance analysis, modeling and optimization of computer systems and networks.
Within this context,
\mss{the}
general classes of structured Markov processes 
\mss{span the broad spectrum of such stochastic models, and thus they}
are of 
particular importance in both theory and practice,
\mss{comprising}
\emph{M/G/$1$-type} processes, \emph{G/M/$1$-type} processes and \emph{quasi-birth-and-death} (QBD) processes whose transitions on
\mss{a}
two-dimensional lattice are respectively skip-free to the left, skip-free to the right, and skip-free to the left and to the right, with no restrictions upward or downward~\cite{Neut81,Neut89}.
Our focus in this paper is on 
\mss{these very}
general classes of
structured Markov processes.

More formally, consider a general discrete-time structured Markov process $\{ X(n) \, ; \, n \in \Ints_+ \}$ on the two-dimensional state space $\hat{\Omega} = \{ (i,j) : i \in \Ints_+, j \in [M] \}$ with transition probability matrix $\bP$, where 
$\Ints_+ := \Ints_{\geq 0}$, $\Ints_{\geq k} := \{ k, k+1, k+2, \ldots \}$, 
$[M] := \{ 1, \ldots, M \}$, and $M$ can be finite or infinite.
The 
transition probability matrix $\bP$
of the M/G/$1$-type
process 
and
the 
G/M/$1$-type process 
has the block Toeplitz-like structure
\begin{equation}
\bP_{\scriptscriptstyle M/G/1} = \begin{bmatrix}
    \B_0&\B_1&\B_2&\B_3&\cdots\\
    \A_{-1}&\A_0&\A_1&\A_2&\cdots\\
    \bzero&\A_{-1}&\A_0&\A_1&\ddots\\
    \vdots&\vdots&\ddots&\ddots&\ddots\\
    \end{bmatrix} 
\qquad \mbox{ and } \qquad
\bP_{\scriptscriptstyle G/M/1} = \begin{bmatrix}
    \B_0&\A_1&\bzero&\bzero&\cdots\\
    \B_{-1}&\A_0&\A_1&\bzero&\cdots\\
    \B_{-2}&\A_{-1}&\A_0&\A_1&\ddots\\
    \vdots&\vdots&\ddots&\ddots&\ddots\\
    \end{bmatrix} ,
\label{eq:P-matrix}
\end{equation}
respectively,
where $\B_i,\, i\in\Ints_+$, and $\A_{i},\, i\in \Ints_{\geq -1}$, for $\bP_{\scriptscriptstyle M/G/1}$ are nonnegative matrices in $\Reals^{M\times M}$ such that (s.t.) $\A = \sum_{i=-1}^\infty \A_{i}$ and $\B = \sum_{i=0}^\infty \B_i$ are stochastic,
and where $\B_{-i},\, i\in\Ints_+$, and $\A_{-i},\, i\in \Ints_{\geq -1}$, for $\bP_{\scriptscriptstyle G/M/1}$ are nonnegative matrices in $\Reals^{M\times M}$
s.t.\
$\B_{-j} + \sum_{i=-1}^{j-1} \A_{-i}$ is stochastic for all $j\in\Ints_+$.
Markov processes embedded at service completion epochs for the
classical 
M/G/$1$ queue have transition probability matrices
of the form of 
$\bP_{\scriptscriptstyle M/G/1}$.
Similarly, Markov processes 
embedded at arrival instant epochs for the
classical 
G/M/$1$ queue have transition probability matrices
of the form of 
$\bP_{\scriptscriptstyle G/M/1}$.
At the intersection of 
these two
processes is the QBD process whose transition probability matrix $\bP_{\scriptscriptstyle QBD}$ has the block tridiagonal Toeplitz-like structure given in~\eqref{eq:P-matrix} with $\A_i = \A_{-i} = \B_i = \B_{-i} = \bzero$ for 
$i\in\Ints_{\geq 2}$, 
$\B_1 = \A_1$ and $\B_{-1} = \A_{-1}$
s.t.\
$\A_{-1}+\A_0+\A_1$ and $\B_0+\A_1$ are stochastic.

Our objective is to obtain the stationary distribution of the 
\mss{very}
general classes of M/G/$1$-type, G/M/$1$-type and QBD processes, namely those with transition probability matrices of the form~\eqref{eq:P-matrix}.
Define
\[
{\bpi} := ( {\bpi}_0 , {\bpi}_1 , {\bpi}_2 , \ldots ) , \qquad
{\bpi}_i := ( \hat{\pi} (i,1), \hat{\pi} (i,2), \ldots, \hat{\pi} (i,M) ) , \qquad
\hat{\pi} (i,j) := \lim_{n \rightarrow \infty} \pr[ X (n) = (i,j) ] , \quad \forall (i,j) \in \hat{\Omega} .
\]
The limiting probability vector ${\bpi}$ is the stationary distribution for the
stochastic 
process $\{ X(n) \, ; \, n \in \Ints_+ \}$.
Assuming this process to be irreducible and ergodic, its invariant 
probability 
vector $\bpi$ exists and is uniquely determined as the solution of $\bpi = \bpi \bP$ and $\bpi\ones =1$, where $\ones = (1,\ldots,1)^\top$~\cite{LevPer2017,Bremaud2020}.
Once the stationary distribution is obtained, various functions of 
the probability vector 
$\bpi$ 
can be used to obtain
performance
measures and quantities 
of interest such as those associated with 
so-called
\emph{levels} $\cL(i) := \{ (i,j) : j \in [M] \}$,
$i \in \Ints_+$,
and 
so-called
\emph{phases} $\cP(j) : = \{ (i,j) : i \in \Ints_+ \}$, $j \in [M]$.
Within the context of queueing systems, this includes the queue-length tail distribution, the expected queue length, the expected sojourn time, and so on.
The stationary distribution together with these 
associated 
measures and quantities can then be leveraged as a core component of various forms of mathematical performance analysis, modeling and optimization.

There are two primary forms of mathematical analysis to determine the stationary distribution.
One 
primary
form concerns analytical approaches based in part on exploiting probabilistic
interpretations~\cite{Neut81,Neut89}
of structural properties of the invariant probability vector
$\bpi$~\cite{rank1,liu,SquNel91,NelSqu96b,Squi98,LeeWin06,LeSqWi09}.
The second primary form
of mathematical analysis 
concerns computational approaches based in part on numerical analysis and
numerical 
methods.
The best-known and most-efficient algorithms of this approach for computing
the 
invariant probability vector 
$\bpi$ of structured Markov processes
consist mostly of 
different variants of cyclic reduction
(CR)
methods.
Buzbee, Golub and Nielsen~\cite{BuGoNi70} 
originally proposed the method of
CR
to solve certain block tridiagonal systems that arise in the numerical solution of elliptical equations.
Motivated by problems in queueing theory, Bini and Meini subsequently developed various adaptations and extensions of
CR
for solving structured Markov
processes~\cite{BinMei95,BinMei96,BinMei97}, 
followed by 
further extensions
in the book of 
Bini, Latouche and Meini~\cite{BiLaMe05}.
This
computational 
approach comprising
CR
methods represents the most efficient solutions for computing the stationary distribution $\bpi$ of structured Markov processes on classical digital computers.

Our
{center of attention}
herein is on these computational approaches to determining the stationary distribution of structured Markov processes.
Even with the computational benefits of
CR,
the time required to compute the stationary distribution can still be prohibitive for large complex stochastic 
performance
models of interest.
Quantum computers offer the potential of achieving significant
\mss{performance}
advantages for certain computational problems, with the possibility of delivering exponential speedups over the best-known and most-efficient solutions on classical computers.
However, despite this great potential, a number of important cases have been identified where quantum algorithms provide modest or no benefits over the most-efficient classical algorithms~\cite{peruzzo2014variational,rebentrost2014quantum,biamonte2017quantum,schuld2015introduction,schuld2019quantum,chia2020sampling,chepurko2020quantum,zhou2020quantum,cerezo2021variational,tang2021quantum}.
Moreover,
to the best of our knowledge, 
there are no quantum algorithms 
for computing the stationary distribution of structured Markov processes and thus no quantum algorithms that realize the great potential of 
significant
speedup 
{over any aspect of the best-known and most-efficient classical algorithms}
for
this class of numerical computations.
In 
\mss{the theoretical study of}
this paper, we 
are the first 
to address both of these critically important 
problems,
{with
a
\mss{particular}
interest
on the computation phase in the decision-space of quantum computers from a mathematical perspective.}

\textbf{Our Contributions.}
We
focus 
\mss{our theoretical study}
on methodological solutions of 
\mss{the very}
general classes of structured Markov processes from the novel perspective of quantum computing, making
the following 
important
contributions.
To start, we derive
the first quantum algorithms for computing the stationary distribution of structured Markov processes. 
No previous work, to the best of our knowledge, 
has developed quantum algorithms to address this important
problem
for mathematical performance analysis, modeling and optimization.
Second, we derive from first principles a rigorous mathematical analysis of the computational errors of our quantum
algorithms
and 
related theoretical results.
This includes a detailed derivation of classical 
CR
methods for structured Markov processes, given the complexity of these algorithms and a 
{possible}
lack of their broad familiarity, from which we discovered and corrected important
subtle errors
in the corresponding analysis, theoretical results, and algorithmic details
of
the existing
research 
literature~\cite{BiLaMe05}.
Third, we derive a rigorous mathematical analysis of the computational complexity of our quantum algorithms {for computing the stationary distribution
and} 
related theoretical results.
This includes:
\mss{($i$)}
establishing the potential
exponential speedup over the
best-known and 
most-efficient
classical algorithms for
the computation phase in the decision-space of quantum computers, which is
\mss{of specific}
interest
from a mathematical perspective;
\mss{($ii$)}
establishing the potential 
polynomial-to-exponential speedup over the
best-known and
most-efficient
classical algorithms 
when
the properties of the input matrices together with block encoding allow efficient data loading;
and
\mss{($iii$)}
establishing the potential 
polynomial-to-exponential speedup over the
best-known and
most-efficient
classical algorithms 
when
the properties of the output matrix
\mss{together with sampling-based methods 
allow efficient result readout.}
This 
also 
expands the limited set of quantum algorithms that provably achieve 
significant computational improvements in various settings of importance,
while 
further
making 
\mss{significant}
algorithmic and theoretical contributions.
\mss{It is additionally important \su{to} note that the computation of the stationary distribution of general structured Markov processes may play the role of a subroutine within the context of larger computational problems on the quantum computer, such as performance optimization problems involving the stationary distribution of the associated structured Markov process\su{. In such cases} the computation phase would be more critical than the input and output phases to the overall computational complexity.}
{Finally,}
since
CR
is core to many important numerical methods,
{our quantum algorithms can be exploited to address a much larger class of numerical computation problems.}

The
remainder of this 
paper is organized as follows.
Section~\ref{sec:prelim} presents
technical 
preliminaries related to structured Markov processes,
classical computational methods 
and quantum computing, 
deferring 
\mss{many aspects of}
our detailed derivation of 
classical CR methods and related theoretical
results
to Appendix~\ref{app:classical}
because of space limitations.
In Section~\ref{sec:quantum} we present our 
quantum 
algorithms
for computing the stationary distribution
of structured Markov processes
together with 
our mathematical analysis of these
quantum 
algorithms.
Section~\ref{sec:proofs} provides the proofs of our theoretical results,
followed by
concluding remarks.
Additional technical details for most sections are provided in the appendices.

\section{Technical Preliminaries}
\label{sec:prelim}
Our
{center of attention}
is on the 
\mss{theoretical}
solution of 
\mss{the very}
general classes of structured Markov processes in quantum computational environments
{from a mathematical perspective}.
Therefore, 
due to space limitations, 
we 
\mss{predominantly}
focus in the main body of this paper on the majority of our primary contributions with respect to (w.r.t.) quantum algorithms and corresponding mathematical analysis and theoretical results; 
and thus we confine
\mss{many aspects of}
our detailed derivation of classical 
CR
methods for structured Markov processes
to Appendix~\ref{app:classical}.
%
Moreover, given
the
strong connections between M/G/$1$-type 
processes 
and G/M/$1$-type processes (as well as QBD processes at the 
intersection), 
and in particular the important duality between these two classes of Markov processes~\cite{AsmRam90,Rama90,BiLaMe05},
we henceforth focus without loss of generality on M/G/$1$/-type processes and computational approaches for determining the stationary distribution of such processes, noting that these computational approaches can also be similarly exploited for
G/M/$1$-type 
and QBD processes
{by}
making necessary alterations while not affecting the basic approach.
Furthermore, within a multiserver queueing-theoretic context, the same approach can be readily adapted to support M/G/$c$-type processes by making corresponding changes to
the boundary matrices of the transition probability matrix $\bP_{\scriptscriptstyle M/G/1}$ 
and to the associated invariant probability vector for the boundary
(i.e., expanding
both 
the boundary matrices in~\eqref{eq:P-matrix} and the boundary probability vector for $\cL(0)$ through $\cL(c-1)$),
both to reflect the 
parallel
service of $k$ customers
on individual servers when there are $k$ customers in the queueing system,
$k \in [c-1], \, c < \infty$.



Herein, we now present technical preliminaries on M/G/$1$-type processes and quantum computing,
with additional technical details respectively provided 
in Appendices~\ref{app:mg1} 
and~\ref{app:quantum},
as well as an overview of 
\mss{our detailed derivation of}
classical computation methods based on CR, with
\mss{many additional aspects of}
our detailed derivation
in Appendix~\ref{app:classical}.

\subsection{M/G/$1$-Type Markov Processes}
\label{sec:prelim:mg1}
The transition probability matrix $\bP_{\scriptscriptstyle M/G/1}$ of
M/G/$1$-type
processes
in~\eqref{eq:P-matrix} 
is stochastic in upper block Hessenberg form and is block Toeplitz except for its first block row.
Markov processes having this
form
are also called \emph{skip-free} to the left because the Markov process can only make one-step transitions from states in $\cL(i+1)$ to states in $\cL(i)$,
$i\in\Ints_+$, 
while being able to make one-step transitions from states in $\cL(i)$ to states in any $\cL(i+k)$,
$i\in\Ints_+$, $k \in\Nats := \Ints_{\geq 1}$,
together with arbitrary one-step transitions between states in
$\cL(i)$.
%
The key to our analysis
is the matrix equation
\begin{equation}
    \X \; = \; \sum_{i=-1}^\infty \A_i \X^{i+1} ,
\label{eq:4.4}
\end{equation}
from which the stationary distribution $\bpi$ is obtained as follows.
Define the time $\theta(\ell) := \min\{ n \geq 0 : X(n) \in \cL(\ell) \}$, i.e., the first entry time to level $\ell$, $\ell \in\Ints_+$,
and define the matrices
$$
\G^{(i)}_{jk} := \pr[ \theta(\ell) < \infty, \, X(\theta(\ell)) = (\ell,k) \, | \, X(0) = (i+\ell,j)] ,
$$
whose $jk$-th element denotes the probability that, starting from state
$(i+\ell,j) \in \cL(i+\ell)$ 
at time $0$, the process enters $\cL(\ell)$ for the first time in finite time with $(\ell,k)$ being the first state visited in $\cL(\ell)$.
Observe that any sequence of transitions that take the process from $\cL(i)$ to $\cL(0)$, i.e., $\ell=0$, also take the process from $\cL(i+1)$ to $\cL(1)$, i.e., $\ell=1$, and also take the process from $\cL(i+2)$ to $\cL(2)$, i.e., $\ell=2$, and so on.

The solution of the stationary distribution $\bpi$ 
w.r.t.~\eqref{eq:4.4}
is then given by a set of known theoretical results, which we provide 
in Appendix~\ref{app:mg1}.
In particular, the matrix $\G_{\min}$ is the minimal nonnegative solution of~\eqref{eq:4.4} in Lemma~\ref{lem4.2} and Theorem~\ref{thm4.3},
{where a solution $\X$ is called a minimal nonnegative solution of a matrix equation if $\bzero \leq \X \leq \Y$ for any other nonnegative solution 
$\Y$}.
The stationary distribution of the M/G/$1$-type 
process
$\{ X(n) \, ; \, n \in \Ints_+ \}$ 
is then given by
Theorems~\ref{thm4.4} and~\ref{thm4.8}, 
where Theorem~\ref{thm4.7} provides the desired ergodicity conditions.
%
We
assume throughout
that the M/G/$1$-type
process of interest is irreducible and ergodic and that $\sum_{i=-1}^\infty (i+1) \A_i \ones$ is finite.
Associated with this
process, define the scalar function $a(z) = \det(z\I - \varphi(z))$
w.r.t.\
the generating function $\varphi(z) := \sum_{i=-1}^\infty z^{i+1} \A_i$.
Then, we have
that 
$z=1$ is the only zero of modulus $1$ of the function $a(z)$.

Once the stationary distribution of the M/G/$1$-type Markov process is obtained,
various functions of $\bpi$ can be used to obtain performance measures and quantities of interest.
In particular, 
when the Markov process $\{ X(n) \, ; \, n \in \Ints_+ \}$ models a level-based queue-length process $Q$, then
the tail distribution of the queue-length process $Q$ 
and
the associated expectation
can be 
respectively
expressed as
$$
\pr[ Q > k ] = \sum_{i=k+1}^\infty {\bpi}_i \ones , \quad k \in \Ints_+ , \qquad \mbox{  and  } \qquad
\E[Q] = \sum_{i=1}^\infty i \bpi_i \ones .
$$
The corresponding expected sojourn time can be obtained from the latter expression via Little's law~\cite{Asmu03} as
$$\E[T] = \frac{\E[Q]}{\nu} = \frac{1}{\nu} \sum_{i=1}^\infty i \bpi_i \ones,$$
where $\nu$ is the mean arrival rate to the M/G/$1$-type queueing system.
In addition to
these
representative instances, the stationary distribution $\bpi$ can also be leveraged as a core component of more detailed mathematical 
performance 
analyses of M/G/$1$-type Markov processes
and their broad applications.

\subsection{Classical Computational Methods}
\label{sec:prelim:classical}
The best-known and most-efficient classical computational approaches for determining the stationary distribution $\bpi$ of the general class of M/G/$1$-type processes on digital computers consist
of CR methods based on adaptations and extensions developed in large part by Bini and Meini~\cite{BinMei95,BinMei96,BinMei97,BiLaMe05}. 
This approach concerns computing the minimal nonnegative solution $\G_{\min}$ of the nonlinear matrix equation~\eqref{eq:4.4} for such ergodic M/G/$1$-type processes
as a result of Theorem~\ref{thm4.3},
from which 
the invariant 
probability 
vector 
$\bpi$ is
obtained as a result of Theorems~\ref{thm4.4} and~\ref{thm4.8}.

The general CR approach basically comprises rewriting~\eqref{eq:4.4} in matrix form as
\begin{equation}
    \bH \begin{bmatrix} \G_{\min} \\ \G_{\min}^2 \\ \G_{\min}^3 \\ \G_{\min}^4 \\ \vdots \end{bmatrix} \; = \; \begin{bmatrix} \A_{-1} \\ \bzero \\ \bzero \\ \vdots \end{bmatrix} , \qquad\qquad
    \bH \; = \; \begin{bmatrix}
    \I - \A_0 & -\A_1 & -\A_2 & -\A_3 & \cdots \\
    -\A_{-1} & \I - \A_0 & -\A_1 & -\A_2 & \ddots \\
    \bzero & -\A_{-1} & \I - \A_0 & -\A_1 & \ddots \\
     \bzero & \bzero & -\A_{-1} & \I - \A_0 & \ddots \\
    \vdots & \vdots & \ddots & \ddots & \ddots \\
    \end{bmatrix} 
\label{lem4.2:4.5+4.6}
\end{equation}
from Lemma~\ref{lem4.2}, and applying variants of the CR method in functional form to solve~\eqref{lem4.2:4.5+4.6}.
More precisely, this process transforms the matrix equation~\eqref{eq:4.4} into a semi-infinite linear system that leads to the linear system~\eqref{lem4.2:4.5+4.6}.
In the first iteration ($n=1$) of
CR,
we
apply an even-odd permutation to the block rows and to the block columns of~\eqref{lem4.2:4.5+4.6}, 
deriving in compact form
\begin{equation}
    \begin{bmatrix}
    \I - \U_{11} & -\U_{12} \\
    - \U_{21} & \I -\U_{22}  \\
    \end{bmatrix}
    \begin{bmatrix}
    \x_{+} \\ \x_{-} \\
    \end{bmatrix}
    =
    \begin{bmatrix}
    \bzero \\ \bb \\
    \end{bmatrix} 
\label{eq:compact}
\end{equation}
where 
\begin{align}
    \U_{11} & 
    =
    \begin{bmatrix}
    \A_0 & \A_2 & \A_4 & \cdots \\
    \bzero & \A_0 & \A_2 & \ddots \\
    \bzero & \bzero & \A_0 & \ddots \\
    \vdots & \ddots & \ddots & \ddots \\
    \end{bmatrix} ,
    \quad & 
    \U_{12} =
    \begin{bmatrix}
    \A_{-1} & \A_1 & \A_3 & \cdots \\
    \bzero & \A_{-1} & \A_1 & \ddots \\
    \bzero & \bzero & \A_{-1} & \ddots \\
    \vdots & \ddots & \ddots & \ddots \\
    \end{bmatrix} ,
    \qquad
    \x_{+} & 
    =
    \begin{bmatrix}
    \G_{\min}^2 \\
    \G_{\min}^4 \\
    \G_{\min}^6 \\
    \vdots \\
    \end{bmatrix} , \label{eq:Us+x+} \\
    \U_{21} & 
    =
    \begin{bmatrix}
    \A_1 & \A_3 & \A_5 & \cdots \\
    \A_{-1} & \A_1 & \A_3 & \ddots \\
    \bzero & \A_{-1} & \A_1 & \ddots \\
    \vdots & \ddots & \ddots & \ddots \\
    \end{bmatrix} ,
    \quad
    & 
    \U_{22} =
    \begin{bmatrix}
    \A_0 & \A_2 & \A_4 & \cdots \\
    \bzero & \A_0 & \A_2 & \ddots \\
    \bzero & \bzero & \A_0 & \ddots \\
    \vdots & \ddots & \ddots & \ddots \\
    \end{bmatrix} ,
    \qquad
    \x_{-}
    & 
    =
    \begin{bmatrix}
    \G_{\min} \\
    \G_{\min}^3 \\
    \G_{\min}^5 \\
    \vdots \\
    \end{bmatrix} \label{eq:Us+x-} ,
\end{align}
and $\bb = [ \A_{-1} \: \bzero \: \bzero \: \cdots]^\top$.

We next perform
the first iteration of block Gaussian elimination on the $2\times 2$ block system in~\eqref{eq:compact}.
The semi-infinite matrices $\I-\U_{22}$, $\U_{21}$, $(\I - \U_{11})^{-1}$ and $\U_{12}$ define bounded operators, and therefore the one iteration of block Gaussian elimination yields
\begin{equation}
    \begin{bmatrix}
    \I - \U_{11} & -\U_{12} \\
    \bzero & \bH^{(1)}  \\
    \end{bmatrix}
    \begin{bmatrix}
    \x_{+} \\ \x_{-} \\
    \end{bmatrix}
    =
    \begin{bmatrix}
    \bzero \\ \bb \\
    \end{bmatrix} ,
\label{eq7.34a}
\end{equation}
where 
\begin{equation}
    \bH^{(1)} = \I -\U_{22} - \U_{21}(\I - \U_{11})^{-1} \U_{12} 
\label{eq7.34b}
\end{equation}
is the Schur complement of $\I -\U_{22}$, and superscript $(1)$ indicates the first iteration of block Gaussian elimination.
From this we obtain
\begin{equation}
    \bH^{(1)} \begin{bmatrix} \G_{\min} \\ \G_{\min}^3 \\ \G_{\min}^5 \\ \G_{\min}^7 \\ \vdots \end{bmatrix}
    = 
    \begin{bmatrix}  \A_{-1} \\ \bzero \\ \bzero \\ \vdots \end{bmatrix} , \qquad\qquad
    \bH^{(1)} = \begin{bmatrix}
    \I - \hat{\A}_0^{(1)} & -\hat{\A}_1^{(1)} & -\hat{\A}_2^{(1)} & -\hat{\A}_3^{(1)} & \cdots \\
    -\A_{-1}^{(1)} & \I - \A_0^{(1)} & -\A_1^{(1)} & -\A_2^{(1)} & \ddots \\
    \bzero & -\A_{-1}^{(1)} & \I - \A_0^{(1)} & -\A_1^{(1)} & \ddots \\
    \bzero & \bzero & -\A_{-1}^{(1)} & \I - \A_0^{(1)} & \ddots \\
    \vdots & \vdots & \ddots & \ddots & \ddots \\
    \end{bmatrix} ,
\label{eq7.35}
\end{equation}
where
the block matrices $\A_i^{(1)}$ and $\hat{\A}_{i+1}^{(1)}$, $i\in \Ints_{\geq -1}$, are defined in explicit functional form by means of recursions w.r.t.\ 
formal matrix power series.
Upon observing that the 
transformation~\eqref{eq7.35} 
maintains the same structure of the original system~\eqref{lem4.2:4.5+4.6},
we then establish by
an inductive argument
that the above process of applying a block even-odd permutation and performing one iteration of block Gaussian elimination can be repeated recursively.
Specifically, after $n$ iterations of this
CR 
process, we derive in general the system
\begin{equation}
    \bH^{(n)} \begin{bmatrix} \G_{\min} \\ \G_{\min}^{2^n+1} \\ \G_{\min}^{2\cdot 2^n+1} \\ \G_{\min}^{3\cdot 2^n+1} \\ \vdots  \end{bmatrix} \; = \; \begin{bmatrix} \A_{-1} \\ \bzero \\ \bzero \\ \vdots \end{bmatrix} , \qquad\qquad
    \bH^{(n)} = \begin{bmatrix}
    \I - \hat{\A}_0^{(n)} & -\hat{\A}_1^{(n)} & -\hat{\A}_2^{(n)} & -\hat{\A}_3^{(n)} & \cdots \\
    -\A_{-1}^{(n)} & \I - \A_0^{(n)} & -\A_1^{(n)} & -\A_2^{(n)} & \ddots \\
    \bzero & -\A_{-1}^{(n)} & \I - \A_0^{(n)} & -\A_1^{(n)} & \ddots \\
    \bzero & \bzero & -\A_{-1}^{(n)} & \I - \A_0^{(n)} & \ddots \\
    \vdots & \vdots & \ddots & \ddots & \ddots \\
    \end{bmatrix} ,
\label{eq7.37}
\end{equation}
where 
the matrix $\bH^{(n)}$ of the infinite system at each iteration $n$ is fully characterized by its first and second block rows, and
the block matrices $\A_i^{(n)}$ and $\hat{\A}_{i+1}^{(n)}$, $i\in \Ints_{\geq -1}$, are defined in explicit functional form by means of recursions w.r.t.\ 
formal matrix power series.

Let $\Pi$ denote the permutation matrix associated with the block even-odd permutation of the block Hessenberg matrix $\bH^{(n)}$
s.t.\
\begin{equation}
    \Pi \bH^{(n)} \Pi^\top \; = \;
    \begin{bmatrix}
    \I - \U_{11}^{(n)} & -\U_{12}^{(n)} \\
    - \U_{21}^{(n)} & \I -\U_{22}^{(n)}  \\
    \end{bmatrix} ,
\label{eq7.54}
\end{equation}
where 
\begin{align}
    \U_{11}^{(n)}
    & 
    =
    \begin{bmatrix}
    \A_0^{(n)} & \A_2^{(n)} & \A_4^{(n)} & \cdots \\
    \bzero & \A_0^{(n)} & \A_2^{(n)} & \ddots \\
    \bzero & \bzero & \A_0^{(n)} & \ddots \\
    \vdots & \ddots & \ddots & \ddots \\
    \end{bmatrix},
    \qquad
    & 
    \U_{12}^{(n)} =
    \begin{bmatrix}
    \A_{-1}^{(n)} & \A_1^{(n)} & \A_3^{(n)} & \cdots \\
    \bzero & \A_{-1}^{(n)} & \A_1^{(n)} & \ddots \\
    \bzero & \bzero & \A_{-1}^{(n)} & \ddots \\
    \vdots & \ddots & \ddots & \ddots \\
    \end{bmatrix},
    \label{eqU^na}
    \\
    \U_{21}^{(n)} 
    & 
    =
    \begin{bmatrix}
    \hat{\A}_1^{(n)} & \hat{\A}_3^{(n)} & \hat{\A}_5^{(n)} & \cdots \\
    \A_{-1}^{(n)} & \A_1^{(n)} & \A_3^{(n)} & \ddots \\
    \bzero & \A_{-1}^{(n)} & \A_1^{(n)} & \ddots \\
    \vdots & \ddots & \ddots & \ddots \\
    \end{bmatrix},
    \qquad
    & 
    \U_{22}^{(n)} =
    \begin{bmatrix}
    \hat{\A}_0^{(n)} & \hat{\A}_2^{(n)} & \hat{\A}_4^{(n)} & \cdots \\
    \bzero & \A_0^{(n)} & \A_2^{(n)} & \ddots \\
    \bzero & \bzero & \A_0^{(n)} & \ddots \\
    \vdots & \ddots & \ddots & \ddots \\
    \end{bmatrix}.
    \label{eqU^nb}
\end{align}
Hence, the main computational task of CR consists of computing the first two block rows of the Schur complement matrix
\begin{equation}
    \bH^{(n+1)} = \I -\U_{22}^{(n)} - \U_{21}^{(n)}(\I - \U_{11}^{(n)})^{-1} \U_{12}^{(n)} ,
\label{eq7.55}
\end{equation}
which involves computations of infinite block triangular Toeplitz matrices that we approximate with infinite banded Toeplitz matrices defined by parameters chosen to achieve a desired level of accuracy $\epsilon$.
From this 
together with 
the first block matrix equation of~\eqref{eq7.37}, the minimal nonnegative solution $\G_{\min}$ of the nonlinear matrix equation~\eqref{eq:4.4} can then be computed up to a desired level of accuracy as
\begin{equation}
    \G_{\min} \; = \; (\I - \hat{\A}_0^{(n)})^{-1} (\A_{-1} + \sum_{i=1}^\infty \hat{\A}_i^{(n)} \G_{\min}^{i\cdot 2^n+1}) .
\label{eq7.39}
\end{equation}
It can be shown that
$\sum_{i=1}^\infty \hat{\A}_i^{(n)} \G_{\min}^{i\cdot 2^n+1}$ asymptotically converges quadratically to zero as $n \rightarrow \infty$ and that $(\I - \hat{\A}_0^{(n)})^{-1}$ is bounded and quadratically convergent, both of which
follow from Theorem~\ref{thm7.13} adapted from~\cite{BiLaMe05}.

We summarize the \mss{corresponding} main \mss{algorithmic} steps of classical CR in Algorithm~\ref{algo:classical}
of Appendix~\ref{app:classical} 
where
Algorithm~\ref{alg7.5} is called
to handle each CR iteration~$n$.
The
overall 
computational complexity of CR is 
\mss{easily shown to be
$$O(M^3 d_{\max} + M^2 d_{\max} \log d_{\max}),$$
}
where $d_{\max}$ is the maximum numerical degrees 
of the matrix power series generated by CR
and recalling 
the order $M$
of the block matrices in~\eqref{eq:P-matrix}.

\subsection{Quantum Computing}
\label{sec:prelim:quantum}
Quantum computing is characterized by operations on the quantum state of $p$ quantum bits, or \emph{qubits}, representing a vector in a $2^p$-dimensional complex vector (Hilbert) space.
The quantum operations or measurements correspond to multiplying the quantum state vector by certain $2^p\times 2^p$ matrices.
Quantum circuits represent these operations in terms of a set of quantum gates operating on the qubits, where the number of gates and the depth of the circuit define the circuit complexity of a 
quantum algorithm.
It is this $2^p$-dimensional complex vector 
space that creates the potential for significant speedup, up to as much as exponential speedup, over classical algorithms executed on digital computers.
However, 
despite some
quantum algorithms
provably
achieving
polynomial-to-exponential speedups over the best-known classical methods~\cite{shor1994algorithms,grover1996fast,shor1997algorithms,nielsen2010quantum,gilyen2019quantum, akhalwaya2024topological}, the full breadth of realizing this potential of exponential speedup with quantum computing has been quite limited.

More specifically, many hybrid classical-quantum algorithms have been proposed that heuristically exploit the exponential computation space, including variational quantum eigen-solvers (VQE)~\cite{peruzzo2014variational} and the quantum approximate optimization algorithm (QAOA)~\cite{zhou2020quantum}.
It is important to note, however, that these algorithms do not achieve a provable speedup~\cite{cerezo2021variational}.
Furthermore, recent developments of ``dequantized'' algorithms~\cite{chia2020sampling,chepurko2020quantum,tang2021quantum} have reduced the potential speedups of many of the linear-algebraic quantum machine learning proposals~\cite{rebentrost2014quantum,biamonte2017quantum,schuld2015introduction,schuld2019quantum,chia2020sampling} to
\mss{fall}
between nonexistent and at most a modest polynomial.
Given the difficulty of building quantum computers (preparing and maintaining the quantum states is
\mss{a hard problem)}
together with their very noisy properties, such limited benefits are
outweighed by the principles of quantum error correction proposed to protect the quantum system from information loss and other damages~\cite{gottesman2010introduction}. 
In
contrast, herein we derive the first quantum algorithms for computing the stationary distribution of general structured Markov processes 
\mss{and~---~with a particular focus on the computation phase in the
\mss{$2^p$-dimensional}
decision-space of quantum computers from a mathematical perspective~---~we} 
establish that our quantum algorithms provide 
{the potential for significant computational improvements}
over {that of} the best-known 
and most-efficient
classical algorithms in
{various}
settings of
importance,
thus possibly
expanding the limited set of quantum algorithms that
provably 
achieve 
\mss{the promise of quantum computing, especially under efficient data loading of the input matrices via block encoding and efficient result readout of a sufficiently sparse output matrix via sampling.}

We employ standard quantum notation and algorithmic components, which includes representing quantum states with the \emph{bra-ket} notation
introduced by Paul Dirac.
In particular, a quantum state is represented as a column vector $\ket{\cdot}$ where different characters inside the \emph{ket} indicate different quantum states; e.g., $\ket{x}$ and $\ket{y}$ represent the (column) vector states $x$ and $y$ of a quantum system.
The corresponding \emph{bra}, represented as $\bra{\cdot}$, is the conjugate transpose of the ket (thus a row vector); e.g., $\bra{x}$ and $\bra{y}$ represent the conjugate transpose of the vector states $x$ and $y$ of a quantum system. 
Finally, $\braket{x|y}$ 
represents the inner (scalar) product formed between the two quantum states $x$ and $y$ indexed by the bra and ket vectors $\bra{x}$ and $\ket{y}$.
A $p$-qubit quantum system is assumed to operate on a $p$-dimensional Hilbert space where each element has $2^p$ complex entries, i.e., the state of the quantum system is described by a $p$-fold tensor product. 
In devising our quantum algorithms, we exploit some key algorithmic components operating on this Hilbert {decision-}space, namely aspect of 
Harrow–Hassidim–Lloyd (HHL),
Quantum Phase Estimation (QPE) and
Quantum Fourier Transformations (QFT, iQFT),
each of which are 
summarized
in Appendix~\ref{app:quantum}.
We refer
the reader 
to~\cite{nielsen2010quantum,Dervovic18,Port18,de2019quantum,LipReg21,LinLin2022} for additional technical details on quantum computing and quantum algorithms.

\section{Quantum Algorithms and Theoretical Results}
\label{sec:quantum}
In this section, we present our main contributions w.r.t.\ quantum algorithms
and a corresponding mathematical analysis.
We
first derive quantum algorithms to address the 
computational bottlenecks of classical
CR
methods, thus providing the first quantum algorithms for computing the stationary distribution of 
\mss{the very general classes of}
structured Markov processes.
We then derive a rigorous mathematical analysis of the computational errors and computational complexity of our quantum 
{CR}
algorithms
{together with}
related theoretical
results.
{This
includes, from a mathematical perspective,
establishing the potential of an exponential speedup over the 
most-efficient
classical algorithms for the computation phase in the decision-space of quantum computers, 
one of our 
\mss{specific}
interests, and
establishing the potential of a polynomial-to-exponential speedup over the 
most-efficient
classical algorithms 
when the properties of the input matrices together with block encoding
allow efficient data 
loading
and
when the properties of the output matrix
\mss{together with sampling-based methods}
allow efficient result readout.} 
\mss{We further note that the computation of the stationary distribution of general structured Markov processes may play the role of a subroutine as part of a larger computational problem on the quantum computer, such as an optimization problem involving the stationary distribution of the associated structured Markov process; in these cases the computation phase would be more critical than the input and output phases to the overall computational complexity.}
\mss{Moreover, since}
CR
is core to many important numerical methods far beyond
general structured Markov processes, 
{it may be possible to exploit our quantum algorithms to address a much larger class of numerical computation problems.}

We
first present a formal derivation of our general 
quantum
CR
algorithms, and then present our mathematical analysis of the computational errors and
computational 
complexity of these 
CR
algorithms within {the context of} quantum computing environments {from a mathematical perspective}.
The corresponding classical CR Algorithms~\ref{algo:classical},
\ref{alg7.5},
\ref{alg3.1} 
and~\ref{alg2.1} are provided
in Appendix~\ref{app:classical},
to which we refer the reader for additional technical details on our derivation of these classical CR methods.
Recall that the overall computational complexity of classical CR is $O(M^3 d_{\max} + M^2 d_{\max} \log d_{\max})$, 
where $d_{\max}$ denotes the maximum numerical degrees 
of the matrix power series generated by CR
and $M$ denotes the order of the block matrices in~\eqref{eq:P-matrix}.

\subsection{Quantum Algorithms}
\label{sec:quantum:alg}
The classical
CR
Algorithm~\ref{algo:classical} provides the most efficient algorithm for computing the stationary distribution of general classes of structured Markov processes on digital computers.
At the same time,
its overall computational complexity
can be prohibitive for large values of $M$ and $d_{\max}$, especially in the context of  
mathematical performance analysis, modeling and optimization of computer systems and networks.
The 
primary computational bottlenecks
concern the iterative calls to Algorithm~\ref{alg3.1} in \emph{Step}~$\mathbf{1}$ of Algorithm~\ref{alg7.5} and to Algorithm~\ref{alg2.1} in \emph{Step}~$\mathbf{2}$ and \emph{Step}~$\mathbf{3}$ of Algorithm~\ref{alg7.5}.
Focusing on these
primary 
bottlenecks of classical
CR,
we observe that the iterative calls
to Algorithm~\ref{alg3.1} and Algorithm~\ref{alg2.1} involve various {linear algebraic} operations on Toeplitz matrices, in particular solving Toeplitz linear systems and Toeplitz matrix products.
The computation phase of
our quantum approach is based in part on the concept of approximating Toeplitz matrices by associated circulant matrices.
This general notion has been extensively studied in the classical preconditioning literature since the late $1980$s~\cite{chan1988optimal,chan1989toeplitz,chan1992circulant,chan1992circulant2,serra1997extension}.
Let $f$ denote a so-called generating function of a sequence of $N \times N$ Toeplitz matrices $\T^{[N]}$, where $N = d \cdot M$
and 
$d$ is the numerical degrees of the matrix power series generated by iteration~$n$ of our quantum 
CR.
It can be shown that the circulant matrices associated with the sequence of Toeplitz matrices converge super linearly whenever $f$ is a positive function in the 
Wiener algebra class $\cW$~\cite{chan1989toeplitz,serra1997extension}.
Since the transition probability matrix is nonnegative and stochastic
(row sums equal one), 
and thus the Toeplitz entries have finite sum, the associated circulant matrix is a very accurate approximation to the corresponding Toeplitz matrix for sufficiently large $N$.
We then exploit the resulting circulant matrices and operations thereon within the context of quantum computers to realize the potential 
exponential
speedups
{of the computation phase in}
such computing environments
{for these matrix operations}.

More precisely, given
an $N \times N$ 
Toeplitz matrix $\T^{[N]}$
with entries $t_{ij} = t_{i-j}$, the entries can be viewed as Fourier coefficients of a certain $2\pi$-periodic strictly positive continuous real-valued function $f$ defined on $[0,2\pi]$ as
\[
t_k = \frac{1}{2\pi}\int_{0}^{2\pi} f(\lambda) e^{-ik\lambda} d\lambda ,
\qquad 
k \in \{ 0,\pm1, \pm 2, \ldots, \pm(n-1) \} .
\]
With this generating function $f$ of the sequence of Toeplitz matrices $\T^{[N]}$, 
we can construct a sequence of circulant matrices $\C^{[N]}$ that are asymptotically equivalent to these Toeplitz
matrices $\T^{[N]}$ as $N\rightarrow\infty$. In particular, we define a circulant matrix $\C^{[N]}$ with top row $(c_0,c_1,\ldots, c_{N-1})$ where 
\[
c_k = \frac{1}{N}\sum_{j=0}^{N-1} f(2\pi j/N) e^{2\pi ijk/N} .
\]
The eigenvalues of $\C^{[N]}$ are simply $f(2\pi j/N)$ for $j \in \{0, 1,\ldots, N-1\}$, and the corresponding eigenvectors are the columns of the Fourier matrix $\Phi$. 
Our quantum approach therefore consists of matrix operations on the matrices $\C^{[N]}$, called the associated circulant matrices of the Toeplitz matrices $\T^{[N]}$. 
We have the inverse $(\C^{[N]})^{-1} = \Phi^{\dagger}\Lambda^{-1}\Phi$, where $\Lambda^{-1}$ is a diagonal matrix with eigenvalues $1/f(2\pi j/N)$. 
Given the properties of the Toeplitz matrices as noted above, $f$ is a positive function in $\cW$ and the entries $\{t_j\}$ are absolutely summable.
Hence, the associated circulant matrices $\C^{[N]}$ converge super linearly to the sequence of Toeplitz matrices $\T^{[N]}$~\cite{chan1989toeplitz,serra1997extension} with $\C^{[N]}$ a very accurate approximation of $\T^{[N]}$ for sufficiently large $N$.
We then exploit key algorithmic components and computational features of quantum computers to efficiently perform the required matrix operations on the circulant matrices $\C^{[N]}$ in quantum
computing 
environments, noting some aspects of which are independently considered in~\cite{wan2018asymptotic} to solve Toeplitz linear systems.
Building 
on these quantum matrix operations, we derive an efficient quantum 
CR
algorithm that addresses the computational bottlenecks of classical 
CR,
as presented
in Algorithm~\ref{algo:quantum}.
For brevity and 
expositional elucidation,
we do not include the normalization details of our quantum algorithms,
\mss{which are typically handled with the use of auxiliary qubits}.

\begin{algorithm}[thb!]
\renewcommand\thealgorithm{Q.\arabic{algorithm}}
\caption{Quantum Cyclic Reduction Algorithm for Ergodic M/G/$1$-type Markov Processes} \label{algo:quantum}
   \begin{algorithmic}
 \State {\bf Input:} Positive integer $d$, $M\times M$ block matrices $\A_i$, $i\in \{-1, 0, 1, \ldots, d-1\}$, defining the block Toeplitz, block Hessenberg matrix $\bH$ in~\eqref{lem4.2:4.6}, and error tolerance $\epsilon > 0$.
  \State {\bf Output:} An approximation $\J$ to the matrix $\G_{\min}$.

  \State {\bf 1.} Set $n=0$, consider quantum circuits for matrices
  $\A_{-1}^{(0)} = \A_{-1}$, $\A_{0}^{(0)} = \A_{0}$, $\ldots$, $\A_{d-1}^{(0)} = \A_{d-1}$, and
  $\hat{\A}_0^{(0)} = \I-\A_0$,
  $\hat{\A}_1^{(0)} = -\A_1$, $\ldots$, $\hat{\A}_{d-1}^{(0)} = -\A_{d-1}$.
  
\State {\bf 2.} Call Algorithm~\ref{algo:quantumCR} with inputs $\T_1 = (\I - \U_{11}^{(n)}), \T_2 = -\U_{21}^{(n)}, \T_3 = \U_{12}^{(n)}$ and $\T_4 = (\I-\U_{22}^{(n)})$ to obtain a quantum output state $\ket{\psi^\ast}$ which is a vectorization of matrices $\Z_{-1}, \Z_{0}, \ldots, \Z_{d'-2}$, $\hat{\Z}_{0}, \hat{\Z}_{1}, \ldots, \hat{\Z}_{d'-1}$.

     \State {\bf 3.} Let $\ket{a_{-1}} =  \text{vec}(\A_{-1})$ and $\text{mask}(\ket{\psi^\ast}) = \text{vec}(\hat{\Z}_0)$. Load columns of $\A_{-1}$
     s.t.\
     we form the state $\ket{a_{-1}z_0} = \ket{a_{-1}} + \text{mask}(\ket{\psi^\ast})$. Compute $\| \ones - (\A_{-1}+\hat{\Z}_0)\bm{1}\|_{\infty}$ from $\ket{a_{-1}z_0}$.
     If $\| \ones - (\A_{-1}+\hat{\Z}_0)\bm{1}\|_{\infty}> \epsilon$, set $n=n+1$, set $\A_i^{(n)} = \Z_i$, $i\in \{-1,\ldots,d'-2\}$, set $\hat{\A}_i^{(n)} = \hat{\Z}_i$, $i\in \{0,\ldots,d'-1\}$, and repeat Steps~$\mathbf{2}$ and~$\mathbf{3}$.

    \State {\bf 4.} Load $\ket{a_{-1}} =  \text{vec}(\A_{-1})$ and use the Toeplitz solution approach in Algorithm~\ref{algo:quantumCR} to form the state $\ket{{j}} = \text{vec}(\J)$ with $\J = (\I - \hat{\Z}_0)^{-1}\A_{-1}$.
    \State {\bf 5.} \emph{Output:} $\J$. 
   \end{algorithmic}
\end{algorithm}

After the initialization in \emph{Step}~$\mathbf{1}$ of Algorithm~\ref{algo:quantum} performed on a digital computer, 
we operate on two Toeplitz matrices, i.e., $\T_1 = (\I -\U_{11}^{(n)})$ and $\T_2 = -\U_{21}^{(n)}$, in the call to Algorithm~\ref{algo:quantumCR}, each of whose steps are executed on a quantum computer.
Let $f_1$ and $f_2$ be the generating functions of these two Toeplitz matrices, respectively.
Upon loading the $N$ columns of the matrix $\T_3 = \U_{12}^{(n)}$ onto the quantum computer as an initial state $\ket{\psi_0}$, we can operate on each of the $N$ columns of $\T_3$ independently and in quantum parallel. 
Namely, we consider $N$ sets of $\log(N)$-qubits that independently encode the $N$ columns of $\U_{12}^{(n)}$~\cite{LinLin2022,sunderhauf2024}.
Then, we approximately compute
$$
\T_2\T_1^{-1}\ket{\psi_0} = \Phi^{\dagger}\Lambda(f_2)\Lambda^{-1}(f_1)\Phi \ket{\psi_0}
$$ 
on the quantum computer, where $\Phi$ is the Fourier matrix, and $\Lambda(f_1)$ and $\Lambda(f_2)$ are diagonal matrices respectively corresponding to the eigenvalues of (the associated circulant matrices of) $\T_1$ and $\T_2$. 
With oracle access to the functions $f_1$ and $f_2$ by exploiting appropriate oracle quantum circuit methods~\cite{nielsen2001quantum,bhaskar2016quantum,LiKais2021},
and  
given the Fourier coefficients (the entries of the Toeplitz matrix $\{t_k\}$), we compute these generating functions in a straightforward manner. Moreover, since these functions are periodic, we can design efficient quantum circuits to estimate the functions using the corresponding Fourier expansions; refer to~\cite{li2021universal} for details. Alternatively, if a classical circuit is given for computing these functions, we 
can design a quantum circuit of comparable efficiency; refer to~\cite[Section~3.2.5]{nielsen2001quantum} and~\cite{bhaskar2016quantum} for details.

Next, we apply QFT to the initial state (columns of $\T_3$) to obtain $\ket{\psi_0'} = \Phi\ket{\psi_0}$. In order to encode the eigenvalues of $\T_1$, we exploit the oracle access to the values of $f_1$ to map 
\[
\ket{\psi_0'} = \sum_{j=0}^Nb_j\ket{j} \longrightarrow \sum_{j=0}^Nb_j\ket{j}\ket{f_1(2\pi j/N)} .
\]
Then, in order to invert the eigenvalues, we employ aspects of the approach used in the HHL quantum algorithm,  namely adding a qubit, using controlled-rotations to invert the phases (eigenvalues), and lastly using uncompute to obtain an approximation to the state $\ket{\psi_1}  =  \Lambda^{-1}(f_1)\ket{\psi_0'}$;
see Appendix~\ref{app:quantum}. 
We now encode the eigenvalues of $\T_2$ exploiting the oracle access to the values of $f_2$ and map
\[
\ket{\psi_1^\prime} = \sum_{j=0}^N\widetilde{b}_j\ket{j} \longrightarrow \sum_{j=0}^N\widetilde{b}_j\ket{j}\ket{f_2(2\pi j/N)} .
\]
Finally, we apply iQFT to obtain a state proportional to
$$\ket{\psi_2} = \Phi^{\dagger}\Lambda(f_2)\Lambda^{-1}(f_1)\Phi \ket{\psi_0}.$$ 
A detailed description of each of the above algorithmic steps is presented in Algorithm~\ref{algo:quantumCR}.

  

\begin{algorithm}[thb!]
\renewcommand\thealgorithm{Q.\arabic{algorithm}}
\caption{Single Iteration~$n$ of Quantum Cyclic Reduction Algorithm for Ergodic M/G/$1$-type Markov Process} \label{algo:quantumCR}
   \begin{algorithmic}
 \State {\bf Input:} Four  $N\times N$ Toeplitz matrices $\T_1,\T_2,\T_3$, $\T_4$.
  \State {\bf Output:} A quantum state $\ket{\psi^*} \approx 
  \text{vec}(\T_2\T_1^{-1}\T_3+ \T_4)$.

\State {\bf 1.} Prepare an initial state $\ket{\psi_0}$ which is a vectorization of $\T_3$.

 \State{\indent \textit{/* Compute the Toeplitz matrix inverse $\T_1^{-1} \T_3$ as $\ket{\psi_1} = (\I \otimes \T_1^{-1}) \ket{\psi_0}$, assuming oracle access to the generating \hspace*{0.6in} function $f_1$ of the Toeplitz matrix $\T_1$} */}

  \State {\bf 2.} Compute $\ket{\psi'_0} = QFT(\ket{\psi_0})$.

  \State {\bf 3.} Suppose $\ket{\psi'_0} = \sum_{j=0}^{N-1}b_j\ket{j}$. Then, using the oracle for $f_1$,  for each column of $\T_3$, compute in parallel on the parallel quantum processors
  $\sum_{j=0}^{N-1}b_j\ket{j}\ket{f_1(2\pi j/N)}$.
  \State {\bf 4.}  Add a qubit and perform a controlled-rotation on $\ket{f_1(2\pi j/N)}$ to obtain
  \[
   \sum_{j=0}^{N-1}b_j\ket{j}\ket{f_1(2\pi j/N)}\left( \sqrt{1-\frac{m^2}{f_1^2(2\pi j/N)}}\ket{0}+ \frac{m}{f_1(2\pi j/N)} \ket{1}\right),
  \]
  where $m$ is a constant
  s.t.\
  $m\leq \min_j|\lambda_j|$ and $\lambda_j$ are the eigenvalues of $\C (f_1)$.
  \State {\bf 5.} Uncompute the second qubit and use amplitude
amplification on the last register to obtain $\ket{1}$, and thus with high probability we attain
\[
\ket{\psi_1} = \sqrt{\frac{1}{\sum_{j}m^2|b_j^2|/|f_1(2\pi j/N)|^2}}\sum_{j=0}^{N-1}b_j \frac{m}{f_1(2\pi j/N)} \ket{j},
\]
which is proportional to $\Lambda^{-1}\ket{\psi_0'} = \sum_{j=0}^{N-1} \frac{b_j}{f_1(2\pi j/N)} \ket{j}$ up to normalization.
  
\State{\indent \textit{/* Compute the Toeplitz matrix-vector product $\ket{\psi_2} = (\I\otimes\T_2)\ket{\psi_1}$, assuming oracle access to the generating \hspace*{0.6in} function $f_2$ of the Toeplitz matrix $\T_2$} */}

\State {\bf 6.} Suppose $\ket{\psi'_1} = \sum_{j=0}^{N-1}\widetilde{b}_j\ket{j}$. Then, using the oracle for $f_2$, for each column of $\T_3$, compute in parallel on the quantum processors
  $\ket{\psi_2^\prime} = \sum_{j=0}^{N-1}\widetilde{b}_j\ket{j}\ket{f_2(2\pi j/N)}$.

   \State {\bf 7.} Compute  $\ket{\psi_2} = iQFT(\ket{\psi_2^\prime})$.
 
   \State {\bf 8.} Load vectorization of $\T_4$ as a quantum state $\ket{\psi_3}$
   s.t.\
   the amplitudes are added to the current state $\ket{\psi_2}$, i.e., prepare $\ket{\psi^*} = \ket{\psi_2}+\ket{\psi_3}$.
  \State {\bf 9.} \emph{Output:} $\ket{\psi^*}$.
   \end{algorithmic}
\end{algorithm}

Upon computing the state $\ket{\psi^*}$ in Algorithm~\ref{algo:quantumCR}, Algorithm~\ref{algo:quantum} proceeds to check the quality of the approximate solution by computing the quantity $\| \ones - (\A_{-1}+\hat{\Z}_0)\bm{1}\|_{\infty}$ on the quantum computer. 
To this end, let $\ket{a_{-1}} =  \text{vec}(\A_{-1})$ and $\text{mask}(\ket{\psi^\ast}) = \text{vec}(\hat{\Z}_0)$, where $\text{mask}(\ket{\psi^\ast})$ considers only the amplitudes corresponding to $\hat{\Z}_0$ in  $\ket{\psi^*}$. We load the columns of $\A_{-1}$ to form the state
$$\ket{a_{-1}z_0} = \ket{a_{-1}} + \text{mask}(\ket{\psi^\ast}),$$
and compute $\| \ones - (\A_{-1}+\hat{\Z}_0)\bm{1}\|_{\infty}$ from $\ket{a_{-1}z_0}$. If $\| \ones - (\A_{-1}+\hat{\Z}_0)\bm{1}\|_{\infty}> \epsilon$, we repeat the inner loop; else, we proceed to output $\J = (\I - \hat{\Z}_0)^{-1}\A_{-1}$.
For this we again load $\ket{a_{-1}} =  \text{vec}(\A_{-1})$, and then use the above Toeplitz solution approach to form the state $\ket{\widetilde{j}} = \text{vec}(\J)$.

\mss{
The input phase of our quantum algorithm addresses the efficient data loading of the input matrices associated with the general structured Markov process of interest by exploiting a combination of appropriate block encoding quantum circuit methods~\cite{LinLin2022,sunderhauf2024} together with the properties of the Toeplitz matrices $\T_1, \T_2, \T_3, \T_4$ provided to Algorithm~\ref{algo:quantumCR} and appropriate oracle quantum circuit methods~\cite{nielsen2001quantum,bhaskar2016quantum,li2021universal} together with the properties of the corresponding generating functions $f_1$ and $f_2$.
The readout phase of our quantum algorithm addresses the efficient outputting of the result matrix associated with the stationary distribution of the general structured Markov process of interest by exploiting appropriate sampling-based methods together with the various properties of the result matrix $\G_{\min}$.
}

\subsection{Mathematical Analysis}
We now turn to a mathematical analysis of the computational errors and computational complexity of our
CR
Algorithm~\ref{algo:quantum} within quantum computing environments.
Starting with our error analysis, this {quantum} algorithm entails two separate sources of error.
The first source
stems from the truncation and approximation of infinite power series of matrices. The second source of error stems from performing computations (and propagating the results) in environments with finite representation of numbers, using an analog (continuous) concept as in quantum computers. Note that each one of these two sources of error occurs naturally at every
CR
iteration, meaning that errors occurring at each 
CR
iteration accumulate to a global error as the algorithm proceeds.

We formally define in functional form the block matrices $\A_i^{(n)}$ and $\hat{\A}_{i+1}^{(n)}$, $i\in \Ints_{\geq -1}$, of the matrix $\bH^{(n)}$ in~\eqref{eq7.37} by means of the recursions
\begin{align}
    \varphi^{(n+1)}(z) & = z \varphi_{\scriptscriptstyle{odd}}^{(n)}(z) + \varphi_{\scriptscriptstyle{even}}^{(n)}(z)(\I - \varphi_{\scriptscriptstyle{odd}}^{(n)}(z))^{-1} \varphi_{\scriptscriptstyle{even}}^{(n)}(z) ,
    \label{eq7.38a} \\
    \hat{\varphi}^{(n+1)}(z) & = \hat{\varphi}_{\scriptscriptstyle{odd}}^{(n)}(z) + \varphi_{\scriptscriptstyle{even}}^{(n)}(z) (\I - \varphi_{\scriptscriptstyle{odd}}^{(n)}(z))^{-1} \hat{\varphi}_{\scriptscriptstyle{even}}^{(n)}(z) ,
    \label{eq7.38b}
\end{align}
together with
\begin{align}
    \varphi^{(n)}(z) := \sum_{i=-1}^\infty z^{i+1} \A_i^{(n)} ,
    \;\; & \;\;
    \varphi_{\scriptscriptstyle{even}}^{(n)}(z) := \sum_{i=0}^\infty z^i \A_{2i}^{(n)} ,
    \; & \;
    \varphi_{\scriptscriptstyle{odd}}^{(n)}(z) := \sum_{i=-1}^\infty z^{i+1} \A_{2i+1}^{(n)} ,
    \label{eq-varphi} \\
    \hat{\varphi}^{(n)}(z) := \sum_{i=0}^\infty z^i \hat{\A}_i^{(n)} ,
    \;\; & \;\; 
    \hat{\varphi}_{\scriptscriptstyle{even}}^{(n)}(z) := \sum_{i=0}^\infty z^i \hat{\A}_{2i}^{(n)} ,
    \; & \; \hat{\varphi}_{\scriptscriptstyle{odd}}^{(n)}(z) := \sum_{i=0}^\infty z^i \hat{\A}_{2i+1}^{(n)} 
    \label{eq-hatvarphi} ,
\end{align}
following~\eqref{eq:LaurentEven} and \eqref{eq:LaurentOdd}.
This functional form supports the efficient computations reflected in the classical  CR algorithm.
It also follows from~\eqref{eq7.38a} and~\eqref{eq7.38b} that the applicability of CR at each iteration~$n$ relies on the invertibility of the matrix power series $\I - \varphi_{\scriptscriptstyle{odd}}^{(n)}(z)$ for $|z| \leq 1$, as established in Theorem~\ref{thm7.8} adapted from~\cite{BiLaMe05}
where, when $M$ is infinite, we replace the spectral radi in Theorem~\ref{thm7.8} with the spectrum related to the corresponding operator appropriately defined~\cite{DunSch-all} (see Remark~\ref{rem:thm7.8}).

Consider the pair of matrix functions $(\varphi^{(n)}(z),\hat{\varphi}^{(n)}(z))$ that defines the matrix $\bH^{(n)}$ 
and let ${\cT}$ denote the transformation that occurs during the $n+1$ iteration of
CR,
namely
$
    (\varphi^{(n+1)}(z),\hat{\varphi}^{(n+1)}(z)) ={\cT}(\varphi^{(n)}(z),\hat{\varphi}^{(n)}(z))$.
Since the transformation ${\cT}$ is implemented  within a quantum
computing 
environment, let ${\cT}_F$ denote the actual finite implementation of ${\cT}$ that includes the computational errors from quantum computers and the approximation errors from matrix and power series truncation. We then can define the quantum error of a single iteration~$n$ of
CR
as
\begin{equation}
    \cE_S(\varphi^{(n)}(z),\hat{\varphi}^{(n)}(z)) = {\cT}_F(\varphi^{(n)}(z),\hat{\varphi}^{(n)}(z))-{\cT}(\varphi^{(n)}(z),\hat{\varphi}^{(n)}(z)).
\label{eq:local-error}
\end{equation}
Since ${\cT}_F$ denotes a finite implementation on a quantum computer, the matrix pair from ${\cT}_F(\varphi^{(n)}(z)$, $\hat{\varphi}^{(n)}(z))$ will generally disagree with the matrix pair $(\varphi^{(n+1)}(z),\hat{\varphi}^{(n+1)}(z))$ from ${\cT}(\varphi^{(n)}(z),\hat{\varphi}^{(n)}(z))$.
Indeed, let $(\vartheta^{(n+1)}(z),\hat{\vartheta}^{(n+1)}(z))$ be the matrix pair generated by the transformation 
${\cT}_F(\vartheta^{(n)}(z)$, $\hat{\vartheta}^{(n)}(z))$, $n\in\Nats$, where $\vartheta^{(0)}(z)=\varphi^{(0)}(z)$ and 
$\hat{\vartheta}^{(0)}(z)=\hat{\varphi}^{(0)}(z)$. We then can define the global quantum error at iteration~$n$ as
\begin{equation}
    \cE_G^{(n)}(z) = (\bE^{(n)}(z),\hat{\bE}^{(n)}(z)) = 
    (\vartheta^{(n)}(z),\hat{\vartheta}^{(n)}(z)) - (\varphi^{(n)}(z),\hat{\varphi}^{(n)}(z)).
\label{eq:global-error}
\end{equation}
Upon substituting the defining transformations for the matrix pairs $(\vartheta^{(n)}(z),\hat{\vartheta}^{(n)}(z))$ and 
$(\varphi^{(n)}(z)$, $\hat{\varphi}^{(n)}(z))$, we can write 
\begin{align}
    \cE_G^{(n+1)}(z) &= 
    {\cT}_F(\vartheta^{(n)}(z),\hat{\vartheta}^{(n)}(z)) - {\cT}(\varphi^{(n)}(z),\hat{\varphi}^{(n)}(z)) \nonumber \\
    & = {\cT}(\vartheta^{(n)}(z),\hat{\vartheta}^{(n)}(z)) - {\cT}(\varphi^{(n)}(z),\hat{\varphi}^{(n)}(z)) +
    \cE_S(\vartheta^{(n)}(z),\hat{\vartheta}^{(n)}(z)) , \label{eq7.56}
\end{align}
where the latter equality follows from~\eqref{eq:local-error} for $(\vartheta^{(n)}(z),\hat{\vartheta}^{(n)}(z))$.

The global error can also be understood as the forward error, which we can divide into two separate parts:
($a$) the quantity 
$${\cT}(\vartheta^{(n)}(z),\hat{\vartheta}^{(n)}(z)) - {\cT}(\varphi^{(n)}(z),\hat{\varphi}^{(n)}(z))$$ 
representing the difference between applying the ideal transformation on both the exact and approximate matrix functions at the $n$th iteration of
CR;
and
($b$) the difference 
$${\cT}_F(\vartheta^{(n)}(z),\hat{\vartheta}^{(n)}(z))-{\cT}(\vartheta^{(n)}(z),\hat{\vartheta}^{(n)}(z))$$ 
representing the fidelity of the transformation ${\cT}_F$, i.e., how close the implementation ${\cT}_F$ is to ${\cT}$ at a single iteration~$n$ on a quantum computer.
Clearly,
the latter quantity $\cE_S(\vartheta^{(n)}(z),\hat{\vartheta}^{(n)}(z))$ depends on the specific implementation of ${\cT}_F$ on a quantum computer,
whereas
the former quantity of the difference in~\eqref{eq7.56} requires a more detailed mathematical analysis.

To this end, we next derive a first-order analysis of the error expressions in~\eqref{eq7.56}, noting that it is standard in the numerical analysis of iterative processes such as
CR
to focus on a first-order analysis and ignore higher-order terms~\cite{Higham02}.
Such a mathematical analysis is most appropriate when the errors are sufficiently small so that the corresponding products and powers are negligible, which will be the case for sufficiently large iterates of
CR.
With this focus on the terms linear in the errors, let $\doteq$ denote equality up to higher-order error terms and thus $(1+\varepsilon)^n \doteq 1 + n\varepsilon$ and $(1-\varepsilon)^{-1} \doteq 1 + \varepsilon$, where $\varepsilon$ represents the error; and similarly let $\dotleq$ denote the corresponding inequality.
A key part of this analysis is the following theorem that analyzes the propagation of the error at a general iteration~$n$ of
our quantum 
CR
algorithm.
Denote the max norm as $\| \bS(z) \|_{\ast} = \| \sum_{i=0}^\infty | \bS_i | \, \|_\infty$ where $\bS(z) = \sum_{i=0}^\infty z^i \bS_i$ is a matrix power series in the Wiener algebra $\cW$, for which it is readily verified that $\| \cdot \|_{\ast}$ is a norm in $\cW$.
\begin{theorem}
Let $\varphi^{(n)}(z)$ and $\hat{\varphi}^{(n)}(z)$ be the matrix power series defining the first two block rows of $\bH^{(n)}$ in~\eqref{eq7.37}, namely
$$
\varphi^{(n)}(z) := \sum_{i=-1}^\infty z^{i+1} \A_i^{(n)} \qquad \mbox{ and } \qquad \hat{\varphi}^{(n)}(z) := \sum_{i=0}^\infty z^i \hat{\A}_i^{(n)} , 
$$
and let $\vartheta^{(n)}(z)$ and $\hat{\vartheta}^{(n)}(z)$ be approximations of $\varphi^{(n)}(z)$ and $\hat{\varphi}^{(n)}(z)$, respectively.
Define 
$$
\R^{(n)}(z) := \vartheta^{(n)}(z) - \varphi^{(n)}(z) \qquad \mbox{ and } \qquad \hat{\R}^{(n)}(z) := \hat{\vartheta}^{(n)}(z) - \hat{\varphi}^{(n)}(z)
$$ 
to be the corresponding approximation errors, and define the matrix power series
$$
\underline{\R}^{(n+1)}(z) = \underline{\vartheta}^{(n+1)}(z) - \underline{\varphi}^{(n+1)}(z) \qquad \mbox{ and } \qquad \underline{\hat{\R}}^{(n+1)}(z) = \underline{\hat{\vartheta}}^{(n+1)}(z) - \underline{\hat{\varphi}}^{(n+1)}(z) ,
$$
where
$$
(\underline{\vartheta}^{(n+1)}(z),\underline{\hat{\vartheta}}^{(n+1)}(z)) = \cT(\vartheta^{(n)}(z),\hat{\vartheta}^{(n)}(z)) \quad \mbox{ and } \quad (\underline{\varphi}^{(n+1)}(z),\underline{\hat{\varphi}}^{(n+1)}(z)) = \cT(\varphi^{(n)}(z),\hat{\varphi}^{(n)}(z)).
$$
Furthermore, let
$$
\V^{(n)}(z) = \varphi_{\scriptscriptstyle{even}}^{(n)}(z)(\I -\varphi_{\scriptscriptstyle{odd}}^{(n)}(z))^{-1} , 
\quad \W^{(n)}(z) = (\I -\varphi_{\scriptscriptstyle{odd}}^{(n)}(z))^{-1}\varphi_{\scriptscriptstyle{even}}^{(n)}(z) , \quad \hat{\W}^{(n)}(z) = (\I -\varphi_{\scriptscriptstyle{odd}}^{(n)}(z))^{-1}\hat{\varphi}_{\scriptscriptstyle{even}}^{(n)}(z).
$$
We then have the first-order equalities
\begin{align}
    \underline{\R}^{(n+1)}(z) & \; \doteq \; z \R_{\scriptscriptstyle{odd}}^{(n)}(z) + \R_{\scriptscriptstyle{even}}^{(n)}(z)\W^{(n)}(z) + \V^{(n)}(z)\R_{\scriptscriptstyle{even}}^{(n)}(z) + \V^{(n)}(z)\R_{\scriptscriptstyle{odd}}^{(n)}(z)\W^{(n)}(z) , \label{eq:7.57a} \\
    \underline{\hat{\R}}^{(n+1)}(z) & \; \doteq \; \hat{\R}_{\scriptscriptstyle{odd}}^{(n)}(z)  
    + \R_{\scriptscriptstyle{even}}^{(n)}(z) \hat{\W}^{(n)}(z) + \V^{(n)}(z) \hat{\R}_{\scriptscriptstyle{even}}^{(n)}(z) + \V^{(n)}(z) \R_{\scriptscriptstyle{odd}}^{(n)}(z) \hat{\W}^{(n)}(z) , \label{eq:7.57b}
\end{align}
together with the corresponding first-order upper bounds
\begin{align}
    \| \underline{\R}^{(n+1)}(z) \|_{\ast} & \; \dotleq \; 2 \|\R^{(n)}(z)\|_{\ast} (1 + \|\V^{(n)}(1)\|_{\infty}) , \label{eq:7.57-UBa} \\
    \| \underline{\hat{\R}}^{(n+1)}(z) \|_{\ast} & \; \dotleq \; \|\hat{\R}^{(n)}(z)\|_{\ast} (1 + \|\V^{(n)}(1)\|_{\infty}) + \|\R^{(n)}(z)\|_{\ast} (1 + \|\V^{(n)}(1)\|_{\infty}) . \label{eq:7.57-UBb}
\end{align}
\label{lem7.15}
\end{theorem}




The next key part of our {mathematical} analysis exploits the above theorem to establish an upper bound on the global quantum errors of our quantum
CR
Algorithm~\ref{algo:quantum} up to iteration~$n$.
\begin{theorem}
    Suppose
    $$
    \|\cE_L^{(n)}(z)\|_{*}\leq \upsilon \qquad \mbox{ and } \qquad \|\hat{\cE}_L^{(n)}(z)\|_{*}\leq \upsilon
    $$
    for some $\upsilon >0$, and
    $$
    2(1 + \|\V^{(n)}(1)\|_{\infty})= \gamma_n\leq \gamma
    $$
    for some $\gamma > 1$. Then, for the quantum 
    CR
    Algorithm~\ref{algo:quantum}, 
    we have
\begin{align*}
   \|\bE^{(n+1)}(z)\|_{*} &\; \dotleq \; \upsilon (1+  \gamma_n+\gamma_{n-1}\gamma_n+\cdots + \gamma_2\gamma_3\cdots\gamma_{n-1}\gamma_n + \gamma_1\gamma_2\cdots\gamma_{n-1}\gamma_n) \leq \frac{\upsilon (\gamma^{n+1} - 1)}{\gamma - 1} , \\
   \|\hat{\bE}^{(n+1)}(z)\|_{*} &\; \dotleq \; \upsilon (1+  \gamma_n+\gamma_{n-1}\gamma_n+\cdots + \gamma_2\gamma_3\cdots\gamma_{n-1}\gamma_n + \gamma_1\gamma_2\cdots\gamma_{n-1}\gamma_n) \leq \frac{\upsilon (\gamma^{n+1} - 1)}{\gamma - 1} .
\end{align*}
\label{thm:expUB}
\end{theorem}

While the above upper bounds on the error 
$\|\bE^{(n+1)}(z)\|_{*}$ and $\|\hat{\bE}^{(n+1)}(z)\|_{*}$
grow exponentially with $n$, numerical results obtained via different implementations of
{classical} 
CR
{on digital computers}
show in fact that the algorithm is both reliable and numerically stable~\cite{BiLaMe05}.
We note that similar observations can be {generally} made about LU decomposition of a matrix~\cite{golub2013matrix}.
The reason for this contradiction stems primarily from the repeated application of the triangular inequality in the above derivation where the worst possible error bound is assumed at each iteration. 
Although this might generally lead to an apparently discouraging error bound, such behavior is nonetheless rarely observed in practice {for classical 
CR
on digital computers}, as noted above.
We
{also}
note that the number of 
iterations generally required in practice
{will be}
very small, due to the quadratic convergence of
CR
{(along the lines of Theorem~\ref{thm7.13})}. 
Hence, the exponential growth of the error bound becomes less impactful than what the theory predicts in the worst case. 
In addition, we observe that if $\bE^{(n)}(z)$ has null coefficients from degree $0$ up to degree $k$ and if $\cE_L^{(n)}(z)$ shares the same property up to degree $k/2$, it then follows from~\eqref{eq:7.59a}
{in the proof of Theorem~\ref{thm:expUB}}
that $\bE^{(n+1)}(z)$ has null coefficients from degree $0$ up to degree $k/2$. Therefore, if each iteration of
CR
is implemented
s.t.\
$\cE_L^{(j)}(z)$ has null coefficients from degree $0$ up to degree
$k/2^j, j\in \{0,1,\ldots,n\}$, 
then $\bE^{(n+1)}(z)$ has null coefficients from degree $0$ up to degree $k/2^{n}$;
and the same also holds for $\hat{\bE}^{(n+1)}(z)$
w.r.t.~\eqref{eq:7.59b}
{in the proof of Theorem~\ref{thm:expUB}}.
Hence, the lower degree coefficients of $\bE^{(n+1)}(z)$ and $\hat{\bE}^{(n+1)}(z)$ can be kept to zero provided that the previous
CR
iterations are truncated at a sufficiently higher degree.

{Meanwhile, the}
above error analysis can be provably improved by extending Algorithm~\ref{algo:quantum}
w.r.t.\
a particular shifting technique that removes the root $\lambda = 1$ of the function $\chi(z) = z\I - \varphi(z)$ from Theorem~\ref{thm7.13} which corresponds to the eigenvector $\ones$
s.t.\
$\chi(z)\ones = 0$.
In particular, consider the function $\widetilde{\chi}(z) = \chi(z)(\I - z^{-1} \Q)^{-1}$ where $\Q = \ones \bu^\top$ for any vector $\bu$
s.t.\
$\bu^\top \ones = 1$ and $\widetilde{\chi}(z)$ has the same roots as $\chi(z)$ except for $z=1$ which is replaced by the root $z=0$.
It follows that $(\I - z^{-1} \Q)^{-1} = \I + \Q \sum_{i=1}^\infty z^{-i}$ and therefore $\widetilde{\chi}(z) = z\I - \sum_{i=-1}^\infty z^{i+1} \widetilde{\A}_i$ where 
$$
\widetilde{\A}_{-1} = \A_{-1} (\I -\Q)
\qquad \mbox{and} \qquad
\widetilde{\A}_i = \A_i - \left(\sum_{j=-1}^i \A_j - \I\right) \Q,
$$ 
for $i \in\Ints_+$.
From our assumptions on the M/G/$1$-type Markov process, we have that $\chi(z)$ is analytic for $|z| < r > 1$ and at least one root of $\chi(z)$ has modulus greater than $1$.
The roots $\widetilde{\xi}_i$ of $\widetilde{\chi}(z)$, $i\in\Nats$, are then 
s.t.\
$$
\widetilde{\xi}_1 = 0 \leq |\widetilde{\xi}_2| \leq \cdots \leq |\widetilde{\xi}_m| = |\xi_{m-1}| < 1 < |\widetilde{\xi}_{m+1}| = |\xi_{m+1}| \leq \cdots < r ,
$$
where $\xi_i$ are the roots of $\chi(z)$ with 
$$
|\xi_1| \leq |\xi_2| \leq \cdots \leq |\xi_{m-1}| < \xi_m = 1 < |\xi_{m+1}| \leq \cdots < r
$$
and $\xi_m = 1$ is simple.
Hence, the function $\widetilde{\chi}(z)$ and the matrix Laurent power series $z^{-1} \widetilde{\chi}(z)$ are both analytic and invertible in the annulus
$\{ z \in \bbC : |\xi_{m-1}| < |z| < |\xi_{m+1}| \}$, 
and thus the coefficients of both the matrix Laurent power series $z^{-1} \widetilde{\chi}(z)$ and $(z^{-1} \widetilde{\chi}(z))^{-1}$ possess exponential decay, which leads to the faster convergence properties of 
CR
applied to the shifted function $\widetilde{\chi}(z)$.
The minimal nonnegative solution $\G_{\min}$ of~\eqref{eq:4.4} is then given by $\G_{\min} = \widetilde{\G}_{\min} + \Q$, where $\widetilde{\G}_{\min}$ is the solution with minimal spectral radius of the matrix equation $\X = \sum_{i=-1}^\infty \widetilde{\A}_i \X^{i+1}$ associated with $\widetilde{\chi}(z)$.

For consistency with the above analysis, let $\eta = |\xi_{m-1}|$ and $\xi = |\xi_{m+1}|$.
Given that $\widetilde{\chi}(z)$ and $z^{-1} \widetilde{\chi}(z)$ are both analytic and invertible for $\eta < |z| < \xi$, we conclude that $(z^{-1} \widetilde{\chi}(z))^{-1}$ is analytic for $\eta < |z| < \xi$.
The shifting technique then consists of applying 
CR
to the function $\widetilde{\chi}(z)$
w.r.t.\
the corresponding sequences of matrix power series $\widetilde{\chi}^{(n)}(z)$ and $\hat{\widetilde{\chi}}^{(n)}(z)$ in place of $\chi^{(n)}(z)$ and $\hat{\chi}^{(n)}(z)$, respectively.
This application of
CR
can be shown, along the lines of Theorem~\ref{thm7.13}, to be convergent
in terms of 
the sequence $\G^{(n)} = (\I - \hat{\widetilde{\A}}_0^{(n)})^{-1} \A_{-1}$, $n\in\Ints_+$, asymptotically tending to $\G_{\min}$.
More specifically, $\lim_{n \rightarrow \infty} \G^{(n)} = \G_{\min}$
and, for any matrix norm and for any $\epsilon > 0$ with $\eta + \epsilon < 1 < \xi - \epsilon$, there exist $\gamma > 0$ and $\sigma_i > 0$, $i \in\Nats$,
s.t.\
\begin{align}
    \G_{\min} - \G^{(n)} & \; = \; (\I - \hat{\widetilde{\A}}_0^{(n)})^{-1}  \sum_{i=1}^\infty \hat{\widetilde{\A}}_i^{(n)} \widetilde{\G}_{\min}^{i2^n} ,
    \qquad\qquad n \in\Ints_+ , \label{eq:8.20}
    \\
    \| \G_{\min} - \G^{(n)} \| & \; \leq \; \gamma \left( \frac{\eta + \epsilon}{\xi - \epsilon} \right)^{2^n} , 
    \qquad\qquad\qquad\qquad\qquad n \in\Ints_+ , \nonumber 
    \\
    \| \hat{\widetilde{\A}}_i^{(n)} \| & \; \leq \; \sigma_i (\xi - \epsilon)^{-i \cdot 2^n} , 
    \qquad\qquad\qquad\qquad\qquad i \in\Nats . \nonumber
\end{align}

We derive our above quantum 
CR
algorithm with the shifting technique as presented in Algorithm~\ref{algo:quantum2}.
The next key part of our {mathematical} analysis exploits Theorem~\ref{lem7.15} and the above results to establish an improved upper bound on the global quantum errors of our quantum
CR
with shifting technique Algorithm~\ref{algo:quantum2} up to iteration~$n$.
\begin{theorem}
    Suppose
    $$
    \|\cE_L^{(n)}(z)\|_{*}\leq \upsilon \qquad \mbox{ and } \qquad \|\hat{\cE}_L^{(n)}(z)\|_{*}\leq \upsilon
    $$
    for some $\upsilon >0$.
    Then, for the quantum
    CR
    with shifting technique Algorithm~\ref{algo:quantum2} and for $\gamma_n = (1 + \theta\sigma^{2^n})^2$, we have
\begin{align*}
   \|\bE^{(n+1)}(z)\|_{*} &\; \dotleq \; \upsilon (1+  \gamma_n+\gamma_{n-1}\gamma_n+\cdots
   + \gamma_1\gamma_2\cdots\gamma_{n-1}\gamma_n) \leq \upsilon \left( 1 + n \exp ( 2 \theta \frac{\sigma^2}{1-\sigma^2} ) \right) , \\
   \|\hat{\bE}^{(n+1)}(z)\|_{*} &\; \dotleq \; \upsilon (1+  \gamma_n+\gamma_{n-1}\gamma_n+\cdots
   + \gamma_1\gamma_2\cdots\gamma_{n-1}\gamma_n) \leq \upsilon \left( 1 + n \exp ( 2 \theta \frac{\sigma^2}{1-\sigma^2} ) \right)  .
\end{align*}
\label{thm:linUB}
\end{theorem}

\begin{algorithm}[thb!]
\renewcommand\thealgorithm{Q.\arabic{algorithm}}
\caption{Quantum Cyclic Reduction with Shifting Technique Algorithm for Ergodic M/G/$1$-type Markov Processes} \label{algo:quantum2}
   \begin{algorithmic}
 \State {\bf Input:} Positive integer $d$, $M\times M$ block matrices $\A_i$, $i\in \{-1, 0, 1, \ldots, d\}$, defining the block Toeplitz, block Hessenberg matrix $\bH$ in~\eqref{lem4.2:4.6},
 and error tolerance $\epsilon > 0$.
  \State {\bf Output:} An approximation $\J$ to matrix $\G_{\min}$, and a real $\sigma >0$ 
  s.t.\
  $\| \G_{\min} - \J \|_{\infty} \leq \epsilon \sigma$.

  \State {\bf 1.} Select any vector $\bu > 0$ 
  s.t.\
  $\bu^\top \ones = 1$. Simply denote $\D_i^{(n)} = \widetilde{\A}_i^{(n)}$ and $\hat{\D}_i^{(n)} = \hat{\widetilde{\A}}_i^{(n)}$.
  \State {\bf 2.} Set $n=0$, $\Q = \ones \bu^\top$, and consider quantum circuits for matrices ${\D}_{-1}^{(0)} = \A_{-1} (\I -\Q)$, ${\D}_i^{(0)} = \A_i + (\sum_{j=i+1}^d \A_j) \Q, i \in \{ 0, \ldots, d\}$, and $\hat{\D}_{0}^{(0)} = \I - \A_0 - (\sum_{j=1}^d \A_j) \Q$, $\hat{\D}_i^{(0)} = - \A_i - (\sum_{j=i+1}^d \A_j) \Q, i \in [d]$.
  
  \State {\bf 3.} Call Algorithm~\ref{algo:quantumCR} with corresponding inputs $\T_1 = (\I - \U_{11}^{(n)}), \T_2 = -\U_{21}^{(n)}, \T_3 = \U_{12}^{(n)}$ and $\T_4 = (\I-\U_{22}^{(n)})$ to obtain a quantum output state $\ket{\psi^\ast}$ which is a vectorization of matrices $\Z_{-1}, \Z_{0}, \ldots, \Z_{d'-1}$, $\hat{\Z}_{0}, \hat{\Z}_{1}, \ldots, \hat{\Z}_{d'}$.

     \State {\bf 4.} Let $\text{mask}(\ket{\psi^\ast}) = \text{vec}(\{\hat{\Z}_i\}_{i=1}^{d'})$. Set $n=n+1,\D_i^{(n)} = \Z_i$, $i\in \{-1,\ldots,d'-1\}$, and $\hat{\D}_i^{(n)} = \hat{\Z}_i$, $i\in \{0,\ldots,d'\}$. Compute $\sum_{i=1}^{d'} \| \hat{\D}_{i}^{(n)} \|_{\infty}$ from $\text{mask}(\ket{\psi^\ast})$.
    If $\sum_{i=1}^{d'} \| \hat{\D}_{i}^{(n)} \|_{\infty} > \epsilon$, repeat Steps~$\mathbf{3}$ and~$\mathbf{4}$.
     

     \State {\bf 5.} Load $\ket{a_{-1}} =  \text{vec}(\A_{-1})$ and use the Toeplitz solution approach in Algorithm~\ref{algo:quantumCR} to form the state $\ket{{j}} = \text{vec}(\J)$ with $\J = (\I - \hat{\Z}_0)^{-1}\A_{-1}$. Compute $\sigma = 2 \| (\I - \hat{\Z}_0)^{-1} \|_{\infty}$ in a similar manner.
    \State {\bf 6.} \emph{Output:} $\J$ and $\sigma$. 

   \end{algorithmic}
\end{algorithm}

The above upper bounds on the error 
$\|\bE^{(n+1)}(z)\|_{*}$ and $\|\hat{\bE}^{(n+1)}(z)\|_{*}$ are linear in $n$, thus providing a significant improvement over the upper bounds in Theorem~\ref{thm:expUB}.
\emph{Step}~$\mathbf{3}$ of our quantum 
CR
with shifting technique Algorithm~\ref{algo:quantum2} calls
the same quantum 
CR
Algorithm~\ref{algo:quantumCR} 
{for iteration~$n$}
as in \emph{Step}~$\mathbf{2}$ of Algorithm~\ref{algo:quantum},
but w.r.t.\ $\widetilde{\chi}(z)$ and $\hat{\widetilde{\chi}}(z)$
instead of w.r.t.\ ${\chi}(z)$ and $\hat{\chi}(z)$.
With $\J$ an approximation to the matrix $\G_{\min}$, the error bound $\| \G_{\min} - \J \|_{\infty} \leq \epsilon \sigma$ in Algorithm~\ref{algo:quantum2} holds because $\widetilde{\G}_{\min}^i = \G_{\min}^i - \Q$ for $i \in\Nats$ in order that
$$
\| \widetilde{\G}_{\min}^i \|_{\infty} \leq \| \G_{\min}^i \|_{\infty} + \| \Q \|_{\infty} = 2 .
$$
We then conclude from the termination condition in Algorithm~\ref{algo:quantum2} that
$$
\sum_{i=1}^\infty \| \hat{\widetilde{\A}}_i^{(n)} \|_{\infty} \doteq \sum_{i=1}^{d'} \| \hat{\widetilde{\A}}_i^{(n)} \|_{\infty} \leq \epsilon ,
$$
and thus the error bound $\| \G_{\min} - \J \|_{\infty} \leq \epsilon \sigma$ follows 
from~\eqref{eq:8.20}.
Moreover, given that $|\xi_{m-1}| < 1 < |\xi_{m+1}|$ and taking into 
account Theorem~\ref{thm:linUB},
we note that Algorithm~\ref{algo:quantum2}
w.r.t.\
$\widetilde{\chi}(z)$ and $\hat{\widetilde{\chi}}(z)$ has considerably better stability properties than Algorithm~\ref{algo:quantum}
w.r.t.\
${\chi}(z)$ and $\hat{\chi}(z)$, wherein the amplification of errors in each iteration is kept relatively small.
It also follows that Algorithm~\ref{algo:quantum2} has significantly faster convergence in comparison with Algorithm~\ref{algo:quantum}, particularly when the second largest modulus eigenvalue $\xi_{m-1}$ of $\G_{\min}$ is far from the unit circle.


Finally, we turn to consider a mathematical analysis of the computational complexity of our 
CR
Algorithm~\ref{algo:quantum} and Algorithm~\ref{algo:quantum2}.
Since both of these algorithms call our quantum Algorithm~\ref{algo:quantumCR} to handle each iteration~$n$ of
CR,
we first consider the computational complexity of Algorithm~\ref{algo:quantumCR} and then consider the overall computational complexity of Algorithm~\ref{algo:quantum} and Algorithm~\ref{algo:quantum2}.
Consistent with Section~\ref{sec:quantum:alg}, our mathematical analysis is based on $N = d \cdot M$ where $d$ is the numerical degrees of the matrix power series generated by a single iteration~$n$ of our quantum 
CR.

In \emph{Step}~$\mathbf{1}$ of Algorithm~\ref{algo:quantumCR}, we need to encode $N$ columns of the Toeplitz matrix $\T_3$ which requires $N\log N$ qubits. 
We note, however, that each of the columns ($\log N$ qubits) can be operated on independently and in parallel.
Hence, each of the $\log N$ qubits form independent parallel quantum processors, which need not be connected/entangled to each other, and all such columns can be loaded in parallel.
Let $\tau_{\scriptscriptstyle load}$ be the time complexity to load a column of the Toeplitz matrix into a quantum state, which therefore represents the time complexity to prepare the initial state $\ket{\psi_0}$.
In \emph{Step}~$\mathbf{2}$, we apply QFT to this initial state to obtain $\ket{\psi_0^\prime} = \sum_{j=0}^{N-1}b_j\ket{j}$, with a time complexity of $O(\log^2 N)$.
Let $\tau_{\scriptscriptstyle oracle}$ be the time complexity of using the oracle for the generating function $f$ that corresponds to a given Toeplitz matrix and of preparing the state $\ket{f(2\pi j/N)}$.  
Then, for each column of $\T_3$ in \emph{Step}~$\mathbf{3}$, 
$$\sum_{j=0}^{N-1}b_j\ket{j}\ket{f_1(2\pi j/N)}$$
is computed in parallel on the parallel quantum processors with an additional $O(1)$ time, also requiring an additional $\log N$ qubits.
\emph{Steps}~$\mathbf{4}$ and~$\mathbf{5}$ both implement the inversion of a Toeplitz matrix, where we need $m \leq \min_j |\lambda_j |$ for the eigenvalues $\lambda_j$ of $\T_1$.
Hence, for $m = f_{1,\min}$, the state succeeds with probability $\Omega(1/\mu^2)$, where $\mu = f_{1,\max}/f_{1,\min}$, and thus $O(\mu^2)$ measurements are required.
However, by exploiting variable time  amplitude amplification~\cite{ambainis2012variable}, we can reduce the time complexity for each of these steps to $O(\mu)$.
\emph{Step}~$\mathbf{6}$ is similar to \emph{Step}~$\mathbf{3}$, using the oracle for $f_2$ and, for each column of $\T_3$, computing in parallel
$$\ket{\psi_2^\prime} = \sum_{j=0}^{N-1}\widetilde{b}_j\ket{j}\ket{f_2(2\pi j/N)}$$ 
on the parallel quantum processors, and thus has a time complexity of $\tau_{\scriptscriptstyle oracle}$.
In \emph{Step}~$\mathbf{7}$, we apply iQFT to the resulting state to obtain $\ket{\psi_2}$ which, similar to \emph{Step}~$\mathbf{2}$, has a time complexity of $O(\log^2 N)$.
\emph{Step}~$\mathbf{8}$ involves encoding $N$ columns of the Toeplitz matrix $\T_4$ (as a quantum state), similar to \emph{Step}~$\mathbf{1}$, and thus $\tau_{\scriptscriptstyle load}$ represents the time to prepare the state $\ket{\psi^\ast} = \ket{\psi_2}+ \ket{\psi_3}$ (i.e., we  load the columns of $\T_4$ as an amplitude addition to the current state $\ket{\psi_2}$).

We now exploit and extend the above analysis to establish our main theoretical result on the asymptotic computational complexity of our quantum 
CR
algorithms.
Let $\mu = f_{1,\max}/f_{1,\min}$ for the generating function $f_1$ associated with Algorithm~\ref{algo:quantumCR},
let $N^Q = d_{\max}^Q \cdot M$ where 
$d_{\max}^Q$ is the maximum numerical degrees of the matrix power series generated by our quantum 
CR,
{and let $\tau_{\scriptscriptstyle readout}$ be the time complexity to readout the results from the quantum computer}.
Further
let $d_{\max}^C$ be the maximum numerical degrees of the matrix power series generated by the
corresponding classical
CR
algorithms, i.e., Algorithm~\ref{algo:classical} corresponding to Algorithm~\ref{algo:quantum} and the classical
CR
algorithm with the shifting technique corresponding to Algorithm~\ref{algo:quantum2}.
\begin{theorem}
The
overall 
computational complexity of the quantum 
CR
Algorithm~\ref{algo:quantum} and Algorithm~\ref{algo:quantum2} is given by
\begin{equation}
    O( \mu ( \tau_{\scriptscriptstyle load} + \log^2 N^Q + \tau_{\scriptscriptstyle oracle}) + \tau_{\scriptscriptstyle readout}) \qquad \mbox{  with  } \qquad d_{\max}^Q = O( d_{\max}^C ) ,
\label{eq:thm:complexity}
\end{equation}
and requiring $N^Q (2 \log N^Q + 1)$ qubits.
%
Supposing
{$\mu , \tau_{\scriptscriptstyle oracle}$}
to be $O(\poly\log N^Q)$,
we
then 
have that the
computational
{complexity $O( \mu ( \log^2 N^Q + \tau_{\scriptscriptstyle oracle}) )$}
of 
Algorithm~\ref{algo:quantum} and Algorithm~\ref{algo:quantum2} represents
an \emph{exponential speedup} 
{of the computation phase}
over the 
computational complexity 
of 
\begin{equation}
O(M^3 d_{\max}^C + M^2 d_{\max}^C \log d_{\max}^C )
\label{eq:thm:CRcomplexity}
\end{equation}
for
the
corresponding classical
CR
algorithm.
{Further supposing
$\tau_{\scriptscriptstyle load}$ to be $O(\poly\log N^Q)$ or subexponential in $(\log N^Q)$, we then additionally have that the
computational
complexity $O( \mu ( \tau_{\scriptscriptstyle load} + \log^2 N^Q + \tau_{\scriptscriptstyle oracle}) )$
of 
Algorithm~\ref{algo:quantum} and Algorithm~\ref{algo:quantum2} represents
a \emph{polynomial-to-exponential speedup} 
over the 
computational complexity
of~\eqref{eq:thm:CRcomplexity} for
the
corresponding classical
CR
algorithm.}
{Moreover, further supposing
$\tau_{\scriptscriptstyle readout}$ to be $O(\poly\log N^Q)$ or subexponential in $(\log N^Q)$, we then additionally have that the
overall 
computational complexity~\eqref{eq:thm:complexity} of 
Algorithm~\ref{algo:quantum} and Algorithm~\ref{algo:quantum2} represents
a \emph{polynomial-to-exponential speedup} 
over the 
overall 
computational complexity
of~\eqref{eq:thm:CRcomplexity} for
the
corresponding classical
CR
algorithm.}
\label{thm:complexity}
\end{theorem}

\mss{
\subsection{Discussion}
}
\mss{Our mathematical analysis of the previous section considers the two sources of error for each iteration of our quantum CR algorithms, one based on the truncation and approximation of infinite power series of matrices and the other based on the computations in quantum environments with finite representation of numbers.
Both sources of error occur at every CR iteration and accumulate to a global error as the iterations of the quantum algorithms proceed.
Our results of Theorem~\ref{lem7.15} then provide first-order estimates and upper bounds related to the global error expressions in~\eqref{eq7.56}, which we next exploit in Theorem~\ref{thm:expUB} and Theorem~\ref{thm:linUB} to establish upper bounds on the global quantum errors of our quantum CR Algorithm~\ref{algo:quantum} and Algorithm~\ref{algo:quantum2}, respectively.
Our results in Theorem~\ref{thm:complexity} establish various aspects of the computational complexity of our quantum CR algorithms in general as well as under certain conditions.
}

Given the definition of $\mu = f_{1,\max}/f_{1,\min}$ for the generating function $f_1$ associated with Algorithm~\ref{algo:quantumCR}
and given the properties of the Toeplitz matrix $\T_1$ (e.g., nonnegative, stochastic) together with typical ratios $f_{1,\max}/f_{1,\min}$ found in practice for M/G/$1$-type processes, 
we typically have $\mu$ to be $O(\poly\log N^Q)$ 
for sufficiently large $N^Q$.
In particular, for large $N^Q$, we have $\mu \approx \kappa$ where $\kappa$ is the condition number of ~$\T_1$. 
By 
exploiting appropriate oracle quantum circuit methods~\cite{nielsen2001quantum,bhaskar2016quantum,li2021universal} together with the properties of the Toeplitz matrices $\T_1, \T_2, \T_3, \T_4$ input to Algorithm~\ref{algo:quantumCR} and the corresponding generating functions $f_1$ and $f_2$, we can have the oracle time $\tau_{\scriptscriptstyle oracle}$ to be $O(\poly\log N^Q)$ for sufficiently large $N^Q$.
In such cases of theoretical and practical importance, Theorem~\ref{thm:complexity} establishes that our quantum algorithms provide 
{the potential for an exponential speedup of the computation phase in the decision-space on a quantum computer}
over that \mss{in~\eqref{eq:thm:CRcomplexity} for}
the best-known 
and most-efficient
classical algorithms,
{which is 
\mss{of particular}
interest from a mathematical perspective.}
\mss{
We note that oracles and related methods represent an important and standard tool in quantum computation~\cite{nielsen2010quantum,Dervovic18,Port18,de2019quantum,LipReg21,LinLin2022}, often playing a major role in many quantum algorithms in a manner consistent with the use of oracles in our quantum CR algorithms.
We further note once again that the computation phase of our quantum CR algorithms may represent a core component of a larger computational problem on the quantum computer,
such as an optimization problem involving the stationary distribution of an associated structured Markov process,
thus making the complexity of this computation phase more critical than that of the input and output phases.
}

Next,
by exploiting appropriate block encoding quantum circuit methods~\cite{LinLin2022,sunderhauf2024} together with properties of the Toeplitz matrices $\T_1, \T_2, \T_3, \T_4$ input to Algorithm~\ref{algo:quantumCR}
{to support efficient data loading},
we can {additionally} have the load time $\tau_{\scriptscriptstyle load}$ to be $O(\poly\log N^Q)$ {or subexponential in $(\log N^Q)$} for sufficiently large $N^Q$.
{In such cases of theoretical and practical importance, Theorem~\ref{thm:complexity} establishes that our quantum algorithms for both input and computation phases,
{ignoring readout time},
provide 
the potential for a polynomial-to-exponential speedup} over 
\mss{that in~\eqref{eq:thm:CRcomplexity} for}
the best-known 
and most-efficient
classical algorithms.
\mss{
We note that block encoding and related methods represent an important and standard tool in quantum computation~\cite{LinLin2022,nielsen2010quantum,Dervovic18,Port18,de2019quantum,LipReg21}, often playing a major role in many quantum algorithms in a manner consistent with the use of block encoding in our quantum CR algorithms.
}

\mss{
Lastly,
by exploiting properties of the result matrix $\G_{\min}$
such as instances where these 
matrices are sufficiently sparse to support efficient sampling-based readout, 
}
we can {additionally} have the readout time $\tau_{\scriptscriptstyle readout}$ to be $O(\poly\log N^Q)$ or subexponential in $(\log N^Q)$ for sufficiently large $N^Q$.
{In such cases of theoretical and practical importance, Theorem~\ref{thm:complexity} establishes that our quantum algorithms for the input, computation and readout phases
provide 
the potential for an overall polynomial-to-exponential speedup}
over 
\mss{that in~\eqref{eq:thm:CRcomplexity} for}
the best-known 
and most-efficient
classical algorithms.
\mss{We note that sampling and related methods represent an important and standard tool in quantum computation~\cite{nielsen2010quantum,Dervovic18,Port18,de2019quantum,LipReg21,LinLin2022}, often playing a major role in many quantum algorithms in a manner consistent with the use of sampling in our quantum CR algorithms;
some well-known examples include the important role of sampling in search~\cite{grover1996,grover1997} and Monte Carlo methods~\cite{montanaro2015},
as well as in solving linear systems of equations~\cite{HHL09}.
We further note that, when $\G_{\min}$ is not sufficiently sparse, we may be able to additionally exploit properties of the result matrix $\G_{\min}$ together with properties of~\eqref{eq:4.8}, \eqref{eq:4.9} and~\eqref{eq:4.10} to readout tailored approximations of $\G_{\min}$ for the computation of the stationary distribution $\bpi$ on a classical computer.
Moreover, when the computation phase of our quantum CR algorithms represents a component of a larger computational problem on the quantum computer as noted above, then readout of the $\G_{\min}$ matrix is not needed and the overall computational complexity does not include the readout phase.
}

\mss{
A fairly wide variety of general quantum computing architectures are available today and continue to emerge.
This spans the general class of quantum computing architectures based on logical quantum gates that are said to be universal in the sense of a quantum Turing machine, and thus in principle capable of executing any algorithm in the bounded-error quantum polynomial time (BQP) computational complexity class~\cite{yao1993,BernsteinVazirani97}.
Examples of this general class of quantum computing architectures include those grounded on superconducting qubits and developed by IBM, Google and Rigetti, those founded on trapped-ion systems and developed by IonQ and Quantinuum, those grounded on cold-atom devices and developed by QuEra,
and those founded on photonic devices~\cite{Photonics2019} and developed by Xanadu and PsiQuantum.
%
Our focus herein is on this general class of quantum computing architectures based on logical quantum gates.
Despite the broad spectrum of quantum computing architectures within this general class being developed by different companies, the implementation of the input, computation and output phases of our quantum algorithms will be 
consistent across these architectures though possibly with some straightforward 
alterations while not affecting the basic approach.
In particular, the implementation of key aspects of the computation phase of our algorithms on a quantum computer may very well require minor differences in the tailored realizations on different quantum architectures, but each implementation tailored to the quantum architecture of interest will have the computational complexity given by Theorem~\ref{thm:complexity} under the conditions noted above.
While the implementation of key aspects of the input phase of our algorithms on a quantum computer may require minor differences in the tailored realizations of the block encoding quantum circuit methods on different quantum architectures, each implementation tailored to the quantum architecture of interest will have the computational complexity given by Theorem~\ref{thm:complexity} under the conditions noted above.
Lastly, the implementation of key aspects of the output phase of our algorithms on a quantum computer primarily involves the sampling of the computed results, which is supported by the broad spectrum of quantum computing architectures based on logical quantum gates, and therefore any differences in the implementation tailored to the quantum architecture of interest will be minimal and will have the computational complexity given by Theorem~\ref{thm:complexity} under the conditions noted above.
}

\mss{
Finally,
it is important to note that our quantum algorithms can be applied
to provide significant computational improvements in solving
problems associated with the very general classes of structured Markov processes which span the wide variety of stochastic processes arising in the mathematical performance analysis, modeling and optimization of computer systems and networks.
Moreover, since CR is core to many important numerical methods,
our quantum algorithms can be exploited to address a much larger class of numerical computation problems in general.
}

\section{Proofs}
\label{sec:proofs}
In this section, we provide proofs of some of our theoretical results presented in Section~\ref{sec:quantum}, with additional details and more complete proofs provided
in Appendix~\ref{app:proofs}.

\subsection{Proof of Theorem~\ref{lem7.15}}
\label{sec:lem7.15}
%
We present an abbreviated version of our proof, with additional details
in Appendix~\ref{app:lem7.15}.
From~\eqref{eq7.38a} and~\eqref{eq7.38b}, replacing $\varphi^{(n)}(z)$ with its approximation $\vartheta^{(n)}(z)$, we obtain
\begin{align}
    \underline{\vartheta}^{(n+1)}(z) & = z \vartheta_{\scriptscriptstyle{odd}}^{(n)}(z) + \vartheta_{\scriptscriptstyle{even}}^{(n)}(z)(\I -\vartheta_{\scriptscriptstyle{odd}}^{(n)}(z))^{-1} \vartheta_{\scriptscriptstyle{even}}^{(n)}(z) \label{eq:7.58a} \\
    \underline{\hat{\vartheta}}^{(n+1)}(z) & = \hat{\vartheta}_{\scriptscriptstyle{odd}}^{(n)}(z) + \vartheta_{\scriptscriptstyle{even}}^{(n)}(z)(\I -\vartheta_{\scriptscriptstyle{odd}}^{(n)}(z))^{-1} \hat{\vartheta}_{\scriptscriptstyle{even}}^{(n)}(z) \label{eq:7.58b} .
\end{align}
By definition of $\R_{\scriptscriptstyle{odd}}^{(n)}(z)$, we have $\I - \vartheta_{\scriptscriptstyle{odd}}^{(n)}(z) = \I - \varphi_{\scriptscriptstyle{odd}}^{(n)}(z) - \R_{\scriptscriptstyle{odd}}^{(n)}(z)$, 
or equivalently
$\I - \vartheta_{\scriptscriptstyle{odd}}^{(n)}(z) = (\I - \varphi_{\scriptscriptstyle{odd}}^{(n)}(z))\, [\I - (\I - \varphi_{\scriptscriptstyle{odd}}^{(n)}(z))^{-1} \R_{\scriptscriptstyle{odd}}^{(n)}(z)]$,
and therefore
\begin{equation}
(\I - \vartheta_{\scriptscriptstyle{odd}}^{(n)}(z))^{-1} = [\I - (\I - \varphi_{\scriptscriptstyle{odd}}^{(n)}(z))^{-1} \R_{\scriptscriptstyle{odd}}^{(n)}(z)]^{-1} (\I - \varphi_{\scriptscriptstyle{odd}}^{(n)}(z))^{-1} .
\label{eq:Bodd-temp}
\end{equation}
By definition of first-order equality ($\doteq$), we conclude
$[\I - (\I - \varphi_{\scriptscriptstyle{odd}}^{(n)}(z))^{-1} \R_{\scriptscriptstyle{odd}}^{(n)}(z)]^{-1} \; \doteq \; \I + (\I - \varphi_{\scriptscriptstyle{odd}}^{(n)}(z))^{-1} \R_{\scriptscriptstyle{odd}}^{(n)}(z)$,
which upon substitution in~\eqref{eq:Bodd-temp} yields
\begin{align}
(\I - \vartheta_{\scriptscriptstyle{odd}}^{(n)}(z))^{-1} 
& \; \doteq \; (\I - \varphi_{\scriptscriptstyle{odd}}^{(n)}(z))^{-1} + (\I - \varphi_{\scriptscriptstyle{odd}}^{(n)}(z))^{-1} \R_{\scriptscriptstyle{odd}}^{(n)}(z) (\I - \varphi_{\scriptscriptstyle{odd}}^{(n)}(z))^{-1} 
.
\label{eq:1storderBodd}
\end{align}
Substituting~\eqref{eq:1storderBodd} and the definition of
$\R_{\scriptscriptstyle{even}}^{(n)}(z)$ 
into~\eqref{eq:7.58a}, and
subtracting~\eqref{eq7.38a} 
while
removing the higher-order terms of 
our first-order error analysis on $\underline{\R}^{(n+1)}(z)$, we derive
\begin{align}
    \underline{\R}^{(n+1)}(z) 
    & \; \doteq \; z \R_{\scriptscriptstyle{odd}}^{(n)}(z) + \R_{\scriptscriptstyle{even}}^{(n)}(z) (\I - \varphi_{\scriptscriptstyle{odd}}^{(n)}(z))^{-1} \varphi_{\scriptscriptstyle{even}}^{(n)}(z) + \varphi_{\scriptscriptstyle{even}}^{(n)}(z)
    (\I - \varphi_{\scriptscriptstyle{odd}}^{(n)}(z))^{-1} \R_{\scriptscriptstyle{even}}^{(n)}(z) \nonumber \\
    & \qquad\qquad + \varphi_{\scriptscriptstyle{even}}^{(n)}(z) (\I - \varphi_{\scriptscriptstyle{odd}}^{(n)}(z))^{-1} \R_{\scriptscriptstyle{odd}}^{(n)}(z) (\I - \varphi_{\scriptscriptstyle{odd}}^{(n)}(z))^{-1} \varphi_{\scriptscriptstyle{even}}^{(n)}(z) .
\label{eq:1storderR2}
\end{align}
Substituting $\V^{(n)}(z)$ and $\W^{(n)}(z)$
then yields~\eqref{eq:7.57a}.
Repeating 
the same analysis for~\eqref{eq:7.58b} 
and 
substituting $\V^{(n)}(z)$ and $\hat{\W}^{(n)}(z)$
renders~\eqref{eq:7.57b}.

Turning to the corresponding first-order upper bounds and focusing on $\|\underline{\R}^{(n+1)}(z)\|_{\ast}$, we apply the max norm $\| \cdot \|_{\ast}$ to both sides of~\eqref{eq:7.57a} and exploit the triangle inequality to obtain 
\begin{align*}
    \|\underline{\R}^{(n+1)}(z)\|_{\ast} & \; \dotleq \; \|\R_{\scriptscriptstyle{odd}}^{(n)}(z)\|_{\ast} + \|\R_{\scriptscriptstyle{even}}^{(n)}(z)\|_{\ast} \|\W^{(n)}(z)\|_{\ast} + \|\R_{\scriptscriptstyle{even}}^{(n)}(z)\|_{\ast} \|\V^{(n)}(z)\|_{\ast}
    \\
    & \qquad\qquad\qquad\qquad 
    + \|\R_{\scriptscriptstyle{odd}}^{(n)}(z)\|_{\ast} \|\V^{(n)}(z)\|_{\ast} \|\W^{(n)}(z)\|_{\ast} .
\end{align*}
From the monotonicity property of the infinity norm, we know that both $\|\R_{\scriptscriptstyle{odd}}^{(n)}(z)\|_{\ast} \leq \|\R^{(n)}(z)\|_{\ast}$ and $\|\R_{\scriptscriptstyle{even}}^{(n)}(z)\|_{\ast} \leq \|\R^{(n)}(z)\|_{\ast}$, thus rendering
\begin{align}
    \|\underline{\R}^{(n+1)}(z)\|_{\ast} & \; \dotleq \; \|\R^{(n)}(z)\|_{\ast} + \|\R^{(n)}(z)\|_{\ast} \|\W^{(n)}(z)\|_{\ast} + \|\R^{(n)}(z)\|_{\ast} \|\V^{(n)}(z)\|_{\ast} \nonumber \\
    & \qquad\qquad\qquad\qquad + \|\R^{(n)}(z)\|_{\ast} \|\V^{(n)}(z)\|_{\ast} \|\W^{(n)}(z)\|_{\ast} . \label{eq:7.57a-leq2} 
\end{align}
The coefficients of the matrix power series $\V^{(n)}(z)$ and $\W^{(n)}(z)$ are nonnegative.
From Theorem~\ref{thm7.8}, the matrices $\V^{(n)}(1)$ and $\W^{(n)}(1)$ have spectral radius $1$ with $\W^{(n)}(1)$ being stochastic, and thus it follows that $\|\W^{(n)}(z)\|_{\ast} = \|\W^{(n)}(1)\|_{\infty} = 1$.
Upon substitution of these relationships into~\eqref{eq:7.57a-leq2}, we have
\begin{align}
    \|\underline{\R}^{(n+1)}(z)\|_{\ast} & \; \dotleq \; 2 \|\R^{(n)}(z)\|_{\ast} + 2 \|\R^{(n)}(z)\|_{\ast} \|\V^{(n)}(z)\|_{\ast} = 2 \|\R^{(n)}(z)\|_{\ast} (1 + \|\V^{(n)}(1)\|_{\infty}) , \label{eq:7.57a-leq3} 
\end{align}
where the latter equality in~\eqref{eq:7.57a-leq3} follows from the nonnegative coefficients of $\V^{(n)}(z)$, thus yielding~\eqref{eq:7.57-UBa}.
Repeating the same type of analysis renders the desired result~\eqref{eq:7.57-UBb} for $\|\underline{\hat{\R}}^{(n+1)}(z)\|_{\ast}$.
\qed

\subsection{Proof of Theorem~\ref{thm:expUB}}
\label{sec:thm:expUB}
%
We present an abbreviated version of our proof, with additional details
in Appendix~\ref{app:thm:expUB}.
From Theorem~\ref{lem7.15} and  equations~\eqref{eq:local-error}, \eqref{eq:global-error} and~\eqref{eq7.56}, we have
\begin{align}
 \bE^{(n+1)}(z) & \; \doteq \; z\bE^{(n)}_{odd}(z) + \V^{(n)}(z)\bE^{(n)}_{even}(z)+ \bE^{(n)}_{even}(z)\W^{(n)}(z) + \V^{(n)}(z)\bE^{(n)}_{odd}(z)\W^{(n)}(z) 
 + \cE_L^{(n)}(z) , \label{eq:7.59a} \\
 \hat{\bE}^{(n+1)}(z) & \; \doteq \; \hat{\bE}^{(n)}_{odd}(z) +  {\V}^{(n)}(z)\hat{\bE}^{(n)}_{even}(z) + {\bE}^{(n)}_{even}(z)\hat{\W}^{(n)}(z) + {\V}^{(n)}(z)\bE^{(n)}_{odd}(z)\hat{\W}^{(n)}(z) 
 + \hat{\cE}_L^{(n)}(z) , \label{eq:7.59b}
\end{align}
where
$$
\bE^{(0)}(z) = \hat{\bE}^{(0)}(z) = 0 , \qquad 
\V^{(n)}(z) = \varphi_{\scriptscriptstyle{even}}^{(n)}(z)(\I -\varphi_{\scriptscriptstyle{odd}}^{(n)}(z))^{-1} , \qquad
\W^{(n)}(z) = (\I -\varphi_{\scriptscriptstyle{odd}}^{(n)}(z))^{-1}\varphi_{\scriptscriptstyle{even}}^{(n)}(z) ,
$$
$$
\hat{\W}^{(n)}(z) = (\I -\varphi_{\scriptscriptstyle{odd}}^{(n)}(z))^{-1}\hat{\varphi}_{\scriptscriptstyle{even}}^{(n)}(z) , \qquad\qquad  
(\cE_L^{(n)}(z),\hat{\cE}_L^{(n)}(z)) = \cE_S(\vartheta^{(n)}(z),\hat{\vartheta}^{(n)}(z)) .
$$
Focusing on $\|\bE^{(n+1)}(z)\|_{\ast}$, we apply the max norm $\|\cdot\|_{*}$ to both sides of~\eqref{eq:7.59a} and exploit the triangle inequality to obtain
\begin{align*}
 \|\bE^{(n+1)}(z)\|_{\ast} & \; \dotleq \; \|\bE^{(n)}_{odd}(z)\|_{\ast} + \|\V^{(n)}(z)\|_{\ast} \|\bE^{(n)}_{even}(z)\|_{\ast} + \|\bE^{(n)}_{even}(z)\|_{\ast} \|\W^{(n)}(z)\|_{\ast} \\
 & \qquad\qquad\qquad + \|\V^{(n)}(z)\|_{\ast} \|\bE^{(n)}_{odd}(z)\|_{\ast} \|\W^{(n)}(z)\|_{\ast} 
 + \|\cE_L^{(n)}(z)\|_{\ast} .
\end{align*}
From the monotonicity property of the infinity norm, we know that both $\|\bE_{\scriptscriptstyle{odd}}^{(n)}(z)\|_{\ast} \leq \|\bE^{(n)}(z)\|_{\ast}$ and $\|\bE_{\scriptscriptstyle{even}}^{(n)}(z)\|_{\ast} \leq \|\bE^{(n)}(z)\|_{\ast}$, thus rendering
\begin{align}
 \|\bE^{(n+1)}(z)\|_{\ast} & \; \dotleq \; \|\bE^{(n)}(z)\|_{\ast} + \|\V^{(n)}(z)\|_{\ast} \|\bE^{(n)}(z)\|_{\ast} + \|\bE^{(n)}(z)\|_{\ast} \|\W^{(n)}(z)\|_{\ast} \nonumber \\
 & \qquad\qquad\qquad + \|\V^{(n)}(z)\|_{\ast} \|\bE^{(n)}(z)\|_{\ast} \|\W^{(n)}(z)\|_{\ast} 
 + \|\cE_L^{(n)}(z)\|_{\ast} . \label{eq:7.59a-leq2} 
\end{align}

The coefficients of the matrix power series $\V^{(n)}(z)$ and $\W^{(n)}(z)$ are nonnegative.
From Theorem~\ref{thm7.8}, the matrices $\V^{(n)}(1)$ and $\W^{(n)}(1)$ have spectral radius $1$ with $\W^{(n)}(1)$ being stochastic, and thus it follows that $\|\W^{(n)}(z)\|_{\ast} = \|\W^{(n)}(1)\|_{\infty} = 1$.
Upon substitution of these relationships into~\eqref{eq:7.59a-leq2}, we have
\begin{align}
 \|\bE^{(n+1)}(z)\|_{\ast} & \; \dotleq \; \|\bE^{(n)}(z)\|_{\ast} + \|\V^{(n)}(z)\|_{\ast} \|\bE^{(n)}(z)\|_{\ast} + \|\bE^{(n)}(z)\|_{\ast} + \|\V^{(n)}(z)\|_{\ast} \|\bE^{(n)}(z)\|_{\ast} + \|\cE_L^{(n)}(z)\|_{\ast} \nonumber \\
 & = 2(1 + \|\V^{(n)}(1)\|_{\infty})\|\bE^{(n)}(z)\|_{\ast} + \|\cE_L^{(n)}(z)\|_{\ast} , \label{eq:7.59a-leq3} 
\end{align}
where the latter equality in~\eqref{eq:7.59a-leq3} follows from the nonnegative coefficients of $\V^{(n)}(z)$.
Under the suppositions of the theorem, we conclude
\begin{align*}
    \|\bE^{(n+1)}(z)\|_{\ast} & \; \dotleq \; 
    \gamma_n \|\bE^{(n)}(z)\|_{\ast} + \upsilon , 
\end{align*}
from which the desired result for $\|\bE^{(n+1)}(z)\|_{\ast}$ then follows.
Repeating the same type of analysis yields the desired result for $\|\hat{\bE}^{(n+1)}(z)\|_{\ast}$.
\qed

\subsection{Proof of Theorem~\ref{thm:linUB}}
\label{sec:thm:linUB}
%
We present an abbreviated version of our proof, with additional details
in Appendix~\ref{app:thm:linUB}.
By applying
CR
to the function $\widetilde{\chi}(z)$ 
w.r.t.\
the corresponding sequences of matrix power series $\widetilde{\chi}^{(n)}(z)$ and $\hat{\widetilde{\chi}}^{(n)}(z)$,
we once again have~\eqref{eq:7.59a} and~\eqref{eq:7.59b}
from Theorem~\ref{lem7.15} and  equations~\eqref{eq:local-error}, \eqref{eq:global-error} and~\eqref{eq7.56}.
Focusing on $\|\bE^{(n+1)}(z)\|_{\ast}$, we
take the max norm $\|\cdot\|_{*}$ on both sides of~\eqref{eq:7.59a} and exploit the triangle inequality to obtain
\begin{align}
 \|\bE^{(n+1)}(z)\|_{\ast} & \; \dotleq \; \|\bE^{(n)}_{odd}(z)\|_{\ast} + \|\V^{(n)}(z)\|_{\ast} \|\bE^{(n)}_{even}(z)\|_{\ast} + \|\bE^{(n)}_{even}(z)\|_{\ast} \|\W^{(n)}(z)\|_{\ast} \nonumber \\
 & \qquad\qquad\qquad + \|\V^{(n)}(z)\|_{\ast} \|\bE^{(n)}_{odd}(z)\|_{\ast} \|\W^{(n)}(z)\|_{\ast} 
 + \|\cE_L^{(n)}(z)\|_{\ast} . \label{eq:7.59a-bnd}
\end{align}
From the monotonicity property of the infinity norm, we know that both $\|\bE_{\scriptscriptstyle{odd}}^{(n)}(z)\|_{\ast} \leq \|\bE^{(n)}(z)\|_{\ast}$ and $\|\bE_{\scriptscriptstyle{even}}^{(n)}(z)\|_{\ast} \leq \|\bE^{(n)}(z)\|_{\ast}$.
By applying 
CR
to the function $\widetilde{\chi}(z)$ 
w.r.t.\
the corresponding sequences of matrix power series $\widetilde{\chi}^{(n)}(z)$ and $\hat{\widetilde{\chi}}^{(n)}(z)$, we further have that $\| \V^{(n)}(z) \|_{\ast}$ and $\| \W^{(n)}(z) \|_{\ast}$ associated with Theorem~\ref{lem7.15} are upper bounded by $\theta \sigma^{2^n}$ for appropriate $\theta > 0$ and $0 < \sigma < 1$.
Upon substituting
these sets of upper bounds 
into~\eqref{eq:7.59a-bnd}, we obtain
\begin{align*}
\| \bE^{(n+1)}(z) \|_{\ast} & \; \dotleq \; \| \bE^{(n)}(z) \|_{\ast} + \theta \sigma^{2^n} \| \bE^{(n)}(z) \|_{\ast} + (\|\bE^{(n)}(z)\|_{\ast} + \theta \sigma^{2^n} \| \bE^{(n)}(z) \|_{\ast})\theta \sigma^{2^n} + \| \cE_L^{(n)}(z) \|_{\ast} \\
    & \qquad\qquad = ( 1 + \theta \sigma^{2^n} )^2 \| \bE^{(n)}(z) \|_{\ast} + \| \cE_L^{(n)}(z) \|_{\ast} .
\end{align*}
Under the suppositions of the theorem, we conclude 
\begin{align}
\|\bE^{(n+1)}(z)\|_{*} & \; \dotleq \; (1 + \theta\sigma^{2^n})^2 \| \bE^{(n)}(z) \|_{\ast} + \upsilon . \label{eq:E-ub} 
\end{align}

From~\eqref{eq:E-ub} together with $\gamma_n = (1 + \theta\sigma^{2^n})^2$, we obtain 
\begin{align}
\|\bE^{(n+1)}(z)\|_{*} & \; \dotleq \; \gamma_n \| \bE^{(n)}(z) \|_{\ast} + \upsilon , \nonumber \\
&\; \dotleq \; \upsilon (1+  \gamma_n+\gamma_{n-1}\gamma_n+\cdots + \gamma_2\gamma_3\cdots\gamma_{n-1}\gamma_n + \gamma_1\gamma_2\cdots\gamma_{n-1}\gamma_n) . \label{eq:7.61a} 
\end{align}
For the $(k+1)$st summand on the right hand side 
of~\eqref{eq:7.61a},
we derive
\begin{align*}
    \log\left( \prod_{i=n-k+1}^n \gamma_i \right) & \; = \; \sum_{i=n-k+1}^n \log\left( \gamma_i \right) \; = \; \sum_{i=n-k+1}^n \log\left( (1 + \theta\sigma^{2^i})^2 \right) \\ & \; = \; 2 \sum_{i=n-k+1}^n \log\left(1 + \theta\sigma^{2^i} \right) \; \leq \;  
    2 \theta \sum_{i=n-k+1}^n \sigma^{2^i} \leq 2 \theta \frac{\sigma^2} {1-\sigma^2} ,
\end{align*}
and therefore~\eqref{eq:7.61a} renders
\begin{equation*}
   \|\bE^{(n+1)}(z)\|_{*} \; \dotleq \; \upsilon \left(1 +  \sum_{k=1}^n \exp ( 2 \theta \frac{\sigma^2}{1-\sigma^2} )\right) ,
\end{equation*}
which yields the desired result for $\|\bE^{(n+1)}(z)\|_{*}$.
Repeating the same type of analysis renders the desired result for $\| \hat{\bE}^{(n+1)}(z)\|_{*}$.
\qed

\subsection{Proof of Theorem~\ref{thm:complexity}}
\label{sec:thm:complexity}
%
We present an abbreviated version of our proof, with additional details
in Appendix~\ref{app:thm:complexity}.
Since Algorithm~\ref{algo:quantumCR} is common to the inner loop of both Algorithm~\ref{algo:quantum} and Algorithm~\ref{algo:quantum2}, we first
sum up
the asymptotic time complexity of \emph{Step}~$\mathbf{1}$ through \emph{Step}~$\mathbf{8}$ of Algorithm~\ref{algo:quantumCR}.
This yields a total asymptotic time complexity of $O( \mu (\tau_{\scriptscriptstyle load} + \log^2 N + \tau_{\scriptscriptstyle oracle}) )$ for each quantum
CR
iteration~$n$, requiring $N(2\log N + 1)$ qubits.

Now, focusing on Algorithm~\ref{algo:quantum} and similar to the complexity analysis for Algorithm~\ref{algo:quantumCR},
the asymptotic time complexity of \emph{Step}~$\mathbf{3}$ of Algorithm~\ref{algo:quantum} is given by $O(\mu(\tau_{\scriptscriptstyle load} + \log^2 N))$ while reusing $M \log M$ qubits.
Due to the quadratic convergence of 
CR
{(along the lines of Theorem~\ref{thm7.13}),}
and consistent with the 
analysis of 
{our quantum 
CR
algorithm}
given 
{in}
Theorem~\ref{lem7.15} and Theorem~\ref{thm:expUB},
the total number of iterations~$n$ of the inner loop (Algorithm~\ref{algo:quantumCR} and \emph{Step}~$\mathbf{3}$ of Algorithm~\ref{algo:quantum}) 
does not depend on 
the increasing sequence of $N$ up to the maximum $N^Q$.
The overall asymptotic time complexity of the inner loop then depends on the corresponding maximum numerical degrees $d_{\max}^Q$ of the matrix power series generated by this instance of quantum
CR,
and therefore this time complexity is given by $O( \mu (\tau_{\scriptscriptstyle load} + \log^2 N^Q + \tau_{\scriptscriptstyle oracle}) )$ and requires $N^Q (2\log N^Q + 1)$ qubits.
The asymptotic time complexity of \emph{Step}~$\mathbf{4}$ of Algorithm~\ref{algo:quantum} is relatedly given by $O(\mu(\tau_{\scriptscriptstyle load} + \log^2 M))$ while reusing $M \log M$ qubits.
Combining this with the time complexity for the inner loop and the time complexity for the result readout then yields the overall asymptotic time complexity of Algorithm~\ref{algo:quantum} given in~\eqref{eq:thm:complexity}.
Lastly, in the case of Algorithm~\ref{algo:quantum},
we have $d_{\max}^Q = O( d_{\max}^C )$ in~\eqref{eq:thm:complexity}
following from Theorem~\ref{lem7.15} and Theorem~\ref{thm:expUB}.

Repeating the same type of complexity analysis yields the corresponding set of asymptotic time complexity results for Algorithm~\ref{algo:quantum2}.

Under the suppositions of the theorem
w.r.t.\
{$\mu$ and $\tau_{\scriptscriptstyle oracle}$,}
the time complexity~\eqref{eq:thm:complexity} 
{for the computation phase}
of both Algorithms~\ref{algo:quantum} and~\ref{algo:quantum2}
reduces to $O( \poly\log \normalsize( d_{\max}^Q \cdot M ) )$.
Then, in comparison with 
the time complexity~\eqref{eq:thm:CRcomplexity} for the corresponding classical CR algorithms,
the exponential speedup 
{for the computation phase}
follows from both 
$d_{\max}^Q = O( d_{\max}^C )$ in~\eqref{eq:thm:complexity}
and the time complexity
in~\eqref{eq:thm:complexity}.

Under the 
{additional supposition}
of the theorem
w.r.t.\
{$\tau_{\scriptscriptstyle load}$,}
the 
time complexity~\eqref{eq:thm:complexity} 
of both Algorithms~\ref{algo:quantum} and~\ref{algo:quantum2}
reduces to $O( \poly\log \normalsize(d_{\max}^Q \cdot M) )$
{or subexponential in $\log \normalsize(d_{\max}^Q \cdot M)$}.
Then, in comparison with
{the time complexity~\eqref{eq:thm:CRcomplexity}}
{for the
corresponding classical 
CR
algorithms}, the
{polynomial-to-exponential speedup}
follows from both 
$d_{\max}^Q = O( d_{\max}^C )$ in~\eqref{eq:thm:complexity}
and the time complexity
in~\eqref{eq:thm:complexity}.

Under the 
{additional supposition}
of the theorem
w.r.t.\
{$\tau_{\scriptscriptstyle readout}$,}
the 
{overall}
time complexity~\eqref{eq:thm:complexity} 
of both Algorithms~\ref{algo:quantum} and~\ref{algo:quantum2}
reduces to $O( \poly\log \normalsize(d_{\max}^Q \cdot M) )$
{or subexponential in $\log \normalsize(d_{\max}^Q \cdot M)$}.
Then, in comparison with
{the overall time complexity~\eqref{eq:thm:CRcomplexity}}
{for the
corresponding classical 
CR
algorithms}, the 
{polynomial-to-exponential speedup}
follows from both 
$d_{\max}^Q = O( d_{\max}^C )$ in~\eqref{eq:thm:complexity}
and the time complexity
in~\eqref{eq:thm:complexity}.
\qed

\section{Conclusions}
In this paper,
we present a theoretical study of
the fundamental problem of efficiently computing the stationary distribution for 
\mss{the very}
general classes of structured Markov processes from the perspective of quantum computing,
{making important mathematical contributions 
w.r.t.\
quantum algorithms and the analysis of their computational properties}.
Specifically, we derive the first quantum algorithms for solving this fundamental problem, and
{we} derive a {rigorous} mathematical analysis of the computational errors and computational complexity of our quantum algorithms
{together with}
related theoretical results,
{establishing that our quantum algorithms provide {the potential for significant computational improvements} over the
best-known 
and most-efficient
classical algorithms in {various} settings of both theoretical and practical importance.}
\mss{In particular, our theoretical results include establishing
the potential exponential speedup for the computation phase, 
the potential polynomial-to-exponential speedup with efficient data loading through block encoding based on the properties of the input matrices, and
the potential polynomial-to-exponential speedup with efficient result readout through sampling based on the properties of the output matrix, all in comparison against the
most-efficient classical algorithms.}
Our primary focus is from a mathematical perspective, 
in which
this collection of results
expands the limited set of quantum algorithms that
provably achieve 
\mss{significant computational improvements in various settings of importance,}
while also making important algorithmic and theoretical contributions.
\mss{We further note that our quantum algorithms can be applied to provide significant computational improvements in solving problems associated with the very general classes of structured Markov processes of interest herein which span the wide range of stochastic processes arising in the mathematical performance analysis, modeling and optimization of computer systems and networks.}
Since 
CR
is core to many important numerical methods far beyond the \mss{general classes of} structured Markov 
\mss{processes,}
our quantum algorithms can be exploited 
to address a much larger class of numerical computation problems\mss{, as well as to potentially play the role of a subroutine as part of solving larger computational problems involving the corresponding stationary distribution on a quantum computer}.

%
%

\newpage
\appendix

\renewcommand{\thesection}{\Alph{section}} 
\makeatletter
\renewcommand\@seccntformat[1]{\appendixname\ \csname the#1\endcsname.\hspace{0.5em}}
\makeatother

\section{M/G/$1$-Type Markov Processes}
\label{app:mg1}
In this appendix, we provide additional technical background and details on the preliminaries presented in Section~\ref{sec:prelim:mg1} covering M/G/$1$-type Markov processes.
Specifically, we now present a set of known theoretical results related to the solution of the stationary distribution $\bpi$ for M/G/$1$-type Markov processes.
To start, we have the following lemma and theorem
\mss{associated with the matrix $\G_{\min}$}
where a solution $\X$ is called a minimal nonnegative solution of a matrix equation if $\bzero \leq \X \leq \Y$ for any other nonnegative solution $\Y$, with the inequalities holding componentwise.
\begin{lemma}[{\cite[Lemma~4.2]{BiLaMe05}}]
We have
\begin{equation}
    \bH \begin{bmatrix} \G^{(1)} \\ \G^{(2)} \\ \G^{(3)} \\ \vdots \end{bmatrix} \; = \; \begin{bmatrix} \A_{-1} \\ \bzero \\ \bzero \\ \vdots \end{bmatrix} ,
\tag{\ref{lem4.2:4.5}}
\end{equation}
where
\begin{equation}
    \bH \; = \; \begin{bmatrix}
    \I - \A_0 & -\A_1 & -\A_2 & -\A_3 & \cdots \\
    -\A_{-1} & \I - \A_0 & -\A_1 & -\A_2 & \ddots \\
    \bzero & -\A_{-1} & \I - \A_0 & -\A_1 & \ddots \\
    \bzero & \bzero & -\A_{-1} & \I - \A_0 & \ddots \\
    \vdots & \vdots & \ddots & \ddots & \ddots \\
    \end{bmatrix}
\tag{\ref{lem4.2:4.6}}
\end{equation}
and the matrices $\G^{(1)}, \: \G^{(2)}, \:  \G^{(3)}, \: \dots$ form the minimal nonnegative solution of the system~\eqref{lem4.2:4.5} according to elementwise ordering.
\label{lem4.2}
\end{lemma}

\begin{theorem}[{\cite[Theorem~4.3]{BiLaMe05}}]
Let $\G_{\min}$ denote the matrix of first passage probabilities from level $\ell+1$ to level $\ell$, independent of $\ell \in\Ints_+$.
The matrix $\G_{\min}$ is the minimal nonnegative solution of~\eqref{eq:4.4} and we have $\G^{(i)} = \G^i_{\min}$, for all $i \in\Ints_+$.
Moreover, under our ergodicity assumption, the matrix $\G_{\min}$ is stochastic.
\label{thm4.3}
\end{theorem}

Next, we can obtain the stationary distribution as follows.
\begin{theorem}[{\cite[Theorem~4.4]{BiLaMe05}}]
Assume that the Markov process $\{ X(n) \, ; \, n \in \Ints_+ \}$ is ergodic and let
\begin{equation}
    \A_i^\ast \; = \; \sum_{k=i}^\infty \A_k \G_{\min}^{k-i} \qquad \mbox{  and  } \qquad \B_i^\ast \; = \; \sum_{k=i}^\infty \B_k \G_{\min}^{k-i}, \qquad i \in\Ints_+ .
\label{eq:4.8}
\end{equation}
Then $\I - \A_0^\ast$ is nonsingular,
\begin{equation}
    \bpi_i \; = \; \left( \bpi_0 \B_i^\ast \, + \, \sum_{k=1}^{i-1} \bpi_k \A_{i-k}^\ast \right) \left(\I - \A_0^\ast\right)^{-1}, \qquad i \in\Ints_+ ,
\label{eq:4.9}
\end{equation}
and $\bpi_0$ satisfies
\begin{equation}
    \bpi_0 \B_0^\ast \; = \; \bpi_0 .
\label{eq:4.10}
\end{equation}
\label{thm4.4}
\end{theorem}

\begin{remark}
It is important to note that Lemma~\ref{lem4.2}, Theorem~\ref{thm4.3} and Theorem~\ref{thm4.4} all hold when $M$ is infinite, in which case we simply need to replace the matrix $(\I - \A_0^\ast)^{-1}$ in~\eqref{eq:4.9} with the series $\sum_{m=0}^\infty (\A_0^\ast)^m$.
\end{remark}

To address the ergodicity of the M/G/$1$-type Markov process, 
recalling $\A = \sum_{i=-1}^\infty \A_{i}$,
the first condition of interest is that the matrix
$\A$ 
is irreducible, in addition to being stochastic.
When $M$ is finite, this condition together with Perron–Frobenius theory implies that $\A$ has an invariant probability vector $\balpha$, i.e., $\balpha \A = \balpha$ and $\balpha \ones = 1$. We then can establish the following result.

\begin{theorem}[{\cite[Theorem~4.7]{BiLaMe05}}]
Assuming $M$ is finite, the transition probability matrix $\bP$ is irreducible and aperiodic, and the matrix $\A$ is irreducible as well as stochastic, let us define the vector $\ba = \sum_{i=-1}^\infty i \A_i \ones$ and the drift $\varrho = \balpha \ba$.
Then the Markov process with transition probability matrix $\bP$ is ergodic if and only if $\varrho < 0$ and $\bb = \sum_{i=1}^\infty i \B_i \ones < \infty$.
\label{thm4.7}
\end{theorem}

Theorem~\ref{thm4.7} provides the desired ergodicity conditions for the Markov process $\{ X(n) \, ; \, n \in \Ints_+ \}$, which together with Theorem~\ref{thm4.4} provides the solution of \eqref{eq:4.10} for $\bpi_0$ up to a multiplicative constant.
To address the latter, we can establish the following result that provides a normalizing equation for $\bpi_0$.

\begin{theorem}[{\cite[Theorem~4.8]{BiLaMe05}}]
Assuming $M$ is finite, the transition probability matrix $\bP$ is irreducible and aperiodic, the matrix $\A$ is irreducible as well as stochastic, and the Markov process with transition probability matrix $P$ is ergodic, then $\bpi_0$ is 
such that 
\begin{equation*}
    \bpi_0 \bb - \varrho \bpi_0 \ones + \bpi_0 (\I - \B) (\I - \A)^{\#} \ba \; = \; - \varrho ,
\end{equation*}
where the operator $(\cdot)^{\#}$ denotes the group inverse of a matrix.
\label{thm4.8}
\end{theorem}

\begin{remark}
It is important to note that, when $M$ is infinite, results corresponding to Theorem~\ref{thm4.7} and Theorem~\ref{thm4.8} 
can be established with appropriately defined operators~\cite{DunSch-all} associated with the drift $\varrho$,
{the vectors $\ba$ and $\bb$,}
and the group inverse.
\end{remark}

\section{Classical Computational Methods}
\label{app:classical}
In this appendix, we present the most-efficient classical solution methods for computing the stationary distribution $\bpi$ of the general class of M/G/$1$-type
Markov 
processes on digital computers.
From Theorem~\ref{thm4.3}, the ultimate goal concerns determining the minimal nonnegative solution $\G_{\min}$ of the nonlinear matrix equation~\eqref{eq:4.4} for such ergodic M/G/$1$-type
processes, from which we can obtain
$\bpi$ as a result of Theorems~\ref{thm4.4} and~\ref{thm4.8}.
The most-efficient classical algorithms for computing the matrix $\G_{\min}$ are based on variants of
cyclic reduction~\cite{BinMei95,BinMei96,BinMei97,BiLaMe05}, and thus our focus is on this class of algorithms.
Due to the complexity of these algorithms,
a {possible} lack of their broad familiarity,
and some important subtle errors in the existing research literature,
%
we first present a detailed derivation of the most advanced classical
cyclic reduction 
methods for structured Markov processes,
including theoretical properties and results that 
will be exploited 
as part of the design and analysis of our
corresponding
quantum algorithms.
We then present algorithmic details
of the most-efficient classical variants of
cyclic reduction 
for our purposes.

\subsection{Mathematical Analysis of Cyclic Reduction}
Conceptually speaking, the general
cyclic reduction 
approach consists of rewriting~\eqref{eq:4.4} in matrix form as 
\begin{equation}
    \bH \begin{bmatrix} \G^{(1)} \\ \G^{(2)} \\ \G^{(3)} \\ \vdots \end{bmatrix} \; = \; \begin{bmatrix} \A_{-1} \\ \bzero \\ \bzero \\ \vdots \end{bmatrix}
\label{lem4.2:4.5}
\end{equation}
from Lemma~\ref{lem4.2}, 
and applying variants of the
cyclic-reduction 
method in functional form to solve~\eqref{lem4.2:4.5}.
A single iteration of
cyclic reduction 
comprises applying an even-odd permutation to the block rows and to the block columns of the matrix 
\begin{equation}
    \bH \; = \; \begin{bmatrix}
    \I - \A_0 & -\A_1 & -\A_2 & -\A_3 & \cdots \\
    -\A_{-1} & \I - \A_0 & -\A_1 & -\A_2 & \ddots \\
    \bzero & -\A_{-1} & \I - \A_0 & -\A_1 & \ddots \\
    \bzero & \bzero & -\A_{-1} & \I - \A_0 & \ddots \\
    \vdots & \vdots & \ddots & \ddots & \ddots \\
    \end{bmatrix}
\label{lem4.2:4.6}
\end{equation}
from Lemma~\ref{lem4.2}, 
followed by one iteration of block Gaussian elimination.
Since this transformation maintains the unchanged structure of the system,
cyclic reduction 
is applied recursively which yields a  sequence of block Hessenberg, block Toeplitz-like infinite systems that provably converges quadratically to a limit system whose solution can be explicitly evaluated.
The matrix of the infinite system at each iteration is fully characterized by its first and second block rows whose respective block entries define two formal matrix power series with an explicit functional relation that expresses in functional form the
cyclic reduction 
iteration.

More precisely, consider 
the general M/G/$1$-type
Markov 
process
and the corresponding matrix equation~\eqref{eq:4.4}.
This
process can be transformed into a semi-infinite linear system that, for the matrix equation~\eqref{eq:4.4}, leads to the linear system~\eqref{lem4.2:4.5} and~\eqref{lem4.2:4.6} where $\G^{(i)} = \G_{\min}^i$ for
$i\in\Nats$.
We then apply an even-odd permutation to the block rows and to the block columns of~\eqref{lem4.2:4.5} and~\eqref{lem4.2:4.6}, deriving in compact form
\begin{equation}
    \begin{bmatrix}
    \I - \U_{11} & -\U_{12} \\
    - \U_{21} & \I -\U_{22}  \\
    \end{bmatrix}
    \begin{bmatrix}
    \x_{+} \\ \x_{-} \\
    \end{bmatrix}
    =
    \begin{bmatrix}
    \bzero \\ \bb \\
    \end{bmatrix} 
\tag{\ref{eq:compact}}
\end{equation}
where 
\begin{align}
    \U_{11} & 
    =
    \begin{bmatrix}
    \A_0 & \A_2 & \A_4 & \cdots \\
    \bzero & \A_0 & \A_2 & \ddots \\
    \bzero & \bzero & \A_0 & \ddots \\
    \vdots & \ddots & \ddots & \ddots \\
    \end{bmatrix} ,
    \quad & 
    \U_{12} =
    \begin{bmatrix}
    \A_{-1} & \A_1 & \A_3 & \cdots \\
    \bzero & \A_{-1} & \A_1 & \ddots \\
    \bzero & \bzero & \A_{-1} & \ddots \\
    \vdots & \ddots & \ddots & \ddots \\
    \end{bmatrix} ,
    \qquad
    \x_{+} & 
    =
    \begin{bmatrix}
    \G_{\min}^2 \\
    \G_{\min}^4 \\
    \G_{\min}^6 \\
    \vdots \\
    \end{bmatrix} , \tag{\ref{eq:Us+x+}} \\
    \U_{21} & 
    =
    \begin{bmatrix}
    \A_1 & \A_3 & \A_5 & \cdots \\
    \A_{-1} & \A_1 & \A_3 & \ddots \\
    \bzero & \A_{-1} & \A_1 & \ddots \\
    \vdots & \ddots & \ddots & \ddots \\
    \end{bmatrix} ,
    \quad
    & 
    \U_{22} =
    \begin{bmatrix}
    \A_0 & \A_2 & \A_4 & \cdots \\
    \bzero & \A_0 & \A_2 & \ddots \\
    \bzero & \bzero & \A_0 & \ddots \\
    \vdots & \ddots & \ddots & \ddots \\
    \end{bmatrix} ,
    \qquad
    \x_{-}
    & 
    =
    \begin{bmatrix}
    \G_{\min} \\
    \G_{\min}^3 \\
    \G_{\min}^5 \\
    \vdots \\
    \end{bmatrix} \tag{\ref{eq:Us+x-}} ,
\end{align}
and $\bb = [ \A_{-1} \: \bzero \: \bzero \: \cdots]^\top$.
Observe that $\U_{11}=\U_{22}$, $\U_{12}$ and $\U_{21}$ are block triangular Toeplitz matrices.

Let $\rho(\bS)$ denote the spectral radius of the matrix $\bS$ and $T_{\infty}[\bS(z)] = (\bS_{j-i})_{i,j \in \Nats}$ denote the semi-infinite block Toeplitz matrix associated with the matrix Laurent power series $\bS(z) = \sum_{i=-\infty}^{+\infty} z^i \bS_i$.
Define the even part $\bS_{\scriptscriptstyle{even}}(z)$ and the odd part $\bS_{\scriptscriptstyle{odd}}(z)$ of the matrix Laurent power series $\bS(z)$ as
\begin{align}
    \bS_{\scriptscriptstyle{even}}(z) & = \frac{1}{2} (\bS(\sqrt{z}) + \bS(-\sqrt{z})) = \sum_{i=-\infty}^{+\infty} z^i \bS_{2i} , \label{eq:LaurentEven} \\
    \bS_{\scriptscriptstyle{odd}}(z) & = \frac{1}{2\sqrt{z}} (\bS(\sqrt{z}) - \bS(-\sqrt{z})) = \sum_{i=-\infty}^{+\infty} z^i \bS_{2i+1} , \label{eq:LaurentOdd}
\end{align}
and thus $\bS(z) = \bS_{\scriptscriptstyle{even}}(z^2) + z \bS_{\scriptscriptstyle{odd}}(z^2)$.
Next, define the generating function $\varphi(z) := \sum_{i=-1}^\infty z^{i+1} \A_i$ associated with the M/G/$1$-type
process and, following~\eqref{eq:LaurentEven} and \eqref{eq:LaurentOdd}, define the related matrix power series
$$
\varphi_{\scriptscriptstyle{even}}(z) := \sum_{i=0}^\infty z^i \A_{2i}
\qquad\qquad \mbox{ and } \qquad\qquad
\varphi_{\scriptscriptstyle{odd}}(z) := \sum_{i=-1}^\infty z^{i+1} \A_{2i+1} .
$$
We now can express the matrices in~\eqref{eq:compact} as
\begin{equation}
    \U_{11} = \U_{22} = T_{\infty}[\varphi_{\scriptscriptstyle{odd}}(z)] , \quad \U_{12} = T_{\infty}[\varphi_{\scriptscriptstyle{even}}(z)] , \quad \U_{21} = T_{\infty}[z^{-1}\varphi_{\scriptscriptstyle{even}}(z)] .
\label{eq7.33}
\end{equation}
Since $\A_i \geq \bzero$, $i\in \Ints_{\geq -1}$, and $\varphi(1) = \sum_{i=-1}^\infty \A_i < \infty$, we conclude that the function $\varphi(z)$ belongs to the Wiener algebra $\cW_z$ of matrix power series in $z$ and thus $\varphi_{\scriptscriptstyle{even}}(z)$ and $\varphi_{\scriptscriptstyle{odd}}(z)$ also belong to 
$\cW_z$.
We then have the following result adapted from~\cite{BiLaMe05}.
\begin{theorem}[{\cite[Theorem~7.7]{BiLaMe05}}]
Assuming $M$ is finite and the transition probability matrix $\bP$ is irreducible and aperiodic, then $\I - \varphi_{\scriptscriptstyle{odd}}(z)$ is nonsingular for $|z| \leq 1$ and $\rho(\A_0) < 1$.
\label{thm7.7}
\end{theorem}

\begin{remark}
When $M$ is infinite, we simply need to replace the spectral radius $\rho(\A_0)$ in Theorem~\ref{thm7.7} with the spectrum related to the operator $\A_0$ appropriately defined~\cite{DunSch-all}.
\label{rem:thm7.7}
\end{remark}

We next perform
the first iteration of block Gaussian elimination on the $2\times 2$ block system in~\eqref{eq:compact}.
Since the matrix functions $\varphi_{\scriptscriptstyle{even}}(z)$,  $\varphi_{\scriptscriptstyle{odd}}(z)$ and $(\I - \varphi_{\scriptscriptstyle{odd}}(z))^{-1}$ belong to the Wiener algebra $\cW_z$, the semi-infinite matrices $\I-\U_{22}$, $\U_{21}$, $(\I - \U_{11})^{-1}$ and $\U_{12}$ define bounded operators, and therefore the one iteration of block Gaussian elimination yields
\begin{equation}
    \begin{bmatrix}
    \I - \U_{11} & -\U_{12} \\
    \bzero & \bH^{(1)}  \\
    \end{bmatrix}
    \begin{bmatrix}
    \x_{+} \\ \x_{-} \\
    \end{bmatrix}
    =
    \begin{bmatrix}
    \bzero \\ \bb \\
    \end{bmatrix} ,
\tag{\ref{eq7.34a}}
\end{equation}
where 
\begin{equation}
    \bH^{(1)} = \I -\U_{22} - \U_{21}(\I - \U_{11})^{-1} \U_{12} 
\tag{\ref{eq7.34b}}
\end{equation}
is the Schur complement of $\I -\U_{22}$, and superscript $(1)$ indicates the first iteration of block Gaussian elimination.
Moreover, given that $(\I - \varphi_{\scriptscriptstyle{odd}}(z))^{-1}$ belongs to the Wiener algebra $\cW_z$, we have that $(\I - \U_{11})^{-1}$ and $(\I - \U_{11})^{-1} \U_{12}$ are both block upper triangular Toeplitz.
We further have $\U_{21} (\I - \U_{11})^{-1} \U_{12}$ to be a block Hessenberg matrix, and thus $\bH^{(1)}$ is block Hessenberg as well as block Toeplitz except for the first block row, 
from which we obtain
\begin{equation}
    \bH^{(1)} \begin{bmatrix} \G_{\min} \\ \G_{\min}^3 \\ \G_{\min}^5 \\ \G_{\min}^7 \\ \vdots \end{bmatrix}
    = 
    \begin{bmatrix}  \A_{-1} \\ \bzero \\ \bzero \\ \vdots \end{bmatrix} , \qquad\qquad
    \bH^{(1)} = \begin{bmatrix}
    \I - \hat{\A}_0^{(1)} & -\hat{\A}_1^{(1)} & -\hat{\A}_2^{(1)} & -\hat{\A}_3^{(1)} & \cdots \\
    -\A_{-1}^{(1)} & \I - \A_0^{(1)} & -\A_1^{(1)} & -\A_2^{(1)} & \ddots \\
    \bzero & -\A_{-1}^{(1)} & \I - \A_0^{(1)} & -\A_1^{(1)} & \ddots \\
    \bzero & \bzero & -\A_{-1}^{(1)} & \I - \A_0^{(1)} & \ddots \\
    \vdots & \vdots & \ddots & \ddots & \ddots \\
    \end{bmatrix} .
\tag{\ref{eq7.35}}
\end{equation}

In addition to the matrix power series $\varphi(z)$, $\varphi_{\scriptscriptstyle{even}}(z)$ and $\varphi_{\scriptscriptstyle{odd}}(z)$ above, define the matrix power series
$$
\varphi^{(1)}(z) := \sum_{i=-1}^\infty z^{i+1} \A_i^{(1)} 
\qquad\qquad \mbox{ and } \qquad\qquad
\hat{\varphi}^{(1)}(z) := \sum_{i=0}^\infty z^i \hat{\A}_i^{(1)}
$$
which completely define in functional form the block Hessenberg matrix $\bH^{(1)}$.
Further define the functions
$$
\hat{\varphi}(z) := \sum_{i=0}^\infty z^i \A_i ,
\qquad
\hat{\varphi}_{\scriptscriptstyle{even}}(z) := \sum_{i=0}^\infty z^i \A_{2i} ,
\qquad \hat{\varphi}_{\scriptscriptstyle{odd}}(z) := \sum_{i=0}^\infty z^i \A_{2i+1} ,
$$
once again following~\eqref{eq:LaurentEven} and \eqref{eq:LaurentOdd}.
Then, from~\eqref{eq7.33}, \eqref{eq7.34a},
\eqref{eq7.34b}
and
the
equivalence between block triangular Toeplitz matrices and matrix power series, we have
the block matrices $\A_i^{(1)}$ and $\hat{\A}_{i+1}^{(1)}$, $i\in \Ints_{\geq -1}$, formally defined in functional form by means of the recursions
\begin{align}
    \varphi^{(1)}(z) & = z \varphi_{\scriptscriptstyle{odd}}(z) + \varphi_{\scriptscriptstyle{even}}(z)(\I - \varphi_{\scriptscriptstyle{odd}}(z))^{-1} \varphi_{\scriptscriptstyle{even}}(z) ,
    \label{eq7.36a} \\
    \hat{\varphi}^{(1)}(z) & = \hat{\varphi}_{\scriptscriptstyle{odd}}(z) + \varphi_{\scriptscriptstyle{even}}(z) (\I - \varphi_{\scriptscriptstyle{odd}}(z))^{-1} \hat{\varphi}_{\scriptscriptstyle{even}}(z) ,
    \label{eq7.36b}
\end{align}
where $\varphi^{(1)}(z)$ and $\hat{\varphi}^{(1)}(z)$ belong to the Wiener algebra $\cW_z$.

Since this transformation \mss{rendering~\eqref{eq7.35}} maintains the same structure of the 
\mss{original} system
\mss{in~\eqref{lem4.2:4.5} and~\eqref{lem4.2:4.6},}
we then establish by
induction
that the above process of applying a block even-odd permutation and performing one iteration of block Gaussian elimination can be repeated recursively.
Specifically, after $n$ iterations of this cyclic reduction process, we derive in general the system
\begin{equation}
    \bH^{(n)} \begin{bmatrix} \G_{\min} \\ \G_{\min}^{2^n+1} \\ \G_{\min}^{2\cdot 2^n+1} \\ \G_{\min}^{3\cdot 2^n+1} \\ \vdots \end{bmatrix}
    = \begin{bmatrix}  \A_{-1} \\ \bzero \\ \bzero \\ \vdots \end{bmatrix} , \qquad\qquad
    \bH^{(n)} = \begin{bmatrix}
    \I - \hat{\A}_0^{(n)} & -\hat{\A}_1^{(n)} & -\hat{\A}_2^{(n)} & -\hat{\A}_3^{(n)} & \cdots \\
    -\A_{-1}^{(n)} & \I - \A_0^{(n)} & -\A_1^{(n)} & -\A_2^{(n)} & \ddots \\
    \bzero & -\A_{-1}^{(n)} & \I - \A_0^{(n)} & -\A_1^{(n)} & \ddots \\
    \bzero & \bzero & -\A_{-1}^{(n)} & \I - \A_0^{(n)} & \ddots \\
    \vdots & \vdots & \ddots & \ddots & \ddots \\
    \end{bmatrix} ,
\tag{\ref{eq7.37}}
\end{equation}
where the block matrices $\A_i^{(n)}$ and $\hat{\A}_{i+1}^{(n)}$, $i\in \Ints_{\geq -1}$, are formally defined in functional form by means of the recursions
\begin{align}
    \varphi^{(n+1)}(z) & = z \varphi_{\scriptscriptstyle{odd}}^{(n)}(z) + \varphi_{\scriptscriptstyle{even}}^{(n)}(z)(\I - \varphi_{\scriptscriptstyle{odd}}^{(n)}(z))^{-1} \varphi_{\scriptscriptstyle{even}}^{(n)}(z) ,
    \tag{\ref{eq7.38a}} \\
    \hat{\varphi}^{(n+1)}(z) & = \hat{\varphi}_{\scriptscriptstyle{odd}}^{(n)}(z) + \varphi_{\scriptscriptstyle{even}}^{(n)}(z) (\I - \varphi_{\scriptscriptstyle{odd}}^{(n)}(z))^{-1} \hat{\varphi}_{\scriptscriptstyle{even}}^{(n)}(z) ,
    \tag{\ref{eq7.38b}}
\end{align}
together with
\begin{align}
    \varphi^{(n)}(z) := \sum_{i=-1}^\infty z^{i+1} \A_i^{(n)} ,
    \qquad & \qquad
    \varphi_{\scriptscriptstyle{even}}^{(n)}(z) := \sum_{i=0}^\infty z^i \A_{2i}^{(n)} ,
    \; & \;
    \varphi_{\scriptscriptstyle{odd}}^{(n)}(z) := \sum_{i=-1}^\infty z^{i+1} \A_{2i+1}^{(n)} ,
    \tag{\ref{eq-varphi}} \\
    \hat{\varphi}^{(n)}(z) := \sum_{i=0}^\infty z^i \hat{\A}_i^{(n)} ,
    \qquad & \qquad
    \hat{\varphi}_{\scriptscriptstyle{even}}^{(n)}(z) := \sum_{i=0}^\infty z^i \hat{\A}_{2i}^{(n)} ,
    \; & \; \hat{\varphi}_{\scriptscriptstyle{odd}}^{(n)}(z) := \sum_{i=0}^\infty z^i \hat{\A}_{2i+1}^{(n)} 
    \tag{\ref{eq-hatvarphi}} ,
\end{align}
following~\eqref{eq:LaurentEven} and \eqref{eq:LaurentOdd}.
This functional form supports the efficient computations reflected in the classical  cyclic-reduction algorithm.
It also follows from~\eqref{eq7.38a} and~\eqref{eq7.38b} that the applicability of cyclic reduction at each iteration~$n$ relies on the invertibility of the matrix power series $\I - \varphi_{\scriptscriptstyle{odd}}^{(n)}(z)$ for $|z| \leq 1$, as established in the next result adapted from~\cite{BiLaMe05}.
\begin{theorem}[{\cite[Theorem~7.8]{BiLaMe05}}]
Assuming $M$ is finite and the transition probability matrix $\bP$ is irreducible and aperiodic, then for any $n\in\Ints_+$ we have:
\begin{enumerate}
\item The functions $\varphi^{(n)}(z)$ and $\hat{\varphi}^{(n)}(z)$ belong to the Wiener algebra $\cW_z$;
\item The matrices $\A_i^{(n)}$, $\hat{\A}_{i+1}^{(n)}$, $i\in \Ints_{\geq -1}$, are nonnegative, and the power series $\varphi^{(n)}(1)$,
$\A_{-1} + \hat{\varphi}^{(n)}(1)$ are stochastic;
\item The function $\I - \varphi_{\scriptscriptstyle{odd}}^{(n)}(z)$ is invertible for $|z| \leq 1$, its inverse belongs to the Wiener algebra $\cW_z$, thus ensuring the ability to apply cyclic reduction, and $\rho(\varphi_{\scriptscriptstyle{odd}}^{(n)}(1)) < 1$;
\item The matrices $\varphi_{\scriptscriptstyle{even}}^{(n)}(1) (\I - \varphi_{\scriptscriptstyle{odd}}^{(n)}(1))^{-1}$ and $(\I - \varphi_{\scriptscriptstyle{odd}}^{(n)}(1))^{-1} \varphi_{\scriptscriptstyle{even}}^{(n)}(1)$ have spectral radius $1$, with the latter being stochastic.
\end{enumerate}
\label{thm7.8}
\end{theorem}
\begin{remark}
When $M$ is infinite, we simply need to replace the spectral radi in Theorem~\ref{thm7.8} (as well as in Theorem~\ref{thm7.13} below) with the spectrum related to the corresponding operator appropriately defined~\cite{DunSch-all}.
\label{rem:thm7.8}
\end{remark}

Now consider the computation of the block Hessenberg matrix $\bH^{(n)}$ for each iteration~$n$ of cyclic reduction given in~\eqref{eq7.37}.
From the first block matrix equation of~\eqref{eq7.37}, we have
\begin{equation}
    \G_{\min} \; = \; (\I - \hat{\A}_0^{(n)})^{-1} (\A_{-1} + \sum_{i=1}^\infty \hat{\A}_i^{(n)} \G_{\min}^{i\cdot 2^n+1}) ,
\tag{\ref{eq7.39}}
\end{equation}
which provides the means to compute the minimal nonnegative solution $\G_{\min}$ up to a desired level of accuracy.
It can be shown that the summation $\sum_{i=1}^\infty \hat{\A}_i^{(n)} \G_{\min}^{i\cdot 2^n+1}$ asymptotically converges quadratically to zero as $n \rightarrow \infty$ and that $(\I - \hat{\A}_0^{(n)})^{-1}$ is bounded and quadratically convergent, both of which
follow from the next set of convergence results adapted from~\cite{BiLaMe05}.
\begin{theorem}[{\cite[Theorem~7.13]{BiLaMe05}}]
Let $\varphi(z)$ be the generating function associated with an ergodic M/G/$1$-type Markov process.
Assume that $\varphi(z)$ is analytic for $|z| < r$, with $1 < r$, and that there exists $\zeta$ 
such that 
$1 < |\zeta| < r$ and $\det(\zeta\I - \varphi(\zeta)) = 0$.
Denote $\chi(z) = z\I - \varphi(z)$, $\eta = \max\{|z| : |z| < 1, \det \chi(z) = 0\}$, and $\xi = \min\{|z| : |z| > 1, \det \chi(z) = 0\}$.
Let $\epsilon$ be a positive number
such that 
$\eta + \epsilon < 1$ and $\xi - \epsilon > 1$, and let $\| \cdot \|$ be any fixed matrix norm.
Then the matrix power series $\varphi^{(n)}(z) = \sum_{i=-1}^\infty z^{i+1} \A_i^{(n)}$ and $\hat{\varphi}^{(n)}(z) = \sum_{i=0}^\infty z^i \hat{\A}_i^{(n)}$, $n \in\Ints_+$, generated by the recurrences~\eqref{eq7.38a} and~\eqref{eq7.38b} satisfy the following properties.
\begin{enumerate}
    \item There exists a positive $\gamma$
    such that,
    for any $n \in \Ints_+$,
    \begin{equation}
        \|\A_i^{(n)}\| \leq \gamma \xi^{2^n} (\xi - \epsilon)^{-(i+1){2^n}} \quad \mbox{ and } \quad
        \|\hat{\A}_i^{(n)}\| \leq \gamma \xi^{2^n} (\xi - \epsilon)^{-(i+1){2^n}} , \qquad i \in \Nats.
    \label{eq7.49}
    \end{equation}
    \item There exist $\A_0^{(\infty)} = \lim_{n \rightarrow \infty} \A_0^{(n)}$, $\hat{\A}_0^{(\infty)} = \lim_{n \rightarrow \infty} \hat{\A}_0^{(n)}$ and $\A_{-1}^{(\infty)} = \lim_{n \rightarrow \infty} \A_{-1}^{(n)}$ where $\A_{-1}^{(\infty)} = (\I-\A_0^{(\infty)}) \ones \bg^\top$, $\hat{\A}_0^{(\infty)} = \sum_{i=0}^\infty \A_i \G_{\min}^i$, and $\bg \geq \bzero$ is
    such that 
    $\bg^\top \G_{\min} = \bg^\top$ and $\bg^\top\ones = 1$; moreover,
    \begin{equation*}
        \| \hat{\A}_0^{(n)} - \hat{\A}_0^{(\infty)} \| \leq \gamma \xi^{2^n} (\xi - \epsilon)^{-{2^{n+1}}} , \qquad n \in \Ints_+.
    \end{equation*}
    \item The sequence $\{ \chi^{(n)}(z) \}_n$, where $\chi^{(n)}(z) = z\I - \varphi^{(n)}(z)$ for $n \in\Ints_+$, uniformly converges to $\chi^{(\infty)}(z) = -\A_{-1}^{(\infty)} - z(\A_0^{(\infty)} - \I)$ over any compact subset of the disk $\{ z \in \Complex : |z| < \xi \}$; moreover, $\rho(\hat{\A}_0^{(\infty)}) \leq \rho(\A_0^{(\infty)}) < 1$.
    \item For any $n \in \Ints_+$,
    \begin{equation*}
        \| \A_0^{(n)} - \A_0^{(\infty)} \| \leq \gamma \xi^{2^n} (\xi - \epsilon)^{-{2^{n+1}}} \quad \mbox{ and } \quad
        \| \A_{-1}^{(n)} - \A_{-1}^{(\infty)} \| \leq \gamma ( \xi^{2^n} (\xi - \epsilon)^{-{2^{n+1}}} + (\eta + \epsilon)^{2^n} ) .
    \end{equation*}
    \item For any $n \in \Ints_+$,
    \begin{equation*}
        \| \G_{\min} - \widetilde{\G}^{(n)} \| \leq \gamma \xi^{2^n} (\xi - \epsilon)^{-{2^{n+1}}} ,
    \end{equation*}
    where $\widetilde{\G}^{(n)} = (\I - \hat{\A}_0^{(n)})^{-1} \A_{-1}$.
\end{enumerate}
\label{thm7.13}
\end{theorem}

Let $\Pi$ denote the permutation matrix associated with the block even-odd permutation of the block Hessenberg matrix $\bH^{(n)}$ such that 
\begin{equation}
    \Pi \bH^{(n)} \Pi^\top \; = \;
    \begin{bmatrix}
    \I - \U_{11}^{(n)} & -\U_{12}^{(n)} \\
    - \U_{21}^{(n)} & \I -\U_{22}^{(n)}  \\
    \end{bmatrix} ,
\tag{\ref{eq7.54}}
\end{equation}
where
\begin{align}
    \U_{11}^{(n)}
    & 
    =
    \begin{bmatrix}
    \A_0^{(n)} & \A_2^{(n)} & \A_4^{(n)} & \cdots \\
    \bzero & \A_0^{(n)} & \A_2^{(n)} & \ddots \\
    \bzero & \bzero & \A_0^{(n)} & \ddots \\
    \vdots & \ddots & \ddots & \ddots \\
    \end{bmatrix},
    \qquad
    & 
    \U_{12}^{(n)} =
    \begin{bmatrix}
    \A_{-1}^{(n)} & \A_1^{(n)} & \A_3^{(n)} & \cdots \\
    \bzero & \A_{-1}^{(n)} & \A_1^{(n)} & \ddots \\
    \bzero & \bzero & \A_{-1}^{(n)} & \ddots \\
    \vdots & \ddots & \ddots & \ddots \\
    \end{bmatrix},
    \tag{\ref{eqU^na}}
    \\
    \U_{21}^{(n)} 
    & 
    =
    \begin{bmatrix}
    \hat{\A}_1^{(n)} & \hat{\A}_3^{(n)} & \hat{\A}_5^{(n)} & \cdots \\
    \A_{-1}^{(n)} & \A_1^{(n)} & \A_3^{(n)} & \ddots \\
    \bzero & \A_{-1}^{(n)} & \A_1^{(n)} & \ddots \\
    \vdots & \ddots & \ddots & \ddots \\
    \end{bmatrix},
    \qquad
    & 
    \U_{22}^{(n)} =
    \begin{bmatrix}
    \hat{\A}_0^{(n)} & \hat{\A}_2^{(n)} & \hat{\A}_4^{(n)} & \cdots \\
    \bzero & \A_0^{(n)} & \A_2^{(n)} & \ddots \\
    \bzero & \bzero & \A_0^{(n)} & \ddots \\
    \vdots & \ddots & \ddots & \ddots \\
    \end{bmatrix}.
    \tag{\ref{eqU^nb}}
\end{align}
The main computational task of cyclic reduction then consists of computing the first two block rows of the Schur complement matrix
\begin{equation}
    \bH^{(n+1)} = \I -\U_{22}^{(n)} - \U_{21}^{(n)}(\I - \U_{11}^{(n)})^{-1} \U_{12}^{(n)} .
\tag{\ref{eq7.55}}
\end{equation}
This involves computations of infinite block triangular Toeplitz matrices, which we approximate with infinite banded Toeplitz matrices defined by parameters $d$ and $q$ chosen to achieve the desired level of accuracy $\epsilon$, as discussed in the next section.

\subsection{Classical Cyclic Reduction Algorithm}
Based on
\mss{our}
above derivation,
we turn to the algorithmic details of the main computational task of~\eqref{eq7.55} at the heart of
{the} classical cyclic reduction {method}.
We start by exploiting infinite block triangular Toeplitz matrix inversion to compute $T_{\infty}[\K(z)] = (\I - \U_{11}^{(n)})^{-1}$ where 
$$
\K(z) = \sum_{i=0}^\infty z^i \K_i = (\I - \varphi_{\scriptscriptstyle{odd}}^{(n)}(z))^{-1} .
$$
From Theorem~\ref{thm7.8}, the functions $\varphi^{(n)}(z)$ and $\hat{\varphi}^{(n)}(z)$ belong to the Wiener algebra $\cW_z$ and $\I - \varphi_{\scriptscriptstyle{odd}}^{(n)}(z)$ is nonsingular, with its inverse $\K(z) = (\I - \varphi_{\scriptscriptstyle{odd}}^{(n)}(z))^{-1}$ also belonging to 
$\cW_z$.
Hence,
the coefficients of $\varphi^{(n)}(z)$, $\hat{\varphi}^{(n)}(z)$ and $\K(z)$ decay to zero, i.e., 
for any $\epsilon > 0$, 
there exist integers $s>0$ and $q>0$
such that 
$$
\sum_{i=s}^\infty \A_i^{(n)} < \epsilon \bO ,
\qquad 
\sum_{i=s+1}^\infty \hat{\A}_i^{(n)} < \epsilon \bO ,
\qquad \sum_{i=q}^\infty \K_i < \epsilon \bO , 
$$
where 
$\K_0, \K_1, \ldots, \K_{q-1}$
are the first $q$ block matrices of the first block row of $(\I - \U_{11})^{-1}$, 
$\epsilon$ is the desired level of accuracy, and $\bO$ is the matrix whose elements are all one.
Moreover, it follows that the matrices $\bH^{(n)}$ can be replaced with block banded matrices defined by the matrix polynomials
$$
\zeta_s^{(n)}(z) := \sum_{i=-1}^{s-1} z^{i+1} \A_i^{(n)}
\qquad\qquad \mbox{ and } \qquad\qquad
\hat{\zeta}_s^{(n)}(z) := \sum_{i=0}^s z^i \hat{\A}_i^{(n)} ,
$$
where $s$ is chosen to satisfy the desired accuracy 
$\epsilon$.
Supposing
without loss of generality
$s+1$ is even, 
then the number of block coefficients of the matrix polynomials $\zeta_s^{(n)}(z)$ and $\hat{\zeta}_s^{(n)}(z)$ is given by the even number $d=s+1$.

From
this
analysis
and Theorem~\ref{thm7.8} 
together with the equivalence of block banded Toeplitz matrices and matrix polynomials, we can approximate $(\I - \U_{11}^{(n)})^{-1}$ with a block banded, block Toeplitz matrix $T_{\infty}[\K_q(z)]$ where $\K_q(z) = \K(z)\mod z^q$ is a matrix polynomial of degree at most $q-1$ with $q$ chosen
such that 
$$
\| T_{\infty}[\K_q(z)] - T_{\infty}[\K(z)] \|_{\infty} \leq \epsilon .
$$
By truncating $T_{\infty}[\K(z)]$ at the corresponding finite band $q$, we obtain the matrix $T_{\infty}[\K_q(z)]$ that coincides with $(\I - \U_{11}^{(n)})^{-1}$ in the nonzero part, thus yielding the first $q$ block matrix elements $\K_0, \K_1, \ldots, \K_{q-1}$ 
of the first block row of $(\I - \U_{11}^{(n)})^{-1}$
such that 
$\sum_{i=q}^\infty \K_i < \epsilon \bO$ for the remaining block matrix elements $\K_q, \K_{q+1}, \ldots$ of the first block row of $(\I - \U_{11}^{(n)})^{-1}$.

Noting
that the matrix $(\I - \U_{11}^{(n)})$ is block upper triangular Toeplitz,
we can compute its inverse using  $k$ block matrices of the first row as follows.
Set
the first 
$k$ blocks of $(\I - \U_{11}^{(n)})$ to
$$
\C_0 = (\I - \A_0)^\top , \;\;\;
\C_1 = -\A_2^\top , \;\;\;
\ldots , \;\;\;
\C_{k-1} = -\A_{d/2}^\top ,
$$
and also set $\y_1 = \C_0^{-1}$. 
Next
set $q=1$ 
and compute the Toeplitz matrix-vector products $\bw = \bS_q\y_q$ and $\bu = -\bL(\y_q)\bw$, where $\bS_q$ is the bottom left $q\times q$ block of the block Toeplitz matrix $(\I - \U_{11}^{(n)})^\top$ and $\bL(\y_q)$ is the block lower triangular Toeplitz matrix having first block column $\y_q$.
Lastly 
set $\y_{2q} = [\y_q; \bu]$ and $q = 2q$, and repeat these two operations on the bigger block.

To further elucidate
the basic idea of this iterative process,
\mss{we}
want to compute the inverse of a block lower triangular Toeplitz matrix $\T_{2q}$ of the form 
\begin{equation}\label{eq:sq}
 \T_{2q} = 
\begin{bmatrix}
    \T_q & 0\\
    \bS_q & \T_q
\end{bmatrix} ,
\end{equation}
where the first column $\y_{2q}$ of $\T_{2q}^{-1}$ is given by
\begin{equation}\label{y2q}
\y_{2q} = \T_{2q}^{-1}\be_1 = \begin{bmatrix}
    \y_q\\
    - \bL(\y_q) \bS_q\y_q
\end{bmatrix}
\end{equation}
for $\y_q = [\Y_1; \ldots; \Y_q]$, 
and where $\bL(\y_q) = \T_q^{-1}$ is given by
\begin{equation}\label{eq:Lq}
\bL(\y_q)  = 
\begin{bmatrix}
   \Y_1 & & & 0 \\
    \Y_2 & \Y_1 && \\
    \vdots & \ddots & \ddots & \\
    \Y_q & \cdots & \Y_2 & \Y_1
\end{bmatrix} .
\end{equation}
Then, letting
$\W = \left(\sum_{i=0}^{k-1}\C_i\right)^{-1}$, we
check the criterion of whether $\W - \sum_{i=0}^{q-1}\Y_i  \leq \epsilon \bO$ to stop the iterations and return $\K_i = \Y_i^\top, \, i\in \{0, \ldots, q-1\}$, as the first row of $(\I - \U_{11}^{(n)})^{-1}$.

Let $\hat{\bu}^\top = [\hat{\A}_1^{(n)}, \hat{\A}_3^{(n)}, \hat{\A}_5^{(n)}, \ldots]$ and $\bu^\top = [\A_{-1}^{(n)}, \A_1^{(n)}, \A_3^{(n)}, \ldots]$ respectively denote the first and second block rows of $\U_{21}^{(n)}$.
We then follow along similar lines to compute the first and second block rows of the matrix product $\U_{21}^{(n)}(\I - \U_{11}^{(n)})^{-1}$ as $\hat{\bv}^\top = \hat{\bu}^\top (\I - \U_{11}^{(n)})^{-1}$ and $\bv^\top = \bu^\top (\I - \U_{11}^{(n)})^{-1}$, respectively, in terms of the matrices 
$$
\A_{-1}^{(n)}, \; \A_1^{(n)}, \; \ldots, \; \A_{d-3}^{(n)}, \; \hat{\A}_1^{(n)}, \qquad \hat{\A}_3^{(n)}, \; \ldots, \; \hat{\A}_{d-1}^{(n)} , \qquad
\K_0, \K_1, \ldots, \K_{q-1} .
$$
Hence,
for
the parameter $d$ appropriately chosen above, we have that the nonnull elements of $\bv^\top = [\V_0, \V_1, \ldots]$ and $\hat{\bv}^\top = [\hat{\V}_0, \hat{\V}_1, \ldots]$ are given by the block coefficients of the matrix polynomials
\begin{equation}
    \sum_{i=0}^{d/2+q-2} z^i \V_i = \left( \sum_{i=0}^{d/2-1} z^i \A_{2i-1}^{(n)} \right) \left( \sum_{i=0}^{q-1} z^i \K_i \right) , \qquad
    \sum_{i=0}^{d/2+q-2} z^i \hat{\V}_i = \left( \sum_{i=0}^{d/2-1} z^i \hat{\A}_{2i+1}^{(n)} \right) \left( \sum_{i=0}^{q-1} z^i \K_i \right) ,
\label{eq7.64}
\end{equation}
respectively.
Similarly, we compute the first and second block rows of the matrix product $\U_{21}^{(n)}(\I - \U_{11}^{(n)})^{-1} \U_{12}^{(n)}$ as $\hat{\y}^\top = \hat{\bv}^\top \U_{12}^{(n)}$ and $\y^\top = \bv^\top \U_{12}^{(n)}$, respectively, in terms of the matrices
$$
\V_0, \; \V_1, \; \ldots, \; \V_{d/2+q-2} , \qquad
\hat{\V}_0, \; \hat{\V}_1, \; \ldots, \; \hat{\V}_{d/2+q-2} , \qquad
\A_{-1}^{(n)}, \; \A_1^{(n)}, \; \ldots, \; \A_{d-3}^{(n)} .
$$
In particular, the nonnull elements of $\y^\top = [\Y_0, \Y_1, \ldots]$ and $\hat{\y}^\top = [\hat{\Y}_0, \hat{\Y}_1, \ldots]$ are given by the block coefficients of the matrix polynomials
\begin{equation}
    \sum_{i=0}^{d+q-3} z^i \Y_i = \left( \sum_{i=0}^{d/2+q-2} z^i \V_i \right) \left( \sum_{i=0}^{d/2-1} z^i \A_{2i-1}^{(n)} \right) , \qquad
    \sum_{i=0}^{d+q-3} z^i \hat{\Y}_i = \left( \sum_{i=0}^{d/2+q-2} z^i \hat{\V}_i \right) \left( \sum_{i=0}^{d/2-1} z^i \A_{2i-1}^{(n)} \right) ,
\label{eq7.65}
\end{equation}
respectively.

The last element of the main computational task of~\eqref{eq7.55} consists of computing the first and second block rows of the matrix equation
$$
(\I - \U_{22}^{(n)}) - \U_{21}^{(n)}(\I - \U_{11}^{(n)})^{-1} \U_{12}^{(n)}
$$
as $\hat{\bz}^\top = (\I - \U_{22}^{(n)}) - \hat{\y}$ and $\bz^\top = (\I - \U_{22}^{(n)}) - \y$, respectively, in terms of the matrices
$$
\Y_0, \; \Y_1, \; \ldots, \; \Y_{d+q-3} , \qquad
\hat{\Y}_0, \; \hat{\Y}_1, \; \ldots, \; \hat{\Y}_{d+q-3} , \qquad
\A_{-1}^{(n)}, \; \A_1^{(n)}, \; \ldots, \; \A_{d-3}^{(n)} .
$$
Letting $d^\prime = \max(d/2,q)$, the nonnull elements of $\bz^\top = [\Z_0, \Z_1, \ldots]$ and $\hat{\bz}^\top = [\hat{\Z}_0, \hat{\Z}_1, \ldots]$ are given by 
\begin{equation}
    \Z_i = \A_{2i}^{(n)} + \Y_{i+1}
    \qquad\quad \mbox{ and } \qquad\quad 
    \hat{\Z}_j = \hat{\A}_{2j}^{(n)} + \hat{\Y}_j ,
\label{eq:Z}
\end{equation}
for $i \in \{ -1, \ldots, d^\prime - 1\}$ and 
$j \in \{ 0, \ldots, d^\prime - 1\}$, 
where $\A_{i}^{(n)} = 0$ if $i < -1$ or if $i > d-2$ and $\hat{\A}_{j}^{(n)} = 0$ if $j > d-1$.


Finally, the algorithm terminates once it is determined that a sufficient number of cyclic reduction iterations have been performed. 
In particular, the cyclic reduction algorithm checks whether the matrix $\A_{-1} + \hat{\varphi}^{(n)}(1)$ is close to being  stochastic (row sums equal one), as indicated by the second point of Theorem~\ref{thm7.8}. Then, from the first point of Theorem~\ref{thm7.13}, we have that $\|\hat{\A}_i^{(n)}\| \leq \gamma \xi^{2^n} (\xi - \epsilon)^{-(i+1){2^n}}$ for $i \in \Nats$, which implies that the row sums of $\A_{-1} + \hat{\A}_0^{(n)}$ should converge to one as the number of cyclic reduction iterations increases. Hence, a check of the condition $\| \ones - (\A_{-1}+\hat{\A}_0^{(n+1)})\bm{1}\|_{\infty} > \epsilon$ is made to determine whether the algorithm has converged.

The above main \mss{algorithmic} steps of classical cyclic reduction are presented in Algorithm~\ref{algo:classical} where, after the initialization in \emph{Step}~$\mathbf{1}$ of Algorithm~\ref{algo:classical}, 
Algorithm~\ref{alg7.5} is called in \emph{Step}~$\mathbf{2}$ of Algorithm~\ref{algo:classical} 
to handle each cyclic reduction iteration~$n$.
In particular, \emph{Step}~$\mathbf{1}$ of Algorithm~\ref{alg7.5} calls Algorithm~\ref{alg3.1} to compute the first block row of $(\I - \U_{11}^{(n)})^{-1}$ in terms of the $q$ block matrices $\K_0, \ldots, \K_{q-1}$ following \eqref{eq:sq}~--~\eqref{eq:Lq}.
\emph{Step}~$\mathbf{2}$ of Algorithm~\ref{alg7.5} then calls Algorithm~\ref{alg2.1} to compute the first and second block rows of $\U_{21}^{(n)} (\I - \U_{11}^{(n)})^{-1}$ in terms of the block vectors $\bv$ and $\hat{\bv}$ following~\eqref{eq7.64}.
Similarly, \emph{Step}~$\mathbf{3}$ of Algorithm~\ref{alg7.5} calls Algorithm~\ref{alg2.1} to compute the first and second block rows of $\U_{21}^{(n)} (\I - \U_{11}^{(n)})^{-1} \U_{12}^{(n)}$ in terms of the block vectors $\y$ and $\hat{\y}$ following~\eqref{eq7.65}.
\emph{Step}~$\mathbf{4}$ of Algorithm~\ref{alg7.5} then computes the first and second block rows of $(\I - \U_{22}^{(n)}) - \U_{21}^{(n)} (\I - \U_{11}^{(n)})^{-1} \U_{12}^{(n)}$ in terms of the block vectors $\bz$ and $\hat{\bz}$ following~\eqref{eq:Z}.
The computational complexity of each inner loop call to Algorithm~\ref{alg7.5} of the overall Algorithm~\ref{algo:classical} is easily shown to be $O(M^3 d^\prime + M^2 d^\prime \log d^\prime)$~\cite{BiLaMe05}.
Termination of this inner loop is handled in \emph{Step}~$\mathbf{3}$ of Algorithm~\ref{algo:classical} as described above according to Theorem~\ref{thm7.8}, followed by the computation in \emph{Step}~$\mathbf{4}$ of the approximation $\J$ to the matrix $\G_{\min}$ according to Theorem~\ref{thm7.13}.
The overall computational complexity of Algorithm~\ref{algo:classical} is then easily shown to be $O(M^3 d_{\max} + M^2 d_{\max} \log d_{\max})$~\cite{BiLaMe05}, where $d_{\max}$ is the maximum numerical degree of the matrix power series $\varphi^{(n)}(z)$ and $\hat{\varphi}^{(n)}(z)$ generated by classical cyclic reduction.

\begin{algorithm}[thb!]
\renewcommand\thealgorithm{C.\arabic{algorithm}}
\caption{Classical Cyclic Reduction Algorithm for Ergodic M/G/$1$-type Markov Processes}\label{algo:classical}
   \begin{algorithmic}
 \State {\bf Input:} Positive even integer $d$, $M\times M$ block matrices $\A_i$, $i\in \{-1, 0, 1, \ldots, d-1\}$, defining the block Toeplitz, block Hessenberg matrix $\bH$ in~\eqref{lem4.2:4.6}, and error tolerance $\epsilon > 0$.
  \State {\bf Output:} An approximation $\J$ to the matrix $\G_{\min}$\\

  \State{\indent \textit{/* Initialization */}}
  \State {\bf 1.} Set $n=0$, 
  $\A_{-1}^{(0)} = \A_{-1}$, $\A_{0}^{(0)} = \A_{0}$, $\ldots$, $\A_{d-1}^{(0)} = \A_{d-1}$,
  $\widehat{\A}_0^{(0)} = \I-\A_0$,
  $\widehat{\A}_1^{(0)} = -\A_1$, $\ldots$, $\widehat{\A}_{d-1}^{(0)} = -\A_{d-1}$\\

  \State{\indent \textit{/* Compute $\bH^{(n+1)} = (\I-\U_{22}^{(n)}) - \U_{21}^{(n)}(\I - \U_{11}^{(n)})^{-1} \U_{12}^{(n)}$ 
  in~\eqref{eq7.55} */ }}
  \State {\bf 2.} Call Algorithm~\ref{alg7.5}
  \begin{itemize}
      \item[ {\bf 2.a}] Input: integer $d$, $\A_{-1}^{(n)}, \A_{0}^{(n)}, \ldots, \A_{d-2}^{(n)}$,  $\widehat{\A}_{0}^{(n)}, \widehat{\A}_{1}^{(n)}, \ldots \widehat{\A}_{d-1}^{(n)}$, error tolerance $\epsilon > 0$
      \item[ {\bf 2.b}] Output: integers $q$ and $d'$, $\Z_{-1}, \Z_{0}, \ldots, \Z_{d'-2}$, $\widehat{\Z}_{0}, \widehat{\Z}_{1}, \ldots, \widehat{\Z}_{d'-1}$
  \end{itemize} \\

   \State{\indent \textit{/* Test for desired level of accuracy $\epsilon$ */}}
    \State {\bf 3.} If $\| \ones - (\A_{-1}+\widehat{\Z}_0)\bm{1}\|_{\infty} > \epsilon$, set $n=n+1$, set $\A_i^{(n)} = \Z_i$, $i\in \{-1,\ldots,d'-2\}$, set $\widehat{\A}_i^{(n)} = \widehat{\Z}_i$, $i\in \{0,\ldots,d'-1\}$,
    and repeat Step~$\mathbf{2}$.\label{step6}\\

     \State{\indent \textit{/* Compute and return approximation $\J$ to the matrix $\G_{\min}$ from Theorem~\ref{thm7.13} */}}
    \State {\bf 4.} \emph{Output:} $\J = (\I - \widehat{\Z}_0)^{-1}\A_{-1}$.\\
   \end{algorithmic}
\end{algorithm}

\begin{algorithm}[thb!]
\renewcommand\thealgorithm{C.\arabic{algorithm}}
\caption{Single Iteration $n$ of Classical Cyclic Reduction Algorithm for Ergodic M/G/$1$-type Markov Processes}\label{alg7.5}
   \begin{algorithmic}
 \State {\bf Input:} Positive even integer $d$, $M\times M$ block matrices $\A_i^{(n)}$ and $\widehat{\A}_{i+1}^{(n)}$, $i\in \{-1, 0, \ldots, d-2\}$, defining the first two block rows of the matrix $\bH^{(n)}$ in~\eqref{eq7.37} at cyclic reduction iteration $n$, and error tolerance $\epsilon > 0$.
  \State {\bf Output:} Positive integers $q$ and $d'$, and $M\times M$ block matrices $\Z_i$ and $\widehat{\Z}_{i+1}$ approximating the block matrices $\A_i^{(n+1)}$ and $\widehat{\A}_{i+1}^{(n+1)}$, $i\in \{-1, 0, \ldots, d'-2\}$, in the first two block rows of the matrix $\bH^{(n+1)}$ in~\eqref{eq7.37} after the single iteration $n$ of cyclic reduction
  such that 
  $\| \ones - \sum_{i=-1}^{d'-2} \Z_i\ones \|_{\infty} \leq \epsilon$, $\| (\sum_{i=-1}^{d'-2} \widehat{\Z}_{i+1} - \sum_{i=-1}^{d'-2} \widehat{\A}_{i+1}^{(n+1)})\ones \|_{\infty} \leq \epsilon$ with $\Z_i = \A_i^{(n+1)}$ and $\widehat{\Z}_{i+1} = \widehat{\A}_{i+1}^{(n+1)}$, $i=-1, 0, \ldots, q-2$, $\| \varphi^{(n+1)}(z) - \sum_{i=-1}^{d'-2} z^{i+1} \Z_i \|_{\ast} \leq \epsilon$, $\| \widehat{\varphi}^{(n+1)}(z) - \sum_{i=-1}^{d'-2} z^{i+1} \widehat{\Z}_{i+1} \|_{\ast} \leq \epsilon$. \\

  \State{\indent \textit{/* Compute $(\I - \U_{11}^{(n)})^{-1}$ in~\eqref{eq7.55}
  with respect to 
  $T_{\infty}[\K_q(z)]$ and $\K_0, \K_1, \ldots, \K_{q-1}$ */}}
  \State {\bf 1.} Call Algorithm~\ref{alg3.1}
  \begin{itemize}
      \item[ {\bf 1.a}] Input: integer $d/2$, matrices $\I-\A_0^{(n)}, -\A_2^{(n)}, \ldots, -\A_{d-2}^{(n)}$, and error bound $\epsilon$
      \item[ {\bf 1.b}] Output: integer $q$, and matrices $\K_0, \K_1, \ldots, \K_{q-1}$
  \end{itemize} \\

 \State{\indent \textit{/* Compute $\U_{21}^{(n)}(\I - \U_{11}^{(n)})^{-1}$ in~\eqref{eq7.55}
 with respect to 
 $\A_i^{(n)}$, 
 $\widehat{\A}_j^{(n)}$, 
 $\K_l$
 in~\eqref{eq7.64} */}}
  \State {\bf 2.} Call Algorithm~\ref{alg2.1}
  \begin{itemize}
      \item[ {\bf 2.a}] Input: integers $d$ and $q$, and matrices $\A_{-1}^{(n)}, \A_1^{(n)}, \ldots, \A_{d-3}^{(n)}$, $\widehat{\A}_1^{(n)}, \widehat{\A}_3^{(n)}, \ldots, \widehat{\A}_{d-1}^{(n)}$,
      and $\K_0, \K_1, \ldots, \K_{q-1}$
      \item[ {\bf 2.b}] Output: integer $d_q$, and block vectors $\bv$ and $\widehat{\bv}$ of~\eqref{eq7.64}
  \end{itemize} \\
  
  \State{\indent \textit{/* Compute $\U_{21}^{(n)}(\I - \U_{11}^{(n)})^{-1} \U_{12}^{(n)}$ 
  in~\eqref{eq7.55}
  with respect to
  $\V_i$, 
  $\widehat{\V}_i$,
  $\A_j^{(n)}$
  in~\eqref{eq7.65} */}}
  \State {\bf 3.} Call Algorithm~\ref{alg2.1}
  \begin{itemize}
      \item[ {\bf 3.a}] Input: integers $d$ and $d_q$, and matrices $\A_{-1}^{(n)}, \A_1^{(n)}, \ldots, \A_{d-3}^{(n)}$, $\V_0, \V_1, \ldots, \V_{d_q}$, and $\widehat{\V}_0, \widehat{\V}_1, \ldots, \widehat{\V}_{d_q}$
      \item[ {\bf 3.b}] Output: block vectors $\y$ and $\widehat{\y}$ of~\eqref{eq7.65}
  \end{itemize} \\

  \State{\indent \textit{/* Compute $(\I-\U_{22}^{(n)}) - \U_{21}^{(n)}(\I - \U_{11}^{(n)})^{-1} \U_{12}^{(n)}$ 
  in~\eqref{eq7.55} 
  with respect to
  $\Y_i$,
  $\widehat{\Y}_i$,
  $\A_j^{(n)}$,
  $\widehat{\A}_l^{(n)}$
  in~\eqref{eq:Z}
  */ }}
   \State {\bf 4.} Set $d' = \max(d/2,q)$. From~\eqref{eq:Z}, compute $\Z_i = \A_{2i}^{(n)} + \Y_{i+1}$, $i\in \{-1, \ldots, d^\prime - 1\}$,
   and $\widehat{\Z}_i = \widehat{\A}_{2i}^{(n)} + \widehat{\Y}_i$, $i\in \{0, \ldots, d^\prime - 1\}$, where $\A_{i}^{(n)} = 0$ if $i < -1$ or if $i > d-2$ and $\widehat{\A}_{i}^{(n)} = 0$ if $i > d-1$ .\\

   \end{algorithmic}
\end{algorithm}

\begin{algorithm}[thb!]
\renewcommand\thealgorithm{C.\arabic{algorithm}}
\caption{Classical Infinite Block Triangular Toeplitz Matrix Inversion} \label{alg3.1}
   \begin{algorithmic}
 \State {\bf Input:} Positive integer $d/2$, block matrices $\I-\A_0^{(n)}, -\A_2^{(n)}, \ldots, -\A_{d-2}^{(n)}$, and error bound $\epsilon$ 
  \State {\bf Output:} Positive integer $q$, and block matrices $\K_0, \K_1, \ldots, \K_{q-1}$\\
  
\State {\bf 1.} Set $\C_0 = (\I - \A_0^{(n)})^\top$, $\C_1 = -\A_2^{(n)\top}$, $\ldots$, $\C_{k-1} = -\A_{d/2}^{(n)\top}$.
  \State {\bf 2.} Compute $\W = \left(\sum_{i=0}^{k-1}\C_i\right)^{-1}$.
  \State {\bf 3.}  Set $q=1, \Y_0 = \y_1 = \C_0^{-1}$.
  \State {\bf 4.} With $\y_q = (\Y_i)_{\in \{0,\ldots,q-1\}}$, compute the Toeplitz matrix-vector products $\bw = \bS_q\y_q$, $\bu = -\bL(\y_q)\bw$, where $\bS_q$ is the bottom left $q\times q$ block of the block Toeplitz matrix as shown in~\eqref{eq:sq} and $\bL(\y_q)$ is the block lower triangular Toeplitz matrix having first block column $\y_q$ as shown in~\eqref{eq:Lq}.
   \State {\bf 5.} Set $\y_{2q} = [\y_q; \bu]$ and $q = 2q$.
    \State {\bf 6.} If $\W - \sum_{i=0}^{q-1}\Y_i > \epsilon \bO$, then repeat Steps~$\mathbf{4}$ and $\mathbf{5}$; else continue with $q$ and $\K_i = \Y_i^\top, i\in \{0,\ldots,q-1\}$, where $\K_i$ are the first $q$ block elements of the first block row of $(\I - \U_{1,1})^{-1}$.
   \end{algorithmic}
\end{algorithm}

\begin{algorithm}[thb!]
\renewcommand\thealgorithm{C.\arabic{algorithm}}
\caption{Classical Toeplitz Matrix Product} \label{alg2.1}
   \begin{algorithmic}
 \State {\bf Input:} Positive integers $d$ and $q$, and block matrices $\A_{-1}^{(n)}, \A_1^{(n)}, \ldots, \A_{d-3}^{(n)}$, $\widehat{\A}_1^{(n)}, \widehat{\A}_3^{(n)}, \ldots, \widehat{\A}_{d-1}^{(n)}$, and $\K_0, \K_1, \ldots, \K_{q-1}$
  \State {\bf Output:} Positive integer $d_q$, and Block vectors $\bv$ and $\widehat{\bv}$ of~\eqref{eq7.64} (or Block vectors $\y$ and $\widehat{\y}$ of~\eqref{eq7.65})\\
  
  \State {\bf 1.} Compute the block vectors $\bv$ and $\widehat{\bv}$ using the following steps:
  
      \State {\bf 2.} Compute the minimum integer power of two $N > d/2 +q$.
      \State {\bf 3.} Let $\ba$ be the $N$-dimensional block column vector with elements 
    $\A_{-1}^{(n)}, \A_1^{(n)}, \ldots, \A_{d-3}^{(n)}$, and $\bk$ the $N$-dimensional block column vector with elements $\K_0, \K_1, \ldots, \K_{q-1}$, and null components elsewhere.
      \State {\bf 4.} Compute $\widetilde{\ba} = \text{IDFT}_N(\ba)$ and $\widetilde{\bk} = \text{IDFT}_N(\bk)$.
     \State {\bf 5.} Compute $\widetilde{\f} = \widetilde{\ba} * \widetilde{\bk}$ and $\f = (\F_i)_{i=1,N} = \text{DFT}_N(\widetilde{\f})$.
     \State {\bf 6.} Set $\V_i = \F_{i+1}$, $i\in \{ 0, \ldots, d_q\}, \, d_q := d/2+q-2$, and $\bv = [\V_0,\ldots,\V_{d_q}]$.
      \State {\bf 7.} Compute $\widehat{\bv}$ in similar manner using $\widehat{\A}_1^{(n)}, \widehat{\A}_3^{(n)}, \ldots, \widehat{\A}_{d-1}^{(n)}$ and $\K_0, \K_1, \ldots, \K_{q-1}$.


   \end{algorithmic}
\end{algorithm}

\section{Quantum Computing}
\label{app:quantum}
In this appendix, we provide additional technical background and details on the preliminaries presented in Section~\ref{sec:prelim:quantum} covering quantum computing.
Specifically, we now present a summary of some key algorithmic components in the context of quantum computing, referring the reader
\mss{to~\cite{nielsen2010quantum,Dervovic18,Port18,de2019quantum,LipReg21,LinLin2022}}
for additional technical details on quantum computing and quantum algorithms.

\subsection{Harrow–Hassidim–Lloyd Algorithm}
The Harrow–Hassidim–Lloyd (HHL) algorithm~\cite{HHL09} is a quantum algorithm for solving a system of linear equations 
$\A\x=\bb$ via quantum computers
such that 
the obtained solution is a scalar multiple of $\x$, i.e., HHL 
prepares a state $\ket{\x}$ whose amplitude equals those of $\x$. In essence, HHL encodes the vector $\bb$ 
into a quantum state $\ket{\bb}$ and proceeds by applying a series of quantum operators, e.g., Quantum Phase 
Estimation (QPE), to find the quantum state $\ket{\x}$ corresponding to the solution vector $\x$. 
The HHL algorithm can be used within 
a statistical process in order to compute the expectation $\braopket{\x}{\M}{\x}$ for some measurable operator 
$\M$. Under the assumption that the matrix $\A$ is sparse and well-conditioned, the HHL algorithm has a runtime of 
$O(log(N)\kappa^2)$ where $\kappa$ denotes the condition number of $\A$ and $N$ denotes the matrix dimension. 
This complexity represents an exponential speedup over classical Krylov subspace algorithms such as 
Conjugate Gradient whose execution generally requires $O(N\kappa)$ operations.\footnote{Note that HHL 
computes only a scalar multiple of the solution compared to classical Krylov subspace algorithms.} 

From a linear algebra viewpoint, HHL first writes the right-hand side $\bb$ in a quantum state form  $\ket{\bb}$ and then applies the operator $e^{i\A t}$ onto $\ket{\bb}$ 
at different timestamps $t$ via Hamiltonian simulation~\cite{zaman2023step,qiskitextbook2023}. Note that, since $\A$ is Hermitian, we can express it as 
\begin{equation*}
\A = \sum\limits_{i=0}^{N-1} \lambda_i \ket{\uu_i}\bra{\uu_i},     
\end{equation*}
where $(\lambda_{i-1},\uu_{i-1})$ denotes the $i$th eigenpair of the matrix $\A$. 
It then directly follows that 
\begin{equation*}
    \A^{-1} = \sum\limits_{i=0}^{N-1} \lambda_i^{-1} \ket{\uu_i}\bra{\uu_i}.
\end{equation*}
Now, since $\A$ is invertible by definition, we can decompose the quantum state of the right-hand side as $\ket{\bb} = \sum\limits_{i=1}^{N}\beta_i \ket{\uu_i}$. The decomposition of $\ket{\bb}$ into the eigenbasis of $\A$ and the computation of the corresponding eigenvalues  $\lambda_{i}$ is achieved via QPE. Given the above, the mathematical goal of the HHL algorithm is to compute the quantum state 
\begin{equation*}
    \ket{\x} = \sum\limits_{i=0}^{N-1} \lambda_i^{-1} b_i \ket{\uu_i}.
\end{equation*}

We next summarize the main steps performed within a single repetition of the HHL algorithm from a quantum perspective, assuming initial access to three quantum registers: the first storing the quantum state $\ket{\bb}$; the second, with length $l_1 \in \Nats$, storing the binary representation of $\lambda_i$; and the third, with length $l_2 \in \Nats$, containing the vector solution. A proper circuit will also include auxiliary 
registers. 
\begin{enumerate}
    \item Apply the operator $\U = e^{j\A t} = \sum\limits_{i=0}^{N-1} e^{j\lambda_i t} \ket{\uu_i}\bra{\uu_i}$ onto $\ket{\bb}$, i.e., the input register. Now the input register reads as 
    $\sum\limits_{i=0}^{N-1} b_i \ket{\lambda_i}_{l_1} \ket{\uu_i}_{l_2}$, where $l_1 \in \Nats$ denotes the number of bits set for the binary representation of $\lambda_i$ 
    and $l_2 \in \Nats$ denotes the number of qubits used to represent $\ket{\bb}$. 
    \item Apply a rotation of the
    \mss{auxiliary}
    register to normalize the quantum state of the 
    input register, i.e., apply a rotation conditioned on $\ket{\lambda_i}_{l_1}$, 
    which leads to
    $$\sum\limits_{i=0}^{N-1} b_i \ket{\lambda_i}_{l_1} \ket{u_i}_{l_2}\left(\sqrt{1-\frac{C^2}{\lambda_i^2}}\ket{0} + \frac{C}{\lambda_i}\ket{1}\right)$$
    where $C\in \Complex$ is a normalization constant that is less than the smallest eigenvalue of the matrix $\A$.
    \item Apply the inverse QPE which, after ignoring accumulated errors from QPE, leads to 
    the quantum state $$\sum\limits_{i=0}^{N-1} b_i \ket{0}_{l_1} \ket{u_i}_{l_2}\left(\sqrt{1-\frac{C^2}{\lambda_i^2}}\ket{0} + \frac{C}{\lambda_i}\ket{1}\right).$$
    \item Measure the auxiliary register in the computational basis. Then, if the outcome is equal to one, the register in the post-measurement state is equal to
    $$\left(\sqrt{\frac{1}{\sum\limits_{k=0}^{N-1}b_k^2/\lambda_k^2}}\right)\sum\limits_{i=0}^{N-1} b_i \ket{0}_{l_1} \ket{u_i}_{l_2}.$$ 
    This quantity corresponds to the true solution up to a scalar multiple. 
\end{enumerate}

\subsection{Quantum Phase Estimation Algorithm}
Roughly speaking, Quantum Phase Estimation (QPE) is a quantum algorithm that, given a unitary transformation $\U$ with an eigenvector $\ket{\bm{\psi}}$ and a corresponding eigenvalue 
$e^{2\pi i\theta}$ satisfying $\U\ket{\bm{\psi}}=e^{2\pi i\theta} \ket{\bm{\psi}}$, returns $\theta$. QPE is a major workhorse of many quantum algorithms, including the HHL algorithm as noted above. While the goal is to return the value of 
$\theta$, such an approximation should be ideally produced with a small number of gates and with high probability. 
Below we summarize the main circuit details of 
QPE assuming p$+m$ input qubits, where the top $p\in \Nats$ qubits will be referred to as ``counting'' qubits and the bottom $m\in \Nats$ qubits hold the eigenvector $\ket{\bm{\psi}}$ where the dimension of the unitary matrix $\U$ is equal to $N=2^m$.
\begin{enumerate}
\item Let $\ket{\bm{\psi}}$ be stored in a $p$-qubit register and $\ket{\bm{\psi}_0}=\ket{0}^{\otimes p} \ket{\bm{\psi}}$ a register that stores the value of $2^p \theta$.
\item Apply a $p$-qubit Hadamard gate operation $H^{\otimes p}$ on the counting register, leading to 
the updated state $\ket{\bm{\psi}_1} = \dfrac{1}{2^{p/2}}(\ket{0}+\ket{1})^{\otimes p}\ket{\bm{\psi}}$.
\item Apply the unitary operator $\U$ on the target register only if the corresponding control bit is equal to $\ket{1}$. Since $\U\ket{\bm{\psi}}=e^{2\pi i\theta} \ket{\psi}$, it follows that 
$\U^{2^j}\ket{\bm{\psi}}=e^{2\pi i\theta 2^j} \ket{\bm{\psi}}$. Then, applying all the $p$ controlled operations leads to
$$\ket{\bm{\psi}_2} = \dfrac{1}{2^{p/2}} \sum\limits_{j=0}^{2^p-1} e^{2 \pi i \theta j} \ket{j} \otimes \ket{\bm{\psi}},$$ where $j\in \Nats$ denotes the integer representation of a $p$-bit binary number.
\item Apply the inverse Quantum Fourier Transformation (iQFT) to the state $\ket{\bm{\psi}_2}$, yielding 
\begin{equation*}
\dfrac{1}{2^{p/2}} \sum\limits_{j=0}^{2^p-1} e^{2 \pi i \theta j} \ket{j}  \left(\dfrac{1}{2^{p/2}} \sum\limits_{g=0}^{2^p-1} e^{\frac{-2 \pi j g}{2^p}}\ket{g}\right)
 = \dfrac{1}{2^p} \sum\limits_{g=0}^{2^p-1} \sum\limits_{j=0}^{2^p-1} e^{-\frac{2\pi i j}{2^p}(g-2^p \theta)}\ket{g} .
\end{equation*}
Rounding now $2^p\theta$ to the nearest integer implies $2^p \theta =\alpha +2^p \delta$, where $\alpha = {\rm{round}}(2^p \theta)$ and the magnitude of the scalar $2^p\delta$ is less than or equal to half. Following the latter splitting, we can then write
$$\dfrac{1}{2^p} \sum\limits_{j=0}^{2^p-1} e^{-\frac{2\pi i j}{2^p}(g-2^p \theta)} =
 \dfrac{1}{2^p} \sum\limits_{j=0}^{2^p-1} e^{-\frac{2\pi i j}{2^p}(g-\alpha)} e^{2\pi i \delta j}.$$
 \item Finally, QPE performs a measurement in the computational basis  of the top $p$-qubit register yielding the 
 outcome $\ket{h}$ with probability $\left|\dfrac{1}{2^p} \sum\limits_{j=0}^{2^p-1} e^{-\frac{2\pi i j}{2^p}(h-\alpha)} e^{2\pi i \delta j}\right|^2$. Therefore, QPE provides an estimate that is within $1/2^p$ of the correct 
 answer $\theta$ with probability at least $4/\pi^2$.
\end{enumerate}

\subsection{Quantum Fourier Transformation Algorithms}
The quantum Fourier transformation (QFT) operates in an analogous fashion to the classical Fourier transformation.
In particular, QFT acts on a quantum state $\ket{\ff} = \sum\limits_{i=0}^{2^p-1}f_i\ket{i}$, resulting in a new quantum state 
$\ket{\yy} = \sum\limits_{i=0}^{2^p-1}y_i\ket{i}$
such that 
\begin{equation*}
y_i=\dfrac{1}{\sqrt{2^p}}\sum\limits_{j=0}^{2^p-1}x_j\omega_{2^p}^{-ji}, \qquad i\in \{0,1,\ldots,2^p-1\},    
\end{equation*}
where 
$\omega_{2^p} = e^{-\frac{2\pi i}{2^p}}$.
Note here that, since $\omega_{2^p}^{-j}$ is a rotation, 
the inverse quantum Fourier transformation (iQFT) acts in a similar fashion, but instead rendering
\begin{equation*}
y_i=\dfrac{1}{\sqrt{2^p}}\sum\limits_{j=0}^{2^p-1}x_j\omega_{2^p}^{ji}, \qquad i\in \{0,1,\ldots,2^p-1\}.   
\end{equation*}

\section{Proofs}
\label{app:proofs}
In this appendix, we provide additional details and more complete versions
of
the proofs
of our theoretical results presented in Section~\ref{sec:proofs}.

\subsection{Proof of Theorem~\ref{lem7.15}}
\label{app:lem7.15}
%
From~\eqref{eq7.38a} and~\eqref{eq7.38b}, replacing $\varphi^{(n)}(z)$ with its approximation $\vartheta^{(n)}(z)$, we obtain
\begin{align}
    \underline{\vartheta}^{(n+1)}(z) & = z \vartheta_{\scriptscriptstyle{odd}}^{(n)}(z) + \vartheta_{\scriptscriptstyle{even}}^{(n)}(z)(\I -\vartheta_{\scriptscriptstyle{odd}}^{(n)}(z))^{-1} \vartheta_{\scriptscriptstyle{even}}^{(n)}(z) \tag{\ref{eq:7.58a}} \\
    \underline{\hat{\vartheta}}^{(n+1)}(z) & = \hat{\vartheta}_{\scriptscriptstyle{odd}}^{(n)}(z) + \vartheta_{\scriptscriptstyle{even}}^{(n)}(z)(\I -\vartheta_{\scriptscriptstyle{odd}}^{(n)}(z))^{-1} \hat{\vartheta}_{\scriptscriptstyle{even}}^{(n)}(z) \tag{\ref{eq:7.58b}} .
\end{align}
By definition of $\R_{\scriptscriptstyle{odd}}^{(n)}(z)$, we have $\I - \vartheta_{\scriptscriptstyle{odd}}^{(n)}(z) = \I - \varphi_{\scriptscriptstyle{odd}}^{(n)}(z) - \R_{\scriptscriptstyle{odd}}^{(n)}(z)$, 
or equivalently
$$
\I - \vartheta_{\scriptscriptstyle{odd}}^{(n)}(z) = (\I - \varphi_{\scriptscriptstyle{odd}}^{(n)}(z))\, [\I - (\I - \varphi_{\scriptscriptstyle{odd}}^{(n)}(z))^{-1} \R_{\scriptscriptstyle{odd}}^{(n)}(z)] ,
$$
and therefore
\begin{equation}
(\I - \vartheta_{\scriptscriptstyle{odd}}^{(n)}(z))^{-1} = [\I - (\I - \varphi_{\scriptscriptstyle{odd}}^{(n)}(z))^{-1} \R_{\scriptscriptstyle{odd}}^{(n)}(z)]^{-1} (\I - \varphi_{\scriptscriptstyle{odd}}^{(n)}(z))^{-1} .
\tag{\ref{eq:Bodd-temp}}
\end{equation}
By definition of first-order equality ($\doteq$), we conclude
$$
[\I - (\I - \varphi_{\scriptscriptstyle{odd}}^{(n)}(z))^{-1} \R_{\scriptscriptstyle{odd}}^{(n)}(z)]^{-1} \; \doteq \; \I + (\I - \varphi_{\scriptscriptstyle{odd}}^{(n)}(z))^{-1} \R_{\scriptscriptstyle{odd}}^{(n)}(z) ,
$$
which upon substitution in~\eqref{eq:Bodd-temp} yields
\begin{align}
(\I - \vartheta_{\scriptscriptstyle{odd}}^{(n)}(z))^{-1} 
& \; \doteq \; (\I - \varphi_{\scriptscriptstyle{odd}}^{(n)}(z))^{-1} + (\I - \varphi_{\scriptscriptstyle{odd}}^{(n)}(z))^{-1} \R_{\scriptscriptstyle{odd}}^{(n)}(z) (\I - \varphi_{\scriptscriptstyle{odd}}^{(n)}(z))^{-1} 
.
\tag{\ref{eq:1storderBodd}}
\end{align}
Substituting~\eqref{eq:1storderBodd} and the definition of
$\R_{\scriptscriptstyle{even}}^{(n)}(z)$ 
into~\eqref{eq:7.58a}, we derive
\begin{align}
    \underline{\vartheta}^{(n+1)}(z) & \; \doteq \; z \vartheta_{\scriptscriptstyle{odd}}^{(n)}(z)
    \nonumber
    + \vartheta_{\scriptscriptstyle{even}}^{(n)}(z)
    [(\I - \varphi_{\scriptscriptstyle{odd}}^{(n)}(z))^{-1} + (\I - \varphi_{\scriptscriptstyle{odd}}^{(n)}(z))^{-1} \R_{\scriptscriptstyle{odd}}^{(n)}(z) (\I - \varphi_{\scriptscriptstyle{odd}}^{(n)}(z))^{-1}] \vartheta_{\scriptscriptstyle{even}}^{(n)}(z) \nonumber \\
    & \; \doteq \; z \vartheta_{\scriptscriptstyle{odd}}^{(n)}(z) + \R_{\scriptscriptstyle{even}}^{(n)}(z) (\I - \varphi_{\scriptscriptstyle{odd}}^{(n)}(z))^{-1} \R_{\scriptscriptstyle{even}}^{(n)}(z) + \R_{\scriptscriptstyle{even}}^{(n)}(z) (\I - \varphi_{\scriptscriptstyle{odd}}^{(n)}(z))^{-1} \varphi_{\scriptscriptstyle{even}}^{(n)}(z) \nonumber \\
    & \qquad\qquad + \R_{\scriptscriptstyle{even}}^{(n)}(z) (\I - \varphi_{\scriptscriptstyle{odd}}^{(n)}(z))^{-1} \R_{\scriptscriptstyle{odd}}^{(n)}(z) (\I - \varphi_{\scriptscriptstyle{odd}}^{(n)}(z))^{-1} \R_{\scriptscriptstyle{even}}^{(n)}(z) \nonumber \\
    & \qquad\qquad + \R_{\scriptscriptstyle{even}}^{(n)}(z)  (\I - \varphi_{\scriptscriptstyle{odd}}^{(n)}(z))^{-1} \R_{\scriptscriptstyle{odd}}^{(n)}(z) (\I - \varphi_{\scriptscriptstyle{odd}}^{(n)}(z))^{-1} \varphi_{\scriptscriptstyle{even}}^{(n)}(z) \nonumber \\
    & \qquad\qquad + \varphi_{\scriptscriptstyle{even}}^{(n)}(z)
    (\I - \varphi_{\scriptscriptstyle{odd}}^{(n)}(z))^{-1} \R_{\scriptscriptstyle{even}}^{(n)}(z) + \varphi_{\scriptscriptstyle{even}}^{(n)}(z)
    (\I - \varphi_{\scriptscriptstyle{odd}}^{(n)}(z))^{-1} \varphi_{\scriptscriptstyle{even}}^{(n)}(z) \nonumber \\
    & \qquad\qquad + \varphi_{\scriptscriptstyle{even}}^{(n)}(z) (\I - \varphi_{\scriptscriptstyle{odd}}^{(n)}(z))^{-1} \R_{\scriptscriptstyle{odd}}^{(n)}(z) (\I - \varphi_{\scriptscriptstyle{odd}}^{(n)}(z))^{-1} \R_{\scriptscriptstyle{even}}^{(n)}(z) \nonumber \\
    & \qquad\qquad + \varphi_{\scriptscriptstyle{even}}^{(n)}(z) (\I - \varphi_{\scriptscriptstyle{odd}}^{(n)}(z))^{-1} \R_{\scriptscriptstyle{odd}}^{(n)}(z) (\I - \varphi_{\scriptscriptstyle{odd}}^{(n)}(z))^{-1} \varphi_{\scriptscriptstyle{even}}^{(n)}(z).
\label{eq:1storderBodd2}
\end{align} 
%
Upon subtracting~\eqref{eq7.38a} from~\eqref{eq:1storderBodd2}, we obtain
\begin{align}
    \underline{\R}^{(n+1)}(z) 
    & \; \doteq \; z \R_{\scriptscriptstyle{odd}}^{(n)}(z) + \R_{\scriptscriptstyle{even}}^{(n)}(z) (\I - \varphi_{\scriptscriptstyle{odd}}^{(n)}(z))^{-1} \R_{\scriptscriptstyle{even}}^{(n)}(z) + \R_{\scriptscriptstyle{even}}^{(n)}(z) (\I - \varphi_{\scriptscriptstyle{odd}}^{(n)}(z))^{-1} \varphi_{\scriptscriptstyle{even}}^{(n)}(z) \nonumber \\
    & \qquad\qquad + \R_{\scriptscriptstyle{even}}^{(n)}(z) (\I - \varphi_{\scriptscriptstyle{odd}}^{(n)}(z))^{-1} \R_{\scriptscriptstyle{odd}}^{(n)}(z) (\I - \varphi_{\scriptscriptstyle{odd}}^{(n)}(z))^{-1} \R_{\scriptscriptstyle{even}}^{(n)}(z) \nonumber \\
    & \qquad\qquad + \R_{\scriptscriptstyle{even}}^{(n)}(z)  (\I - \varphi_{\scriptscriptstyle{odd}}^{(n)}(z))^{-1} \R_{\scriptscriptstyle{odd}}^{(n)}(z) (\I - \varphi_{\scriptscriptstyle{odd}}^{(n)}(z))^{-1} \varphi_{\scriptscriptstyle{even}}^{(n)}(z) \nonumber \\
    & \qquad\qquad + \varphi_{\scriptscriptstyle{even}}^{(n)}(z)
    (\I - \varphi_{\scriptscriptstyle{odd}}^{(n)}(z))^{-1} \R_{\scriptscriptstyle{even}}^{(n)}(z) \nonumber \\
    & \qquad\qquad + \varphi_{\scriptscriptstyle{even}}^{(n)}(z) (\I - \varphi_{\scriptscriptstyle{odd}}^{(n)}(z))^{-1} \R_{\scriptscriptstyle{odd}}^{(n)}(z) (\I - \varphi_{\scriptscriptstyle{odd}}^{(n)}(z))^{-1} \R_{\scriptscriptstyle{even}}^{(n)}(z) \nonumber \\
    & \qquad\qquad + \varphi_{\scriptscriptstyle{even}}^{(n)}(z) (\I - \varphi_{\scriptscriptstyle{odd}}^{(n)}(z))^{-1} \R_{\scriptscriptstyle{odd}}^{(n)}(z) (\I - \varphi_{\scriptscriptstyle{odd}}^{(n)}(z))^{-1} \varphi_{\scriptscriptstyle{even}}^{(n)}(z) .
\label{eq:1storderR}
\end{align} 
Given our first-order error analysis on $\underline{\R}^{(n+1)}(z)$, we can remove the higher-order terms 
which renders
\begin{align}
    \underline{\R}^{(n+1)}(z) 
    & \; \doteq \; z \R_{\scriptscriptstyle{odd}}^{(n)}(z) + \R_{\scriptscriptstyle{even}}^{(n)}(z) (\I - \varphi_{\scriptscriptstyle{odd}}^{(n)}(z))^{-1} \varphi_{\scriptscriptstyle{even}}^{(n)}(z) + \varphi_{\scriptscriptstyle{even}}^{(n)}(z)
    (\I - \varphi_{\scriptscriptstyle{odd}}^{(n)}(z))^{-1} \R_{\scriptscriptstyle{even}}^{(n)}(z) \nonumber \\
    & \qquad\qquad + \varphi_{\scriptscriptstyle{even}}^{(n)}(z) (\I - \varphi_{\scriptscriptstyle{odd}}^{(n)}(z))^{-1} \R_{\scriptscriptstyle{odd}}^{(n)}(z) (\I - \varphi_{\scriptscriptstyle{odd}}^{(n)}(z))^{-1} \varphi_{\scriptscriptstyle{even}}^{(n)}(z) .
\tag{\ref{eq:1storderR2}}
\end{align}
Substituting $\V^{(n)}(z)$ and $\W^{(n)}(z)$
then yields~\eqref{eq:7.57a}.

Next, we repeat the same analysis for~\eqref{eq:7.58b} by substituting both~\eqref{eq:1storderBodd} and the definition of
$\R_{\scriptscriptstyle{even}}^{(n)}(z)$,
and deriving
\begin{align}
    \underline{\hat{\vartheta}}^{(n+1)}(z) & \; \doteq \; \hat{\vartheta}_{\scriptscriptstyle{odd}}^{(n)}(z) 
    + \vartheta_{\scriptscriptstyle{even}}^{(n)}(z)
    [(\I - \varphi_{\scriptscriptstyle{odd}}^{(n)}(z))^{-1} + (\I - \varphi_{\scriptscriptstyle{odd}}^{(n)}(z))^{-1} \R_{\scriptscriptstyle{odd}}^{(n)}(z) (\I - \varphi_{\scriptscriptstyle{odd}}^{(n)}(z))^{-1}] \hat{\vartheta}_{\scriptscriptstyle{even}}^{(n)}(z) \nonumber \\
    & \; \doteq \; \hat{\vartheta}_{\scriptscriptstyle{odd}}^{(n)}(z) + \R_{\scriptscriptstyle{even}}^{(n)}(z) (\I - \varphi_{\scriptscriptstyle{odd}}^{(n)}(z))^{-1} \hat{\R}_{\scriptscriptstyle{even}}^{(n)}(z) + \R_{\scriptscriptstyle{even}}^{(n)}(z) (\I - \varphi_{\scriptscriptstyle{odd}}^{(n)}(z))^{-1} \hat{\varphi}_{\scriptscriptstyle{even}}^{(n)}(z) \nonumber \\
    & \qquad\qquad + \R_{\scriptscriptstyle{even}}^{(n)}(z) (\I - \varphi_{\scriptscriptstyle{odd}}^{(n)}(z))^{-1} \R_{\scriptscriptstyle{odd}}^{(n)}(z) (\I - \varphi_{\scriptscriptstyle{odd}}^{(n)}(z))^{-1} \hat{\R}_{\scriptscriptstyle{even}}^{(n)}(z) \nonumber \\
    & \qquad\qquad + \R_{\scriptscriptstyle{even}}^{(n)}(z) (\I - \varphi_{\scriptscriptstyle{odd}}^{(n)}(z))^{-1} \R_{\scriptscriptstyle{odd}}^{(n)}(z) (\I - \varphi_{\scriptscriptstyle{odd}}^{(n)}(z))^{-1} \hat{\varphi}_{\scriptscriptstyle{even}}^{(n)}(z) \nonumber \\
    & \qquad\qquad + \varphi_{\scriptscriptstyle{even}}^{(n)}(z)
    (\I - \varphi_{\scriptscriptstyle{odd}}^{(n)}(z))^{-1} \hat{\R}_{\scriptscriptstyle{even}}^{(n)}(z) + \varphi_{\scriptscriptstyle{even}}^{(n)}(z)
    (\I - \varphi_{\scriptscriptstyle{odd}}^{(n)}(z))^{-1} \hat{\varphi}_{\scriptscriptstyle{even}}^{(n)}(z) \nonumber \\
    & \qquad\qquad  + \varphi_{\scriptscriptstyle{even}}^{(n)}(z) (\I - \varphi_{\scriptscriptstyle{odd}}^{(n)}(z))^{-1} \R_{\scriptscriptstyle{odd}}^{(n)}(z) (\I - \varphi_{\scriptscriptstyle{odd}}^{(n)}(z))^{-1} \hat{\R}_{\scriptscriptstyle{even}}^{(n)}(z) \nonumber \\
    & \qquad\qquad + \varphi_{\scriptscriptstyle{even}}^{(n)}(z) (\I - \varphi_{\scriptscriptstyle{odd}}^{(n)}(z))^{-1} \R_{\scriptscriptstyle{odd}}^{(n)}(z) (\I - \varphi_{\scriptscriptstyle{odd}}^{(n)}(z))^{-1} \hat{\varphi}_{\scriptscriptstyle{even}}^{(n)}(z) .
\label{eq:1storderBodd-hat}
\end{align} 
Upon subtracting~\eqref{eq7.38b} from~\eqref{eq:1storderBodd-hat}, we obtain
\begin{align}
    \underline{\hat{\R}}^{(n+1)}(z) & \; \doteq \; \hat{\R}_{\scriptscriptstyle{odd}}^{(n)}(z)
    + \R_{\scriptscriptstyle{even}}^{(n)}(z) (\I - \varphi_{\scriptscriptstyle{odd}}^{(n)}(z))^{-1} \hat{\R}_{\scriptscriptstyle{even}}^{(n)}(z) + \R_{\scriptscriptstyle{even}}^{(n)}(z) (\I - \varphi_{\scriptscriptstyle{odd}}^{(n)}(z))^{-1} \hat{\varphi}_{\scriptscriptstyle{even}}^{(n)}(z) \nonumber \\
    & \qquad\qquad + \R_{\scriptscriptstyle{even}}^{(n)}(z) (\I - \varphi_{\scriptscriptstyle{odd}}^{(n)}(z))^{-1} \R_{\scriptscriptstyle{odd}}^{(n)}(z) (\I - \varphi_{\scriptscriptstyle{odd}}^{(n)}(z))^{-1} \hat{\R}_{\scriptscriptstyle{even}}^{(n)}(z) \nonumber \\
    & \qquad\qquad + \R_{\scriptscriptstyle{even}}^{(n)}(z) (\I - \varphi_{\scriptscriptstyle{odd}}^{(n)}(z))^{-1} \R_{\scriptscriptstyle{odd}}^{(n)}(z) (\I - \varphi_{\scriptscriptstyle{odd}}^{(n)}(z))^{-1} \hat{\varphi}_{\scriptscriptstyle{even}}^{(n)}(z) \nonumber \\
    & \qquad\qquad + \varphi_{\scriptscriptstyle{even}}^{(n)}(z)
    (\I - \varphi_{\scriptscriptstyle{odd}}^{(n)}(z))^{-1} \hat{\R}_{\scriptscriptstyle{even}}^{(n)}(z) \nonumber \\
    & \qquad\qquad  + \varphi_{\scriptscriptstyle{even}}^{(n)}(z) (\I - \varphi_{\scriptscriptstyle{odd}}^{(n)}(z))^{-1} \R_{\scriptscriptstyle{odd}}^{(n)}(z) (\I - \varphi_{\scriptscriptstyle{odd}}^{(n)}(z))^{-1} \hat{\R}_{\scriptscriptstyle{even}}^{(n)}(z) \nonumber \\
    & \qquad\qquad + \varphi_{\scriptscriptstyle{even}}^{(n)}(z) (\I - \varphi_{\scriptscriptstyle{odd}}^{(n)}(z))^{-1} \R_{\scriptscriptstyle{odd}}^{(n)}(z) (\I - \varphi_{\scriptscriptstyle{odd}}^{(n)}(z))^{-1} \hat{\varphi}_{\scriptscriptstyle{even}}^{(n)}(z) .
\label{eq:1storderR-hat}
\end{align}
Given our first-order error analysis on $\underline{\hat{\R}}^{(n+1)}(z)$, we can remove the higher-order terms 
which renders
\begin{align}
    \underline{\hat{\R}}^{(n+1)}(z) & \; \doteq \; \hat{\R}_{\scriptscriptstyle{odd}}^{(n)}(z)  
    + \R_{\scriptscriptstyle{even}}^{(n)}(z) (\I - \varphi_{\scriptscriptstyle{odd}}^{(n)}(z))^{-1} \hat{\varphi}_{\scriptscriptstyle{even}}^{(n)}(z) + \varphi_{\scriptscriptstyle{even}}^{(n)}(z)
    (\I - \varphi_{\scriptscriptstyle{odd}}^{(n)}(z))^{-1} \hat{\R}_{\scriptscriptstyle{even}}^{(n)}(z) \nonumber \\
    & \qquad\qquad + \varphi_{\scriptscriptstyle{even}}^{(n)}(z) (\I - \varphi_{\scriptscriptstyle{odd}}^{(n)}(z))^{-1} \R_{\scriptscriptstyle{odd}}^{(n)}(z) (\I - \varphi_{\scriptscriptstyle{odd}}^{(n)}(z))^{-1} \hat{\varphi}_{\scriptscriptstyle{even}}^{(n)}(z) .
\label{eq:1storderR-hat2}
\end{align} 
Substituting $\V^{(n)}(z)$ and $\hat{\W}^{(n)}(z)$
then yields~\eqref{eq:7.57b}.

Turning to the corresponding first-order upper bounds, we apply the max norm $\| \cdot \|_{\ast}$ to both sides of~\eqref{eq:7.57a} and~\eqref{eq:7.57b} and exploit the triangle inequality to obtain 
\begin{align*}
    \|\underline{\R}^{(n+1)}(z)\|_{\ast} & \; \dotleq \; \|\R_{\scriptscriptstyle{odd}}^{(n)}(z)\|_{\ast} + \|\R_{\scriptscriptstyle{even}}^{(n)}(z)\|_{\ast} \|\W^{(n)}(z)\|_{\ast} + \|\R_{\scriptscriptstyle{even}}^{(n)}(z)\|_{\ast} \|\V^{(n)}(z)\|_{\ast}
    \\
    & \qquad\qquad\qquad\qquad 
    + \|\R_{\scriptscriptstyle{odd}}^{(n)}(z)\|_{\ast} \|\V^{(n)}(z)\|_{\ast} \|\W^{(n)}(z)\|_{\ast} , 
    \\
    \|\underline{\hat{\R}}^{(n+1)}(z)\|_{\ast} & \; \dotleq \; \|\hat{\R}_{\scriptscriptstyle{odd}}^{(n)}(z)\|_{\ast} + \|\R_{\scriptscriptstyle{even}}^{(n)}(z)\|_{\ast} \|\hat{\W}^{(n)}(z)\|_{\ast} + \|\hat{\R}_{\scriptscriptstyle{even}}^{(n)}(z)\|_{\ast} \|\V^{(n)}(z)\|_{\ast}
    \\
    & \qquad\qquad\qquad\qquad 
    + \|\R_{\scriptscriptstyle{odd}}^{(n)}(z)\|_{\ast} \|\V^{(n)}(z)\|_{\ast} \|\hat{\W}^{(n)}(z)\|_{\ast} .
\end{align*}
From the monotonicity property of the infinity norm, we know that $\|\R_{\scriptscriptstyle{odd}}^{(n)}(z)\|_{\ast}$ and $\|\R_{\scriptscriptstyle{even}}^{(n)}(z)\|_{\ast}$ are both less than or equal to $\|\R^{(n)}(z)\|_{\ast}$ and that $\|\hat{\R}_{\scriptscriptstyle{odd}}^{(n)}(z)\|_{\ast}$ and $\|\hat{\R}_{\scriptscriptstyle{even}}^{(n)}(z)\|_{\ast}$ are both less than or equal to $\|\hat{\R}^{(n)}(z)\|_{\ast}$, thus rendering
\begin{align}
    \|\underline{\R}^{(n+1)}(z)\|_{\ast} & \; \dotleq \; \|\R^{(n)}(z)\|_{\ast} + \|\R^{(n)}(z)\|_{\ast} \|\W^{(n)}(z)\|_{\ast} + \|\R^{(n)}(z)\|_{\ast} \|\V^{(n)}(z)\|_{\ast} \nonumber \\
    & \qquad\qquad\qquad\qquad + \|\R^{(n)}(z)\|_{\ast} \|\V^{(n)}(z)\|_{\ast} \|\W^{(n)}(z)\|_{\ast} , \tag{\ref{eq:7.57a-leq2}} \\
    \|\underline{\hat{\R}}^{(n+1)}(z)\|_{\ast} & \; \dotleq \; \|\hat{\R}^{(n)}(z)\|_{\ast} + \|\R^{(n)}(z)\|_{\ast} \|\hat{\W}^{(n)}(z)\|_{\ast} + \|\hat{\R}^{(n)}(z)\|_{\ast} \|\V^{(n)}(z)\|_{\ast} \nonumber \\
    & \qquad\qquad\qquad\qquad + \|\R^{(n)}(z)\|_{\ast} \|\V^{(n)}(z)\|_{\ast} \|\hat{\W}^{(n)}(z)\|_{\ast} . \label{eq:7.57b-leq2}
\end{align}
The coefficients of the matrix power series $\V^{(n)}(z)$, $\W^{(n)}(z)$ and $\hat{\W}^{(n)}(z)$ are nonnegative.
From Theorem~\ref{thm7.8}, the matrices $\V^{(n)}(1)$, $\W^{(n)}(1)$ and $\hat{\W}^{(n)}(1)$ have spectral radius $1$ with $\W^{(n)}(1)$ and $\hat{\W}^{(n)}(1)$ being stochastic, and thus it follows that $\|\W^{(n)}(z)\|_{\ast} = \|\W^{(n)}(1)\|_{\infty} = 1$ and $\|\hat{\W}^{(n)}(z)\|_{\ast} = \|\hat{\W}^{(n)}(1)\|_{\infty} = 1$.
Upon substitution of these relationships into~\eqref{eq:7.57a-leq2} and~\eqref{eq:7.57b-leq2}, we respectively have
\begin{align}
    \|\underline{\R}^{(n+1)}(z)\|_{\ast} & \; \dotleq \; 2 \|\R^{(n)}(z)\|_{\ast} + 2 \|\R^{(n)}(z)\|_{\ast} \|\V^{(n)}(z)\|_{\ast} = 2 \|\R^{(n)}(z)\|_{\ast} (1 + \|\V^{(n)}(1)\|_{\infty}) , \tag{\ref{eq:7.57a-leq3}} \\
    \|\underline{\hat{\R}}^{(n+1)}(z)\|_{\ast} & \; \dotleq \; \|\hat{\R}^{(n)}(z)\|_{\ast} + \|\R^{(n)}(z)\|_{\ast} + \|\hat{\R}^{(n)}(z)\|_{\ast} \|\V^{(n)}(z)\|_{\ast} + \|\R^{(n)}(z)\|_{\ast} \|\V^{(n)}(z)\|_{\ast} \nonumber \\
    & \qquad\qquad = \|\hat{\R}^{(n)}(z)\|_{\ast} (1 + \|\V^{(n)}(1)\|_{\infty}) + \|\R^{(n)}(z)\|_{\ast} (1 + \|\V^{(n)}(1)\|_{\infty}) , \label{eq:7.57b-leq3}
\end{align}
where the latter equalities in both~\eqref{eq:7.57a-leq3} and~\eqref{eq:7.57b-leq3} follow from the nonnegative coefficients of $\V^{(n)}(z)$, thus respectively yielding~\eqref{eq:7.57-UBa} and~\eqref{eq:7.57-UBb}.
\qed

\subsection{Proof of Theorem~\ref{thm:expUB}}
\label{app:thm:expUB}
%
From Theorem~\ref{lem7.15} and  equations~\eqref{eq:local-error}, \eqref{eq:global-error} and~\eqref{eq7.56}, we have
\begin{align}
 \bE^{(n+1)}(z) & \; \doteq \; z\bE^{(n)}_{odd}(z) + \V^{(n)}(z)\bE^{(n)}_{even}(z)+ \bE^{(n)}_{even}(z)\W^{(n)}(z) + \V^{(n)}(z)\bE^{(n)}_{odd}(z)\W^{(n)}(z) 
 + \cE_L^{(n)}(z) , \tag{\ref{eq:7.59a}} \\
 \hat{\bE}^{(n+1)}(z) & \; \doteq \; \hat{\bE}^{(n)}_{odd}(z) +  {\V}^{(n)}(z)\hat{\bE}^{(n)}_{even}(z) + {\bE}^{(n)}_{even}(z)\hat{\W}^{(n)}(z) + {\V}^{(n)}(z)\bE^{(n)}_{odd}(z)\hat{\W}^{(n)}(z) 
 + \hat{\cE}_L^{(n)}(z) , \tag{\ref{eq:7.59b}}
\end{align}
where
\begin{align*}
\bE^{(0)}(z) = \hat{\bE}^{(0)}(z) = 0 , \quad & \quad
\V^{(n)}(z) = \varphi_{\scriptscriptstyle{even}}^{(n)}(z)(\I -\varphi_{\scriptscriptstyle{odd}}^{(n)}(z))^{-1} , \\
\W^{(n)}(z) = (\I -\varphi_{\scriptscriptstyle{odd}}^{(n)}(z))^{-1}\varphi_{\scriptscriptstyle{even}}^{(n)}(z) , \quad & \quad
\hat{\W}^{(n)}(z) = (\I -\varphi_{\scriptscriptstyle{odd}}^{(n)}(z))^{-1}\hat{\varphi}_{\scriptscriptstyle{even}}^{(n)}(z) , 
\end{align*}
and
$$(\cE_L^{(n)}(z),\hat{\cE}_L^{(n)}(z)) = \cE_S(\vartheta^{(n)}(z),\hat{\vartheta}^{(n)}(z)).$$
Upon applying the max norm $\|\cdot\|_{*}$ to both sides of~\eqref{eq:7.59a} and~\eqref{eq:7.59b}, and exploiting the triangle inequality, we obtain
\begin{align*}
 \|\bE^{(n+1)}(z)\|_{\ast} & \; \dotleq \; \|\bE^{(n)}_{odd}(z)\|_{\ast} + \|\V^{(n)}(z)\|_{\ast} \|\bE^{(n)}_{even}(z)\|_{\ast} + \|\bE^{(n)}_{even}(z)\|_{\ast} \|\W^{(n)}(z)\|_{\ast} \\
 & \qquad\qquad\qquad + \|\V^{(n)}(z)\|_{\ast} \|\bE^{(n)}_{odd}(z)\|_{\ast} \|\W^{(n)}(z)\|_{\ast} 
 + \|\cE_L^{(n)}(z)\|_{\ast} , \\
 \|\hat{\bE}^{(n+1)}(z)\|_{\ast} & \; \dotleq \; \|\hat{\bE}^{(n)}_{odd}(z)\|_{\ast} + \|{\V}^{(n)}(z)\|_{\ast} \|\hat{\bE}^{(n)}_{even}(z)\|_{\ast} + \|{\bE}^{(n)}_{even}(z)\|_{\ast} \|\hat{\W}^{(n)}(z)\|_{\ast} \\
 & \qquad\qquad\qquad + \|{\V}^{(n)}(z)\|_{\ast} \|\bE^{(n)}_{odd}(z)\|_{\ast} \|\hat{\W}^{(n)}(z)\|_{\ast} 
 + \|\hat{\cE}_L^{(n)}(z)\|_{\ast} .
\end{align*}
From the monotonicity property of the infinity norm, we know that $\|\bE_{\scriptscriptstyle{odd}}^{(n)}(z)\|_{\ast}$ and $\|\bE_{\scriptscriptstyle{even}}^{(n)}(z)\|_{\ast}$ are both less than or equal to $\|\bE^{(n)}(z)\|_{\ast}$ and that $\|\hat{\bE}_{\scriptscriptstyle{odd}}^{(n)}(z)\|_{\ast}$ and $\|\hat{\bE}_{\scriptscriptstyle{even}}^{(n)}(z)\|_{\ast}$ are both less than or equal to $\|\hat{\bE}^{(n)}(z)\|_{\ast}$, thus rendering
\begin{align}
 \|\bE^{(n+1)}(z)\|_{\ast} & \; \dotleq \; \|\bE^{(n)}(z)\|_{\ast} + \|\V^{(n)}(z)\|_{\ast} \|\bE^{(n)}(z)\|_{\ast} + \|\bE^{(n)}(z)\|_{\ast} \|\W^{(n)}(z)\|_{\ast} \nonumber \\
 & \qquad\qquad\qquad + \|\V^{(n)}(z)\|_{\ast} \|\bE^{(n)}(z)\|_{\ast} \|\W^{(n)}(z)\|_{\ast} 
 + \|\cE_L^{(n)}(z)\|_{\ast} , \tag{\ref{eq:7.59a-leq2}}\\
 \|\hat{\bE}^{(n+1)}(z)\|_{\ast} & \; \dotleq \; \|\hat{\bE}^{(n)}(z)\|_{\ast} + \|{\V}^{(n)}(z)\|_{\ast} \|\hat{\bE}^{(n)}(z)\|_{\ast} + \|{\bE}^{(n)}(z)\|_{\ast} \|\hat{\W}^{(n)}(z)\|_{\ast} \nonumber \\
 & \qquad\qquad\qquad + \|{\V}^{(n)}(z)\|_{\ast} \|\bE^{(n)}(z)\|_{\ast} \|\hat{\W}^{(n)}(z)\|_{\ast} 
 + \|\hat{\cE}_L^{(n)}(z)\|_{\ast} \label{eq:7.59b-leq2} .
\end{align}

The coefficients of the matrix power series $\V^{(n)}(z)$, $\W^{(n)}(z)$ and $\hat{\W}^{(n)}(z)$ are nonnegative.
From Theorem~\ref{thm7.8}, the matrices $\V^{(n)}(1)$, $\W^{(n)}(1)$ and $\hat{\W}^{(n)}(1)$ have spectral radius $1$ with $\W^{(n)}(1)$ and $\hat{\W}^{(n)}(1)$ being stochastic, and thus it follows that $\|\W^{(n)}(z)\|_{\ast} = \|\W^{(n)}(1)\|_{\infty} = 1$ and $\|\hat{\W}^{(n)}(z)\|_{\ast} = \|\hat{\W}^{(n)}(1)\|_{\infty} = 1$.
Upon substitution of these relationships into~\eqref{eq:7.59a-leq2} and~\eqref{eq:7.59b-leq2}, we respectively have
\begin{align}
 \|\bE^{(n+1)}(z)\|_{\ast} & \; \dotleq \; \|\bE^{(n)}(z)\|_{\ast} + \|\V^{(n)}(z)\|_{\ast} \|\bE^{(n)}(z)\|_{\ast} + \|\bE^{(n)}(z)\|_{\ast} + \|\V^{(n)}(z)\|_{\ast} \|\bE^{(n)}(z)\|_{\ast} + \|\cE_L^{(n)}(z)\|_{\ast} \nonumber \\
 & \qquad\qquad = 2(1 + \|\V^{(n)}(1)\|_{\infty})\|\bE^{(n)}(z)\|_{\ast} + \|\cE_L^{(n)}(z)\|_{\ast} , \tag{\ref{eq:7.59a-leq3}} \\
 \|\hat{\bE}^{(n+1)}(z)\|_{\ast} & \; \dotleq \; \|\hat{\bE}^{(n)}(z)\|_{\ast} + \|{\V}^{(n)}(z)\|_{\ast} \|\hat{\bE}^{(n)}(z)\|_{\ast} + \|{\bE}^{(n)}(z)\|_{\ast} + \|{\V}^{(n)}(z)\|_{\ast} \|\bE^{(n)}(z)\|_{\ast} + \|\hat{\cE}_L^{(n)}(z)\|_{\ast} \nonumber \\
 & \qquad\qquad = \|\hat{\bE}^{(n)}(z)\|_{\ast} (1+\|{\V}^{(n)}(1)\|_{\infty}) + 
 \|\bE^{(n)}(z)\|_{\ast} (1+\|{\V}^{(n)}(1)\|_{\infty}) +\|\hat{\cE}_L^{(n)}(z)\|_{\ast} , \label{eq:7.59b-leq3} 
\end{align}
where the latter equalities in both~\eqref{eq:7.59a-leq3} and~\eqref{eq:7.59b-leq3} follow from the nonnegative coefficients of $\V^{(n)}(z)$.
Under the suppositions of the theorem, we conclude
\begin{align*}
    \|\bE^{(n+1)}(z)\|_{\ast} & \; \dotleq \; 
    \gamma_n \|\bE^{(n)}(z)\|_{\ast} + \upsilon , \\
    \|\hat{\bE}^{(n+1)}(z)\|_{\ast} & \; \dotleq \; 
    \frac{1}{2} \gamma_n \|\hat{\bE}^{(n)}(z)\|_{\ast} + \frac{1}{2} \gamma_n \|\bE^{(n)}(z)\|_{\ast} + \upsilon \\
    & \; \dotleq \; \gamma_n \|\bE^{(n)}(z)\|_{\ast} + \upsilon , 
\end{align*}
where the last inequality follows from an induction argument on the upper bounds for $\|\bE^{(n+1)}(z)\|_{*}$ and $\|\hat{\bE}^{(n+1)}(z)\|_{*}$.
In particular,
$\| \hat{\bE}^{(0)}(z) \|_{\ast} \, \dotleq \, \| \bE^{(0)}(z) \|_{\ast} = 0$ in~\eqref{eq:7.59a} and~\eqref{eq:7.59b} for the base case, and then we have for the induction step
\begin{align*}
   \|\hat{\bE}^{(n+1)}(z)\|_{\ast} & \; \dotleq \; \frac{1}{2} \gamma_n \|\hat{\bE}^{(n)}(z)\|_{\ast} + \frac{1}{2} \gamma_n \|\bE^{(n)}(z)\|_{\ast} + \upsilon \\
    & \; \dotleq \; \frac{1}{2} \gamma_n \|{\bE}^{(n)}(z)\|_{\ast} + \frac{1}{2} \gamma_n \|\bE^{(n)}(z)\|_{\ast} + \upsilon .
\end{align*}
The desired results then follow.
\qed

\subsection{Proof of Theorem~\ref{thm:linUB}}
\label{app:thm:linUB}
%
We apply
cyclic reduction 
to the function $\widetilde{\chi}(z)$ 
with respect to 
the corresponding sequences of matrix power series $\widetilde{\chi}^{(n)}(z)$ and $\hat{\widetilde{\chi}}^{(n)}(z)$.
From Theorem~\ref{lem7.15} and  equations~\eqref{eq:local-error}, \eqref{eq:global-error} and~\eqref{eq7.56}, we once again have~\eqref{eq:7.59a} and~\eqref{eq:7.59b}.
Taking the max norm $\|\cdot\|_{*}$ on both sides of~\eqref{eq:7.59a} and~\eqref{eq:7.59b}, and exploiting the triangle inequality, we obtain
\begin{align}
 \|\bE^{(n+1)}(z)\|_{\ast} & \; \dotleq \; \|\bE^{(n)}_{odd}(z)\|_{\ast} + \|\V^{(n)}(z)\|_{\ast} \|\bE^{(n)}_{even}(z)\|_{\ast} + \|\bE^{(n)}_{even}(z)\|_{\ast} \|\W^{(n)}(z)\|_{\ast} \nonumber \\
 & \qquad\qquad\qquad + \|\V^{(n)}(z)\|_{\ast} \|\bE^{(n)}_{odd}(z)\|_{\ast} \|\W^{(n)}(z)\|_{\ast} 
 + \|\cE_L^{(n)}(z)\|_{\ast} , \tag{\ref{eq:7.59a-bnd}} \\
 \|\hat{\bE}^{(n+1)}(z)\|_{\ast} & \; \dotleq \; \|\hat{\bE}^{(n)}_{odd}(z)\|_{\ast} + \|{\V}^{(n)}(z)\|_{\ast} \|\hat{\bE}^{(n)}_{even}(z)\|_{\ast} + \|{\bE}^{(n)}_{even}(z)\|_{\ast} \|\hat{\W}^{(n)}(z)\|_{\ast} \nonumber \\
 & \qquad\qquad\qquad + \|{\V}^{(n)}(z)\|_{\ast} \|\bE^{(n)}_{odd}(z)\|_{\ast} \|\hat{\W}^{(n)}(z)\|_{\ast} 
 + \|\hat{\cE}_L^{(n)}(z)\|_{\ast} . \label{eq:7.59b-bnd}
\end{align}
From the monotonicity property of the infinity norm, we know that $\|\bE_{\scriptscriptstyle{odd}}^{(n)}(z)\|_{\ast}$ and $\|\bE_{\scriptscriptstyle{even}}^{(n)}(z)\|_{\ast}$ are both less than or equal to $\|\bE^{(n)}(z)\|_{\ast}$ and that $\|\hat{\bE}_{\scriptscriptstyle{odd}}^{(n)}(z)\|_{\ast}$ and $\|\hat{\bE}_{\scriptscriptstyle{even}}^{(n)}(z)\|_{\ast}$ are both less than or equal to $\|\hat{\bE}^{(n)}(z)\|_{\ast}$.
By applying 
cyclic reduction 
to the function $\widetilde{\chi}(z)$ 
with respect to 
the corresponding sequences of matrix power series $\widetilde{\chi}^{(n)}(z)$ and $\hat{\widetilde{\chi}}^{(n)}(z)$, we further have that $\| \V^{(n)}(z) \|_{\ast}$, $\| \W^{(n)}(z) \|_{\ast}$ and $\| \hat{\W}^{(n)}(z) \|_{\ast}$ associated with Theorem~\ref{lem7.15} are upper bounded by $\theta \sigma^{2^n}$ for appropriate $\theta > 0$ and $0 < \sigma < 1$.
Upon substituting
both of these sets of upper bounds 
into~\eqref{eq:7.59a-bnd} and~\eqref{eq:7.59b-bnd}, we respectively obtain
\begin{align*}
\| \bE^{(n+1)}(z) \|_{\ast} & \; \dotleq \; \| \bE^{(n)}(z) \|_{\ast} + \theta \sigma^{2^n} \| \bE^{(n)}(z) \|_{\ast} + (\|\bE^{(n)}(z)\|_{\ast} + \theta \sigma^{2^n} \| \bE^{(n)}(z) \|_{\ast})\theta \sigma^{2^n} + \| \cE_L^{(n)}(z) \|_{\ast} \\
    & \qquad\qquad = ( 1 + \theta \sigma^{2^n} )^2 \| \bE^{(n)}(z) \|_{\ast} + \| \cE_L^{(n)}(z) \|_{\ast} , \\
\| \hat{\bE}^{(n+1)}(z) \|_{\ast} & \; \dotleq \; \| \hat{\bE}^{(n)}(z) \|_{\ast} + \theta \sigma^{2^n} \| \hat{\bE}^{(n)}(z) \|_{\ast} + (\| {\bE}^{(n)}(z) \|_{\ast} + \theta \sigma^{2^n} \| \bE^{(n)}(z) \|_{\ast} ) \theta \sigma^{2^n} + \| \hat{\cE}_L^{(n)}(z) \|_{\ast} \\
& \qquad\qquad = (1 + \theta \sigma^{2^n} ) \| \hat{\bE}^{(n)}(z) \|_{\ast} 
+ ( \theta \sigma^{2^n} + (\theta \sigma^{2^n})^2 ) \| {\bE}^{(n)}(z) \|_{\ast} + \| \hat{\cE}_L^{(n)}(z) \|_{\ast} .
\end{align*}
Under the suppositions of the theorem, we conclude 
\begin{align}
\|\bE^{(n+1)}(z)\|_{*} & \; \dotleq \; (1 + \theta\sigma^{2^n})^2 \| \bE^{(n)}(z) \|_{\ast} + \upsilon , \tag{\ref{eq:E-ub}} \\
   \|\hat{\bE}^{(n+1)}(z)\|_{*} & \; \dotleq \; (1 + \theta \sigma^{2^n} ) \| \hat{\bE}^{(n)}(z) \|_{\ast} 
+ ( \theta \sigma^{2^n} + (\theta \sigma^{2^n})^2 ) \| {\bE}^{(n)}(z) \|_{\ast} + \upsilon \nonumber \\
& \; \dotleq \; (1 + \theta\sigma^{2^n})^2 \| \bE^{(n)}(z) \|_{\ast} + \upsilon , \label{eq:Ehat-ub}
\end{align}
where the last inequality follows from an induction argument on the upper bounds for $\|\bE^{(n+1)}(z)\|_{*}$ and $\|\hat{\bE}^{(n+1)}(z)\|_{*}$.
In particular,
$\| \hat{\bE}^{(0)}(z) \|_{\ast} \, \dotleq \, \| \bE^{(0)}(z) \|_{\ast} = 0$ in~\eqref{eq:7.59a} and~\eqref{eq:7.59b} for the base case, and then we have for the induction step
\begin{align*}
   \|\hat{\bE}^{(n+1)}(z)\|_{*} & \; \dotleq \; (1 + \theta \sigma^{2^n} ) \| \hat{\bE}^{(n)}(z) \|_{\ast} 
+ ( \theta \sigma^{2^n} + (\theta \sigma^{2^n})^2 ) \| {\bE}^{(n)}(z) \|_{\ast} + \upsilon \\
& \; \dotleq \; (1 + \theta \sigma^{2^n} ) \| \bE^{(n)}(z) \|_{\ast} 
+ ( \theta \sigma^{2^n} + (\theta \sigma^{2^n})^2 ) \| {\bE}^{(n)}(z) \|_{\ast} + \upsilon 
.
\end{align*}

From~\eqref{eq:E-ub} and~\eqref{eq:Ehat-ub} together with $\gamma_n = (1 + \theta\sigma^{2^n})^2$, we obtain 
\begin{align}
\|\bE^{(n+1)}(z)\|_{*} & \; \dotleq \; \gamma_n \| \bE^{(n)}(z) \|_{\ast} + \upsilon , \nonumber \\
&\; \dotleq \; \upsilon (1+  \gamma_n+\gamma_{n-1}\gamma_n+\cdots + \gamma_2\gamma_3\cdots\gamma_{n-1}\gamma_n + \gamma_1\gamma_2\cdots\gamma_{n-1}\gamma_n) , \tag{\ref{eq:7.61a}} \\
   \|\hat{\bE}^{(n+1)}(z)\|_{*} & \; \dotleq \; \gamma_n \| \bE^{(n)}(z) \|_{\ast} + \upsilon \nonumber \\
&\; \dotleq \; \upsilon (1+  \gamma_n+\gamma_{n-1}\gamma_n+\cdots + \gamma_2\gamma_3\cdots\gamma_{n-1}\gamma_n + \gamma_1\gamma_2\cdots\gamma_{n-1}\gamma_n) \label{eq:7.61b} .
\end{align}
For the $(k+1)$st summand on the right hand side 
of~\eqref{eq:7.61a} and~\eqref{eq:7.61b},
we derive
\begin{align*}
    \log\left( \prod_{i=n-k+1}^n \gamma_i \right) & \; = \; \sum_{i=n-k+1}^n \log\left( \gamma_i \right) \; = \; \sum_{i=n-k+1}^n \log\left( (1 + \theta\sigma^{2^i})^2 \right) \\ & \; = \; 2 \sum_{i=n-k+1}^n \log\left(1 + \theta\sigma^{2^i} \right) \; \leq \;  
    2 \theta \sum_{i=n-k+1}^n \sigma^{2^i} \leq 2 \theta \frac{\sigma^2} {1-\sigma^2} ,
\end{align*}
and therefore~\eqref{eq:7.61a} and~\eqref{eq:7.61b} respectively render
\begin{equation*}
   \|\bE^{(n+1)}(z)\|_{*} \; \dotleq \; \upsilon \left(1 +  \sum_{k=1}^n \exp ( 2 \theta \frac{\sigma^2}{1-\sigma^2} )\right) 
\end{equation*}
and
\begin{equation*}
\| \hat{\bE}^{(n+1)}(z)\|_{*} \; \dotleq \; \upsilon \left(1 +  \sum_{k=1}^n \exp ( 2 \theta \frac{\sigma^2}{1-\sigma^2} )\right) ,
\end{equation*}
which yield the desired results.
\qed

\subsection{Proof of Theorem~\ref{thm:complexity}}
\label{app:thm:complexity}
%
Since Algorithm~\ref{algo:quantumCR} is common to the inner loop of both Algorithm~\ref{algo:quantum} and Algorithm~\ref{algo:quantum2}, we first
sum up
the asymptotic time complexity of \emph{Step}~$\mathbf{1}$ through \emph{Step}~$\mathbf{8}$ of Algorithm~\ref{algo:quantumCR}.
\emph{Step}~$\mathbf{1}$ encodes $N$ columns of the Toeplitz matrix $\T_3$ requiring $N\log N$ qubits, with each column comprising $\log N$ qubits operated on independently and in parallel.
Since each of the $\log N$ qubits form independent parallel quantum processors that need not be connected/entangled to each other, the corresponding columns are loaded in parallel.
The time complexity to prepare the initial state $\ket{\psi_0}$ is therefore given by $\tau_{\scriptscriptstyle load}$.
\emph{Step}~$\mathbf{2}$ applies QFT to this initial state to obtain $\ket{\psi_0^\prime} = \sum_{j=0}^{N-1}b_j\ket{j}$, having a time complexity of $O(\log^2 N)$.
\emph{Step}~$\mathbf{3}$ uses the oracle for the generating function $f$ that corresponds to a given Toeplitz matrix and prepares the state $\ket{f(2\pi j/N)}$, having a time complexity of $\tau_{\scriptscriptstyle oracle}$.
Then \emph{Step}~$\mathbf{3}$ computes $\sum_{j=0}^{N-1}b_j\ket{j}\ket{f_1(2\pi j/N)}$ in parallel on the parallel quantum processors for each column of $\T_3$, requiring an additional $O(1)$ time and an additional $\log N$ qubits.
By exploiting amplitude amplification, the inversion of a Toeplitz matrix in \emph{Steps}~$\mathbf{4}$ and~$\mathbf{5}$ each have a time complexity of $O(\mu)$.
\emph{Step}~$\mathbf{6}$ uses the oracle for the generating function $f_2$ and, for each column of $\T_3$, computes in parallel $\ket{\psi_2^\prime} = \sum_{j=0}^{N-1}\widetilde{b}_j\ket{j}\ket{f_2(2\pi j/N)}$ on the parallel quantum processors, having a time complexity of $\tau_{\scriptscriptstyle oracle}$.
\emph{Step}~$\mathbf{7}$ applies iQFT to the resulting state to obtain $\ket{\psi_2}$ with a time complexity of $O(\log^2 N)$.
\emph{Step}~$\mathbf{8}$ encodes $N$ columns of the Toeplitz matrix $\T_4$ as a quantum state, loading the columns of $\T_4$ as an amplitude addition to the current state $\ket{\psi_2}$, thus preparing the state $\ket{\psi^\ast} = \ket{\psi_2}+ \ket{\psi_3}$ with a time complexity of $\tau_{\scriptscriptstyle load}$.
This yields a total asymptotic time complexity of $O( \mu (\tau_{\scriptscriptstyle load} + \log^2 N + \tau_{\scriptscriptstyle oracle}) )$ for each quantum
cyclic reduction 
iteration~$n$, requiring $N(2\log N + 1)$ qubits.

Now, focusing first on Algorithm~\ref{algo:quantum} and similar to the above complexity analysis for Algorithm~\ref{algo:quantumCR}, 
the asymptotic time complexity of \emph{Step}~$\mathbf{3}$ of Algorithm~\ref{algo:quantum} is given by $O(\mu(\tau_{\scriptscriptstyle load} + \log^2 N))$ while reusing $M \log M$ qubits.
Due to the quadratic convergence of 
cyclic reduction
{along the lines of Theorem~\ref{thm7.13},}
and consistent with the 
analysis of 
{our quantum 
cyclic reduction 
algorithm}
given 
{in}
Theorem~\ref{lem7.15} and Theorem~\ref{thm:expUB},
the total number of iterations~$n$ of the inner loop (Algorithm~\ref{algo:quantumCR} and \emph{Step}~$\mathbf{3}$ of Algorithm~\ref{algo:quantum}) 
does not depend on 
the increasing sequence of $N$ up to the maximum $N^Q$.
The overall asymptotic time complexity of the inner loop then depends on the corresponding maximum numerical degrees $d_{\max}^Q$ of the matrix power series generated by this instance of quantum
cyclic reduction, 
and therefore this time complexity is given by $O( \mu (\tau_{\scriptscriptstyle load} + \log^2 N^Q + \tau_{\scriptscriptstyle oracle}) )$ and requires $N^Q (2\log N^Q + 1)$ qubits.
The asymptotic time complexity of \emph{Step}~$\mathbf{4}$ of Algorithm~\ref{algo:quantum} is relatedly given by $O(\mu(\tau_{\scriptscriptstyle load} + \log^2 M))$ while reusing $M \log M$ qubits.
Combining this with the time complexity for the inner loop and the time complexity for the result readout then yields the overall asymptotic time complexity of Algorithm~\ref{algo:quantum} given in~\eqref{eq:thm:complexity}.
Lastly, in the case of Algorithm~\ref{algo:quantum},
we have $d_{\max}^Q = O( d_{\max}^C )$ in~\eqref{eq:thm:complexity}
following from Theorem~\ref{lem7.15} and Theorem~\ref{thm:expUB}.

Turning next to Algorithm~\ref{algo:quantum2} and along the same lines as the above complexity analysis for Algorithm~\ref{algo:quantum}, the asymptotic time complexity of \emph{Step}~$\mathbf{4}$ of Algorithm~\ref{algo:quantum2} is given by $O(\tau_{\scriptscriptstyle load})$.
Due to the quadratic convergence of 
cyclic reduction
along the lines of Theorem~\ref{thm7.13}, 
and consistent with the 
analysis of 
{our quantum 
cyclic reduction 
algorithm}
given 
{in}
Theorem~\ref{lem7.15} and Theorem~\ref{thm:linUB},
the total number of iterations~$n$ of the inner loop (Algorithm~\ref{algo:quantumCR} and \emph{Step}~$\mathbf{4}$ of Algorithm~\ref{algo:quantum2}) 
does not depend on 
the increasing sequence of $N$ up to the maximum $N^Q$.
The overall asymptotic time complexity of the inner loop then depends on the corresponding maximum numerical degrees $d_{\max}^Q$ of the matrix power series generated by this instance of quantum
cyclic reduction, 
and therefore this time complexity is given by $O( \mu (\tau_{\scriptscriptstyle load} + \log^2 N^Q + \tau_{\scriptscriptstyle oracle}) )$ and requires $N^Q (2\log N^Q + 1)$ qubits.
The asymptotic time complexity of \emph{Step}~$\mathbf{5}$ of Algorithm~\ref{algo:quantum2} is relatedly given by $O(\mu(\tau_{\scriptscriptstyle load} + \log^2 M))$ while reusing $M \log M$ qubits.
Combining this with the time complexity for the inner loop and the time complexity for the result readout then yields the overall asymptotic time complexity of Algorithm~\ref{algo:quantum2} given in~\eqref{eq:thm:complexity}.
Lastly, in the case of Algorithm~\ref{algo:quantum2},
we have 
$d_{\max}^Q = O( d_{\max}^C )$ in~\eqref{eq:thm:complexity}
following from Theorem~\ref{lem7.15} and Theorem~\ref{thm:linUB}.

In terms of both Algorithm~\ref{algo:quantum} and Algorithm~\ref{algo:quantum2}, under the suppositions of the theorem
with respect to 
{$\mu$ and $\tau_{\scriptscriptstyle oracle}$ having complexity $O(\poly\log N^Q)$,}
the time complexity~\eqref{eq:thm:complexity}
{for the computation phase}
reduces to $O( \poly\log \normalsize (d_{\max}^Q \cdot M) )$.
Then,
in comparison with
the
time complexity for the
corresponding classical 
cyclic reduction 
algorithms 
in~\eqref{eq:thm:CRcomplexity},
the exponential speedup 
{for the computation phase}
follows from both 
$d_{\max}^Q = O( d_{\max}^C )$ in~\eqref{eq:thm:complexity}
and the time complexity
in~\eqref{eq:thm:complexity}.

In terms of both Algorithm~\ref{algo:quantum} and Algorithm~\ref{algo:quantum2}, under the 
{additional supposition}
of the theorem
with respect to 
{$\tau_{\scriptscriptstyle load}$
having complexity
$O(\poly\log N^Q)$ or subexponential in $(\log N^Q)$,}
the 
time complexity~\eqref{eq:thm:complexity} reduces to $O( \poly\log \normalsize (d_{\max}^Q \cdot M) )$
{or subexponential in $\log \normalsize (d_{\max}^Q \cdot M)$}.
Then, in comparison with
{the time complexity}
{for the
corresponding classical 
cyclic reduction 
algorithms}
in~\eqref{eq:thm:CRcomplexity}, the
{polynomial-to-exponential speedup}
follows from both 
$d_{\max}^Q = O( d_{\max}^C )$ in~\eqref{eq:thm:complexity}
and the time complexity
in~\eqref{eq:thm:complexity}.

In terms of both Algorithm~\ref{algo:quantum} and Algorithm~\ref{algo:quantum2}, under the 
{additional supposition}
of the theorem
with respect to 
{$\tau_{\scriptscriptstyle readout}$
having complexity $O(\poly\log N^Q)$ or subexponential in $(\log N^Q)$,}
the 
{overall}
time complexity~\eqref{eq:thm:complexity} reduces to $O( \poly\log \normalsize (d_{\max}^Q \cdot M) )$
{or subexponential in $\log \normalsize (d_{\max}^Q \cdot M)$}.
Then, in comparison with
{the overall time complexity}
{for the
corresponding classical 
cyclic reduction 
algorithms
in~\eqref{eq:thm:CRcomplexity}}, the 
{polynomial-to-exponential speedup}
follows from both 
$d_{\max}^Q = O( d_{\max}^C )$ in~\eqref{eq:thm:complexity}
and the time complexity
in~\eqref{eq:thm:complexity}.
\qed

\bibliographystyle{elsarticle-num} 
\bibliography{main}



\end{document}